\documentclass[amsmath,amsfonts,amssymb,aps,
               prfluids,floatfix,onecolumn, 
               10pt]{revtex4-2}

\usepackage[T1]{fontenc}

\usepackage{stix}
\usepackage[text={142mm,215mm},b5paper]{geometry}

\usepackage{graphicx}
\usepackage{siunitx}

\usepackage{hyperref}

\begin{document}

\title{Kolmogorov cascade as the governing mechanism for intervortex spacing in quantum turbulence
}

\author{Clément Bret}
\author{Pantxo Diribarne}
\email{pantxo.diribarne@univ-grenoble-alpes.fr}
\author{Jérôme Duplat}
\author{Bernard Rousset}
\affiliation{dSBT/IRIG CEA, Universit\'{e} Grenoble Alpes - F-38054 Grenoble, France}

\date{\today}

\begin{abstract}
In this paper, we investigate inertially forced isothermal quantum
turbulence (the coflow of normal and superfluid components) at
temperatures of 1.6 and \SI{2}{K}. The experiments are carried out  
in a large optical cryostat, where quasi-isotropic, homogeneous
turbulence is generated using a  double oscillating grid. Turbulence
intensity is tuned by varying the grid stroke and frequency. 
The flow is probed via two-dimensional (2D) and three-dimensional (3D)
reconstructions of quasi-isodense microsphere trajectories, from 
which we extract the large-scale properties of the flow in the fully
turbulent regime for different Reynolds numbers: the turbulent
velocity fluctuations, the energy transfer rate, and the integral length scale.
Additionally, we determine the mean vortex line density $\mathcal{L}$
via attenuation of a second sound standing wave across the entire
measurement volume. Our results confirm with an improved accuracy that
the intervortex spacing $\ell =1/\sqrt{\mathcal{L}}$ scales with the
Reynolds number $\text{Re}_\kappa$  (based on the quantum of
circulation $\kappa$) as  $\ell\propto \text{Re}_\kappa^{-3/4}$, with
a well-defined numerical prefactor and no observed temperature
dependence. This scaling recalls that of the dissipative length scale
in  classical Kolmogorov (K41) turbulence and it lead previous authors
to interpret $\ell$ as an effective dissipation length scale. However,
in our temperature range, this interpretation is not consistent with
the apparent temperature independence of the prefactor. Based on those
arguments, we propose an alternative interpretation that suggests that
the inter-vortex distance in coflow turbulence is the consequence of
the quantum restricted depth of the superfluid component energy
cascade.

\end{abstract}

\maketitle

\section{\label{sec:intro} Inertially Forced Quantum versus Classical Turbulence}

Since the pioneering experiments of the 1920s \cite{Dana26,Wolfke27},
\textsuperscript{4}He has been known to exist in two distinct liquid
phases below and above the so-called $\lambda$-transition ($T_\lambda
= \SI{2.1768}{K}$ at atmospheric pressure), referred to as He~I and
He~II by \textcite{Keesom27}, due to their very different
thermodynamic properties.  In the 1930s, \textcite{Kapitza38} dubbed
the He~II phase a ``superfluid'' because of its ability to flow without
viscosity through microscopic interstices, exhibiting behavior that is
neither laminar nor turbulent, as observed by \textcite{Allen38} in
thin capillaries. These peculiar observations combined with insights
from the Bose-Einstein condensation theory motivated theorists such as
\textcite{London38}, \textcite{Tisza_two_fluids_38}, and
\textcite{Landau_two_fluids_41} to develop the so-called two-fluid
model in the 1940s. This framework describes He~II as a mixture of two
interpenetrating components: a viscous normal component carrying all
entropy and a superfluid component flowing without any viscosity. The
proportion of each component evolves with temperature, with the
superfluid fraction increasing as the temperature decreases toward
\SI{0}{K}.  Further experimental evidence from \textcite{Osborne50} in
the 1950s demonstrated that, despite being free of viscosity, the
superfluid component can sustain circulation. This paradox was
resolved by introducing quantum vortex lines, first proposed by
\textcite{Onsager1949} and later developed by
\textcite{Feynman55}. These singular ``topological defects'' in the
superfluid velocity field can be viewed as Ångström-scale vortex
filaments with quantized circulation, $\kappa = h/m \approx
\SI{9.97e-8}{m^2/s}$, where $h$ is Planck's constant and $m$ is the
atomic mass of \textsuperscript{4}He. Under inertial forcing, where
both components are synchronously driven at large scales by pressure
gradients or mechanical stirring (e.g., in a von Kármán flow), He-II
can enter a coflow Quantum Turbulent state. In this regime, vortex
lines form a tangled structure that serves as the backbone of the
superfluid velocity field, while the normal component organizes into
classical multi-scale eddies in the manner of Richardson. Notably, these
vortex lines also mediate the coupling between the normal and
superfluid velocity fields: as they move relative to the normal
component, they experience a ``mutual friction force'' allowing for
momentum exchange as originally described by \textcite{Hall56a,
Hall56b} and \textcite{Bekarevich61} in the 1950s. More recently,
experiments were conducted to characterize the differences between classical and
quantum turbulent flows. In the late 1990s, driven by applications in
large-scale refrigeration (such as superconducting magnets for fusion
or particle accelerators), measurements of pressure drop in He~II at
high Reynolds numbers (greater than $10^6$) were conducted in pipes
with centimeter-scale diameters and lengths spanning several meters
(see, e.g., Refs.~\cite{Walstrom88a, Rousset94, Fuzier01}). These studies
revealed that, under these conditions, turbulent He~II flow behaves
similarly to classical fluids. The pressure drop through the pipe,
$\Delta P \propto v^2$, holds for velocities $v$ typically determined
from the mass flow rate passing through the pipe and the associated
friction coefficients resemble those found for classical flows. As
mentioned previously, note that this is not the case for He~II flow in
thin capillary \cite{Allen38}.  Global calorimetric measurements in
the \textit{Superfluid High REynolds von Kármán experiment} \cite{Rousset14} further
emphasized this similarity, demonstrating that the energy injected at
large scales is identical in He~I and He~II \cite{SaintMichel14_Kp}.
Local velocity measurements using total head pressure tubes
in various flow configurations
also showed that the $E(k) \propto k^{-5/3}$ energy
spectrum and the $4/5$-th law can be applied in He~II up to
intermediate scales \cite{Maurer98, Salort10, Salort12}.
This confirms the existence of an inertial range
and suggests a locking between the two velocity fields,
$\mathbf{v}_n$ and $\mathbf{v}_s$, within these scales. This locking
argument is further supported by numerical simulations \cite{Roche2009, Salort_2011_equipartition} and theoretical
considerations \cite{Vinen2002, Barenghi14}. The origin of
this coupling is a direct consequence of the mutual friction force
$\mathbf{f}_{mf}$, which tends to smooth out the velocity differences
between the superfluid and normal components. However, the velocity fields $\mathbf{v}_n$
and $\mathbf{v}_s$ cannot be perfectly ``locked'' because, at scales
smaller than the mean inter-vortex distance $\ell$, where nthe superfluid
velocity field is primarily governed by the nearest quantum vortex
lines. Moreover, only the normal component experiences viscous
dissipation. Thus, at some sufficiently small scale He~I and He~II 
turbulence should behave differently and mutual friction may be
responsible for a part of the dissipation. The latter could be
expected, at least within the temperature range from $\SI{2.15}{K}$ to
$\SI{1}{K}$, where both the normal and superfluid components
constitute more than 1\% of the total fluid density. 
Determining the decoupling scale and the effective dissipation
mechanisms below it remains an open question. A promising approach
to addressing this challenge  likely lies in developing  a deeper understanding of
the so-called effective viscosity $\nu_{\text{eff}}$ of quantum
turbulence. This concept was first introduced by \textcite{Stalp99} in
the 1990s to explain the observed classical-like scaling, $\mathcal{L}
\propto t^{-3/2}$, in the decay of the mean vortex line density,
$\mathcal{L}$, as a function of the time $t$ after which a grid was towed in a finite channel. Associating
$\langle (\kappa \mathcal{L})^2 \rangle$ to the variance of the
coarse-grained vorticity field of the superfluid component
$\langle \omega_s^2\rangle\approx \langle \kappa \mathcal{L}^2\rangle$, the authors drew an
analogy to classical turbulence to express the energy dissipation rate
$\epsilon$ as
\begin{equation} \label{nu_eff} \epsilon = \nu_{\text{eff}} \langle
(\kappa \mathcal{L})^2 \rangle.
\end{equation} By further assuming that after a brief initial period,
the integral length scale $L$ of the flow saturates at the channel
diameter $D$, so that the spatial energy distribution follows a
classical Kolmogorov spectrum, $E(k) = C \epsilon^{2/3} k^{-5/3}$ where $C\approx1.6$ is
the Kolmogorov constant, up
to the scale $k = 2\pi/L=2\pi/D$, they finally derived the decay of the
mean vortex line density,
\begin{equation} \mathcal{L}(t) = \frac{(3C)^{3/2}D}{2\pi \kappa
\sqrt{\nu_{\text{eff}}}} t^{-3/2}.
\end{equation} 

These considerations provide a straightforward
framework that effectively matches the experimental observations. Several
attempts have been made to measure $\nu_{\text{eff}}$ and
its temperature dependence in the original towed-grid configuration
\cite{Oregon_ill_grid, Oregon_good_grid},
in a similar configuration where the temporal decay was studied behind a fixed grid \cite{Babuin2013}, 
or in a spin-down experiment at lower temperatures \cite{Walmsley14}. Although the
reported values are in reasonable agreement giving credit to the
approach proposed by \textcite{Stalp99}, Eq.~(\ref{nu_eff}) lacks a
theoretical justification and it remains unclear which dissipation
mechanisms are captured by $\nu_{\text{eff}}$.

In a more recent study, \textcite{Babuin14} inferred
$\nu_{\text{eff}}$ directly from Eq.~(\ref{nu_eff}) in a steady-state
regime. They measured the mean vortex line density $\mathcal{L}$ in a
channel behind a grid (with their sensor probing a region between 16
and 31 mesh sizes downstream of the grid) or directly in the channel
without the grid as a function of the mean velocity $V$ imposed in the
channel. As they did not directly measure the mean dissipation rate
$\epsilon$, they estimated its value indirectly using the classical
relation $\epsilon = u_{\text{rms}}^3 / L$. For the grid
configuration, to extrapolate the turbulent velocity fluctuation from
the mean velocity $V$, they assumed a turbulence intensity of $\tau =
u_{\text{rms}} / V = 9\%$ and they inferred an integral scale of $L =
M$, where $M$ is the grid mesh size. For the no-grid configuration,
they proposed $\tau = 5\%$ and $L = d/2$, where $d$ is the
diameter of the capillaries in the honeycomb-type flow conditioner
placed at the channel inlet. Across all tested temperatures from
\SI{1.18}{K} to \SI{2.16}{K}, they did observe the 
$\mathcal{L}\propto V^{3/2}$ scaling predicted by Eq.~(\ref{nu_eff}) and 
were able to extract a value for
$\nu_{\text{eff}}$. Note that an equivalent formulation of
Eq.~(\ref{nu_eff}) was proposed by
\textcite{Salort_2011_equipartition}:
\begin{equation}\label{delta_L} \frac{\ell}{L} = \beta
\mathrm{Re}_\kappa^{-3/4},
\end{equation} where $\mathrm{Re}_\kappa = u_{\text{rms}}L/\kappa$ is the Reynolds number based on the quantum of circulation $\kappa$, and $\beta$ is related to $\nu_{\text{eff}}$ by
$\nu_{\text{eff}} = \beta^4 \kappa$. The formulation of
Eq.~(\ref{delta_L}) clearly resembles the classical Kolmogorov
relation for the dissipative scale $\eta\propto \text{Re}^{-3/4}$. Based on these
similarities, it is tempting to associate $\ell$ to the dissipative
scale of quantum turbulence and see $\nu_{\text{eff}}$ as the
effective viscosity arising from mutual friction. Assuming that mutual
friction is responsible for the entire dissipation (i.e. no dissipation is due to classical viscosity associated with normal component fraction),
\textcite{Babuin14} proposed the following relation,
\begin{equation}\label{epsilon_mf}
\epsilon=\epsilon_{mf}=\frac{\langle \mathbf{f}_{mf} \cdot
\mathbf{v}_{ns} \rangle}{\rho}=\frac{B}{2}\frac{\rho_s
\rho_n}{\rho^2}(\kappa \mathcal{L})\langle \mathbf{v}_{ns}^{2}
\rangle,
\end{equation} where $B$ is the temperature-dependent mutual friction
parameter, $\rho_s$ and $\rho_n$ are the superfluid and normal
component densities, respectively, and $\rho = \rho_s + \rho_n$ is the
total bulk density.  Assuming that $\langle \mathbf{v}_{ns}^{2}
\rangle$ can be approximated by the classical relation $\langle
\mathbf{v}_{ns}^{2} \rangle = \gamma \kappa^2 / \ell^2$ (where
$\gamma$ is an adjustable dimensionless coefficient), an analytical
expression for $\nu_{\text{eff}}$ (or equivalently $\beta$) follows:
\begin{equation}\label{nu_eff_babuin_f_mf} \nu_{\text{eff}} = \gamma
\frac{B}{2} \frac{\rho_s \rho_n}{\rho^2} \kappa.
\end{equation} 

As the measured values of $\nu_{\text{eff}}$ exhibited
little to no dependence on temperature within the explored range (1.18
to $\SI{2.16}{K}$), it appears that Eq.~(\ref{nu_eff_babuin_f_mf})
does not accurately describe the data. Furthermore, the indirect
estimation of $\epsilon$ introduced an uncertainty of about a factor
of 10 in $\nu_{\text{eff}}$, making any quantitative discussion
relatively delicate. 

From a broader perspective, an examination of the experimental dataset
for $\nu_{\text{eff}}$ in the literature -- derived from both decaying
and steady-state turbulence experiments -- reveals that, while the values
remain within the same order of magnitude, the data appear relatively
scattered (by a factor of $\sim 3$ to 4). A more
accurate and well-characterized experimental determination of
$\nu_{\text{eff}}$, free from the ambiguities related to large-scale
properties, would be a crucial milestone for both gaining a deeper
understanding of the origin of Eq.~(\ref{nu_eff}) and testing the
validity of Eq.~(\ref{nu_eff_babuin_f_mf}). Additionally, such
measurements could contribute to the validation of numerical models.

This paper presents  what we believe to be the first direct experimental measurement of
$\nu_{\text{eff}}$ in a stationary oscillating-grid turbulent flow at temperatures of $1.6$ and 
\SI{2}{K}. By ``direct measurement,'' we refer to the independent evaluation of large-scale flow 
properties, from which we extract both the mean energy dissipation rate $\epsilon$ and the mean 
vortex line density $\mathcal{L}$. We show that these measurements agree quantitatively with the 
traditional scalings of the K41 turbulent cascade. 
This  provides insights into the energy cascade and dissipation mechanisms, leading us to propose 
a new paradigm and adaptation of Richardson's cascade for He~II turbulent flow.

\section{\label{sec:expsetup}Experimental setup}

The measurements presented in this article were conducted within the
Oscillating Grid Experiment in Superfluid (OGRES) experiment, which
consists of a cylindrical cryostat with optical access
windows. Compared to other experiments that study He~II hydrodynamics
through direct visualization \cite{Guo14, Svancara17, Peretti23}, this
setup is relatively large, with the generated flow involving
approximately \SI{5}{L} of liquid helium. The setup has been described
in previous works \cite{Rousset15,Sy16_phd,Sy21_article}, however, due
to recent substantial modifications, this section presents the updated
version along with the operational procedure developed to
maintain stable thermodynamic conditions in both the He~I and He~II
phases.

\subsection{New OGRES setup}

\begin{figure}[ht!]  \centering
\includegraphics[width=\linewidth]{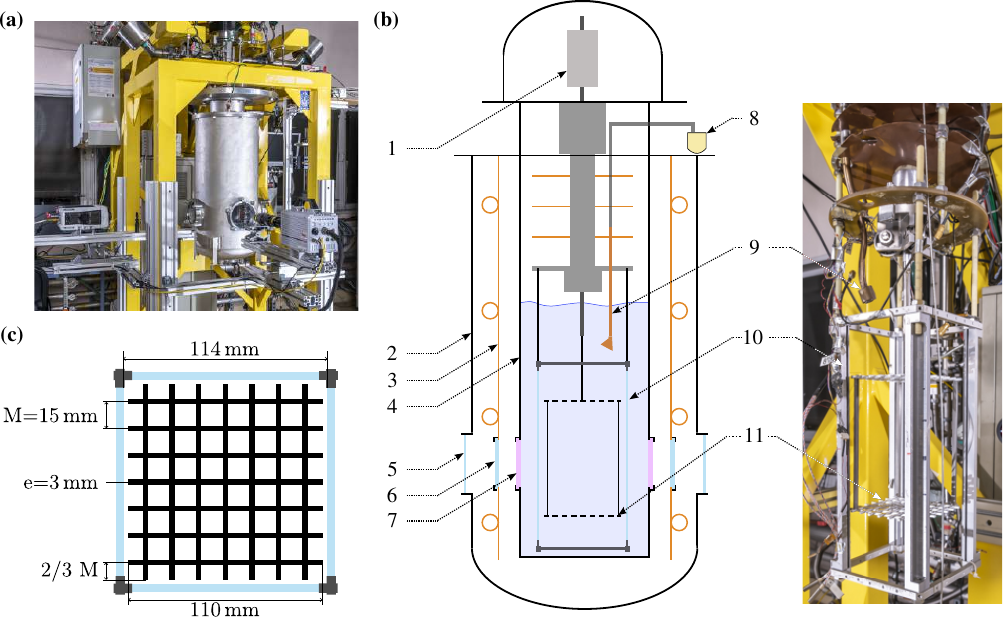}
  \caption{(a) Global view of the updated OGRES experiment. (b)
Detailed schematic: 1- linear motor, 2- outer stainless-steel vessel,
3- copper thermal shield with liquid nitrogen cooling coil, 4- inner
stainless-steel vessel, 5- circular borosilicate window, 6- square
borosilicate window, 7- square sapphire window, 8- seeding particle
reservoir, 9- injector nozzle, 10- confinement tunnel, 11- double
grid. (c) Detailed grid geometry. }
  \label{fig:cryostat}
\end{figure}

As shown in Figs.~\ref{fig:cryostat}(a) and \ref{fig:cryostat}(b), the
cryostat consists 
of two nested stainless-steel vessels (item 2 and 4 in
Fig.~\ref{fig:cryostat}(b)) separated by an insulating vacuum gap. To
reduce radiative heat transfer, a copper thermal shield (item 3) -- wrapped 
in multi-layer insulation (MLI) and cooled to approximately
\SI{100}{K} using liquid nitrogen -- is placed within the vacuum
gap. Since the primary objective of the experiment is to directly
visualize the flow inside the cryostat, each of the three concentric vessels is
fitted with four windows (item 5, 6 and 7), each approximately
\SI{10}{cm} $\times$ \SI{10}{cm} in size, positioned at 90-degree
intervals around the perimeter.

The inner vessel (item 4) has a diameter of \SI{210}{mm} and can hold
up to approximately \SI{600}{mm} of liquid helium when filled to about
two-thirds of its capacity. The temperature of the helium bath can
then be adjusted by controlling the pressure inside the vessel,
allowing for measurements both above and below the
$\lambda$-transition.

The flow under investigation is generated in a \SI{300}{mm} high
confinement tunnel (item 10) with a square cross-section of
\SI{114}{mm} $\times$ \SI{114}{mm}, using an oscillating double-grid
(item 11). The grid geometry, illustrated in
Fig.~\ref{fig:cryostat}(b), is the same as the one utilized in the
previous studies \cite{Sy16_phd, Sy21_article}. The mesh size is $M =
\SI{15}{mm}$, and the grids consist of a $6 \times 6$ matrix of
meshes, with an additional 2/3 mesh at the edges. The bars have a
square cross-section of \SI{3}{mm} thickness, resulting in a solid
fraction of 34.5\%. The two grids are separated by a distance
$H=\SI{167}{mm}$. This configuration, as reported in the literature
\cite{Srdic96, Shy97}, is known to enhance the homogeneity of
turbulence properties in the direction perpendicular to the grids. For
sufficiently large grid spacing (i.e. $H/M > 5$), \textcite{Shy97}
observed that the turbulent velocity fluctuations became homogeneous
over a band approximately $H/2$ wide, centered on the midplane.

The grids are oscillated using a linear motor, which
allows for continuous adjustment of the stroke ($S_{\text{mot}} \in
[0, 32]\,\si{mm}$) and frequency ($f_{\text{mot}} \in [0, 12]\,\si{Hz}$),
leading to a wide range of forcing intensities within a single experimental
campaign ($u_{\text{rms}}$ up to \SI{30}{mm/s}).

\subsection{Operating procedure}

In the He~I phase, temperature
gradients driven by heat inputs (of order \SI{2}{W}) can easily
exceed the spinodal temperature due to the poor thermal conductivity
of He~I, leading to nucleate boiling at the inner vessel
walls. to perform undisturbed measurements, it is thus essential to
subcool the liquid in the confinement tunnel to prevent boiling.
To ensure the reproducibility of our measurements, a
systematic procedure was developed: first, the vessel pressure is
reduced to \SI{52}{mbar}, cooling the bath to a uniform
$T\approx\SI{2.2}{K}$. Once this temperature is reached, the pressure is
rapidly increased to \SI{100}{mbar}, effectively stopping the boiling
in the core of the bath. This allows for several minutes of turbulence
measurements, during which the temperature drift remains
lower than $\SI{0.3}{K}$, ensuring that relative variations in density and 
dynamic viscosity are kept under 1\%. Once the measurements are completed, the pumping cycle is repeated. During each recooling phase, the liquid level gradually drops. The cryostat is refilled as soon as the liquid height approaches the top of the confinement tunnel.

In the He~II phase, as observed nearly a century ago \cite{McLennan32}, 
the bath does not boil below the $\lambda$-transition
because of the very high apparent thermal conductivity of superfluid helium. Consequently, 
turbulence measurements can be performed
directly at saturation temperature. The pressure is adjusted to achieve the desired temperature
(\SI{1.6}{K} or \SI{2}{K} in the reported measurements), and the bath
pressure regulation system is activated to ensure temperature
stability, which is essential for the readability of second sound
measurements (see Sec.~\ref{sec:2S_calib} for a detailed discussion).

\section{\label{sec:metrology}Measurement systems}

We adopt two independent and complementary approaches to obtain a
complete description of the properties of the flow. The first approach
involves visualization systems that capture trajectories of particles
seeded in the flow. Two distinct systems are employed: one
reconstructs 3D trajectories, while the other only reconstructs $2
\times 2D$ trajectories. The second approach, applied exclusively in
He~II, analyzes the attenuation of a stationary second sound wave as
it interacts with quantum vortex lines present in the flow. In this
section, we present the characterization of the particle properties
(size and density), followed by an overview of the two visualization
systems, and finally, a description of the second sound resonant
cavity, from its fundamental principles to its effective operating
procedure.

\subsection{\label{sec:lagrangian_traj}Particle tracking velocimetry}

\subsubsection{Particle characteristics and injection system}

To seed the flow with small, monodisperse tracers, a new technique is
currently being developed in the laboratory to generate solid
spherical deuterium particles, a few microns in diameter, by dropping
liquid droplets above the bath \cite{Bret2023}. For this
study, however, we opted for a simpler alternative: microspheres with
a hollow core and a glass shell. The ones we use, supplied by
Microsphere Technology Ltd., have a specified density of
\SI{140}{kg/m^3} and diameters ranging from 75 to \SI{90}{\micro
m}. To eliminate any potential uncertainties, we conducted direct
measurements to verify these values.

\begin{figure}[ht!]  \centering
\includegraphics[width=\linewidth]{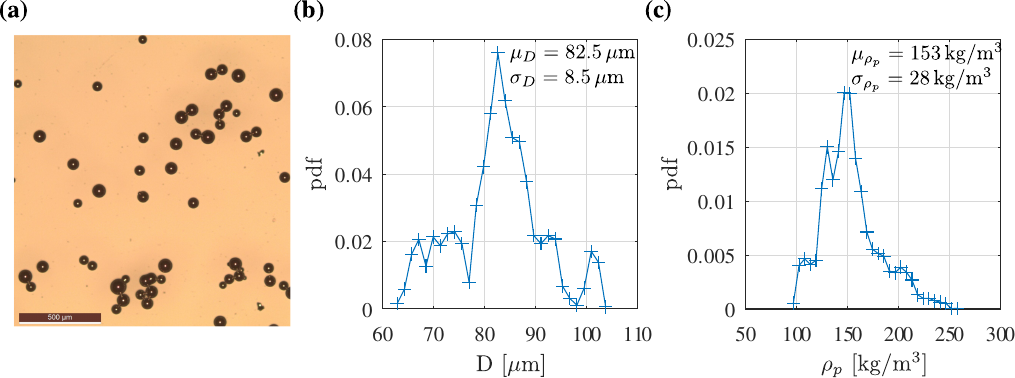}
  \caption{(a) Optical microscope image (top view). Illumination from
above creates a small white dot on the top of each particle due to
light reflection.(b) Distribution of particle diameters.(c)
Distribution of particle densities, deduced from
Eq.~(\ref{v_stokes}). }
  \label{fig:carac_part}
\end{figure}

To assess the properties of the particles, we first examined their
shape under an optical microscope to qualitatively confirm their
sphericity [see Fig.~\ref{fig:carac_part}(a)]. To further validate
their reported properties, we measured both the diameter and the
terminal velocity of a few thousand particles falling freely in an air
cavity. The diameter distribution is shown in
Fig.~\ref{fig:carac_part}(b) and matches relatively well with the range
announced by the manufacturer (we found a mean diameter $\mu_D =
\SI{82.5}{\micro m}$ and a standard deviation $\sigma_D =
\SI{8.5}{\micro m}$).  According to Stokes' law -- assuming that the
flow around the falling particles is laminar -- the sedimentation
velocity $v_{\text{sed}}$ for a particle of density $\rho_p$ falling through
a fluid at rest of density $\rho_f$ and dynamic viscosity $\mu_f$ is
given by:
\begin{equation}\label{v_stokes} v_{\text{sed}} = \frac{1}{18} \frac{(\rho_p
- \rho_f) D^2 g}{\mu_f},
\end{equation} where $g$ is the gravity acceleration. This relationship
allowed us to deduce the particles densities from their
terminal velocity and diameter. The results are reported in
Fig.~\ref{fig:carac_part}(c) and yield a particle density distribution
with a mean of $\mu_{\rho_p} = \SI{153}{kg/m^3}$ and a relatively
large standard deviation of $\sigma_{\rho_p} = \SI{28}{kg/m^3}$.

To seed the particles into the cryogenic helium flow, we designed a
specialized setup (see Fig.~\ref{fig:cryostat}(b)). It consists of an
independent circuit with a \SI{20}{\centi\liter} particle reservoir,
which is connected to a conical nozzle located directly above the
confinement tunnel. The reservoir can be isolated from the cryostat to
be filled with particles and then connected alternatively to a vacuum
pump and a pressurized helium gas bottle to rinse the particles,
ensuring a clean mixture of helium gas and particles without air. Once
prepared, the mixture is pressurized to \SI{300}{\milli bar} above the
bath pressure and injected into the flow by opening an
electromechanical valve. This setup addresses the clogging issues that
limited previous injection methods used in Ref.~\cite{Sy16_phd}.
Furthermore, it allows the particle reservoir to be
refilled without interrupting the experiment and improves
reproducibility by enabling synchronized data acquisition (e.g.,
imaging and second sound measurements) with particle injection.

In practice, after the injection of the particles, we typically wait for about \SI{30}{s}
to let sedimentation remove the heaviest and lightest
particles from the measurement volume.
However, despite those precautions, we observe that the residual
characteristic sedimentation velocity is non-negligible (of the order of a few millimeters per second), and may vary slightly between runs
(for more details refer to
Ref.~\cite{Bret24_PHD}). Therefore, special care must be taken when analyzing
particle motion and velocity along the vertical axis as gravity
introduces a bias.

\subsubsection{\label{sec:Visu3D} 3D visualization system }

\begin{figure}[ht!]  \centering
\includegraphics[width=\linewidth]{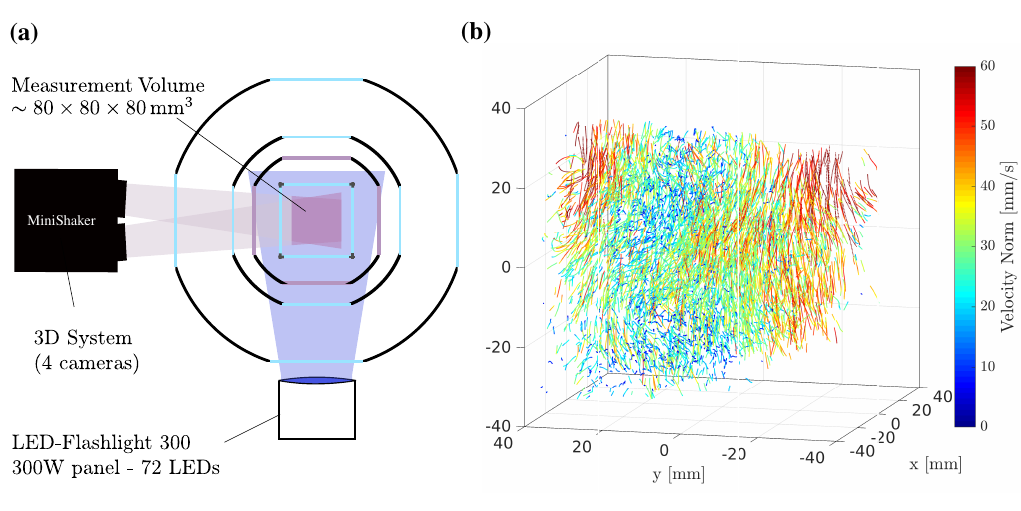}
  \caption{(a) Top view of the 3D setup. An estimate of the
measurement volume at the center of the confinement tunnel is
represented by the intersection of the two visualization beams of the
cameras. (b) Example of reconstructed 3D turbulent trajectories (He~II at \SI{2}{K}). The
norm of the velocity, represented in color, is calculated by
convolution with a Gaussian derivative kernel with a characteristic
time of four time steps ($\sim \SI{18}{ms}$).}
  \label{fig:VisuSystem3D}
\end{figure}

The system used to capture and reconstruct the three-dimensional
trajectories of particles in the flow is the LaVision
$\text{``MiniShaker.''}$ It consists of a box of four pre-aligned cameras
and a high-power LED panel to illuminate the particles within the flow
[see Fig.~\ref{fig:VisuSystem3D}(a)]. The measurement volume is
approximately $80\times80\times\SI{80}{mm^3}$.

Each camera has a resolution of $1216\times \SI{1000}{px^2}$ sensor,
leading to a spatial resolution of \SI{64.4}{\micro m/px}. The frame
rate is set to \SI{220}{Hz} and the exposure time is adjusted to
\SI{200}{\micro s} to avoid motion blur for particles moving at
velocities of up to \SI{100}{mm/s} (displacement $<\SI{0.3}{px}$ during the exposure time). The system can store up to $29
000$ images per run, providing over two minutes of data acquisition,
which is enough to capture the relevant large scale flow dynamics,
considering the flow's characteristic time scale is of the order of one
second.

Calibration is performed using a 3D target placed in the measurement
volume. Each of the four cameras captured images of
the calibration plate, allowing the LaVision software to initialize
the optical transfer function (OTF) and define a unified coordinate
system. Due to spatial constraints, calibration is conducted only at
room temperature. To account for the minor refractive index
differences between air and liquid helium, we take advantage of the
auto-calibration step which iteratively refines the OTF by detecting
the 3D particle positions directly from flow images and
back-projecting the centroid positions onto the two-dimensional (2D)
images, reducing error to a standard deviation of $0.2$ to $\SI{0.3}{px}$.

Particle trajectories are then reconstructed using the ``Shake The Box''
method, which allows us to obtain trajectories for four
runs in He~II at \SI{2}{K}. These trajectories span from 20 time steps
to over 2000, with an average of 50 time steps per trajectory,
capturing around 7000 particles within the 3D measurement volume at
each time step [see Fig.~\ref{fig:VisuSystem3D}(b)].

\subsubsection{\label{sec:Visu2x2D} 2\texorpdfstring{$\times$}{x}2D
visualization system }

\begin{figure}[ht!]  \centering 
\includegraphics[width=\textwidth]{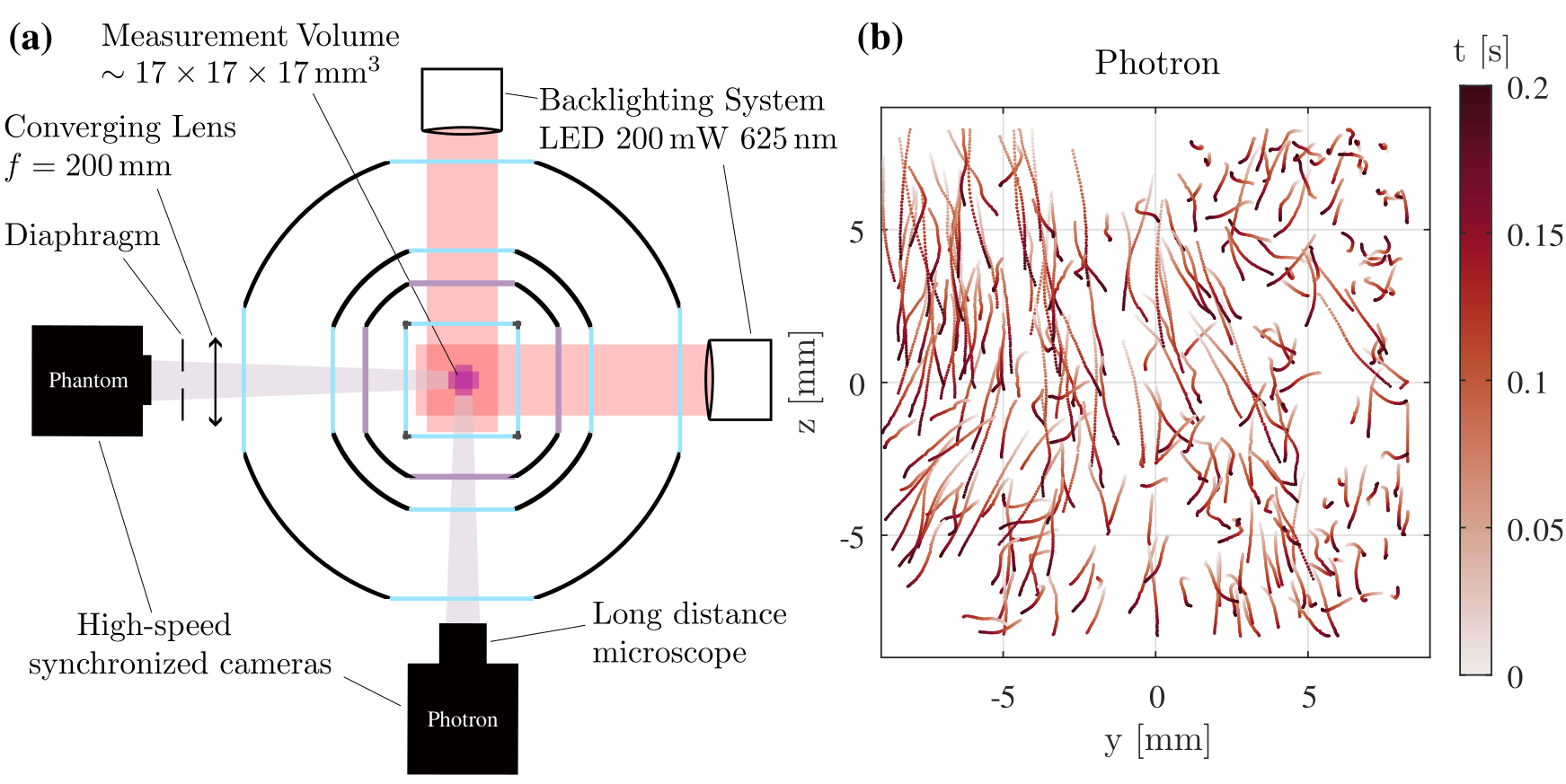}
  \caption{(a) Top view of the 2$\times$2D setup. An estimate of the
measurement volume at the center of the confinement tunnel is
represented by the intersection of the two visualization beams of the
cameras. (b) Example of reconstructed 2D turbulent trajectories. }
  \label{fig:VisuSystem2x2D}
\end{figure}

The second system captures 2$\times$2D particle
trajectories at higher temporal and spatial resolution using two
synchronized high-speed cameras positioned orthogonally (see
Fig.~\ref{fig:VisuSystem2x2D}(a)).

The Phantom V311 is operated without a standard camera lens. Instead,
a plano-convex converging lens with a focal length of \SI{200}{mm} is
placed at an appropriate distance to achieve the desired
magnification. The Photron Fastcam SA5, on the other hand, is equipped
with a long-distance microscope (Infinity Model
K2/SC). Note that a key constraint is ensuring
that the optics' working distance is sufficient to keep them outside
the cryostat.  Both cameras are set to an 800$\times$\SI{800}{px^2}
resolution, yielding a field of view of approximately
17$\times$\SI{17}{mm^2}. Compared to the previous system, lighting is
provided by directly illuminating the camera with
\SI{625}{nm}-red LEDs (M625L4-C4) in a back-light configuration. In
this setup, particles within the depth of field shadow the light,
appearing as sharp black circles on a white background. Note that
the LED power required is significantly lower than that of the 3D
system ($\sim 1000$ times less for a shutter ten times shorter).

This configuration introduces the challenge of determining
whether a particle lies within the depth of field. As a particle moves
out of the core of the depth of field, it gradually shifts from a
sharp black disk to a larger blurred gray disk before fully
disappearing. This gradual defocusing introduces ambiguity. To address
this, we developed an objective method based on \textcite{Helmli01}'s
``Mean Method'' to measure the cameras effective depth of field (for
more details French-speaking readers can refer to
Ref.~\cite{Bret24_PHD}). The diaphragm aperture is adjusted to
achieve a depth of field of approximately \SI{30}{mm}.

At room temperature, to center the fields of view of the two cameras
and measured the scale factors, we suspended a \SI{6}{mm} calibrated
sphere inside the cryostat using the sealing port for the helium
transfer. The sphere is placed at the center of the confinement tunnel
and the cameras and lights are adjusted to focus on it. After fully
opening the apertures to minimize depth of field, the calibration
images are captured. The apertures are then closed back to obtain appropriate depths of field and the sphere
is removed. The resulting scale factors are approximately
\SI{20.4}{\micro m/px} for the Phantom V311 and \SI{20.7}{\micro m/px}
for the Photron Fastcam SA5.

During turbulence measurement, images are captured at a high frame
rate of \SI{3000}{Hz}. To prevent the cameras memory from saturating
in just 4 seconds, the data acquisition is divided into 19 segments of
600 images, each separated by approximately \SI{2}{s}, with the aim of
capturing distinct realizations of the flow. The exposure time is set
to \SI{20}{\micro s} to avoid motion blur for particles moving at
velocities of up to \SI{100}{mm/s} (displacement $<\SI{0.1}{px}$ during the exposure time).

To reconstruct the trajectories from the images, we employ a classic
particle tracking algorithm, which typically consists of a detection
phase followed by a predictive association phase. We obtain 36 runs in
He~I, He~II (\SI{2}{K}), and He~II
(\SI{1.6}{K}). Figure~\ref{fig:VisuSystem2x2D}(b) shows a selection of
long trajectories reconstructed from a segment of a He~I run.

\subsection{\label{sec:secsoundcavity}Second sound resonant cavity}

\subsubsection{\label{sec:2S_principle}Principle}

Deriving the propagation of an isobaric and isochoric temperature
perturbation through an isotropic vortex tangle with a mean vortex
line density $\mathcal{L}$ in the so-called acoustic approximation, we
obtain the following equation (see appendix of Ref.~\cite{Babuin12} for a
step by step derivation):
\begin{equation}\label{2S_propagation} \frac{\partial^2 T}{\partial
t^2} = c_2^2 \nabla^2 T - 2 \alpha_{\mathcal{L}} c_2 \frac{\partial T}{\partial t}
\end{equation} where
\begin{equation}\label{alpha_vld} c_2 =
\sqrt{\frac{\rho_{s,0}}{\rho_{n,0}} \frac{\sigma_0}{C_p}}
\sqrt{\sigma_0 T_0} \quad \text{and} \quad \alpha_{\mathcal{L}} =
\frac{B \kappa}{6 c_2}
\mathcal{L}.
\end{equation}

Equation~(\ref{2S_propagation}) can be read as a d'Alembert equation
with a damping term $\alpha_\mathcal{L}$, indicating that the temperature perturbation propagates as a
wave with velocity $c_2\sim15\text{-}\SI{20}{m/s}$, the so-called ``second sound velocity,'' by analogy to
the propagation velocity of isothermal and isentropic pressure
perturbation (i.e. ordinary sound wave). Moreover, as shown by
Eq.~(\ref{alpha_vld}), the temperature perturbation is damped by the
presence of vortex lines. By measuring this attenuation, one can
deduce the vortex line density.

In practice, we set up a second sound standing wave across
the confinement tunnel at the grids midplane. 
The physics of such a resonant cavity can be described by the
following expression, which gives the complex amplitude of the
standing wave measured by the receptor (see, e.g., Refs.~
\cite{Varga19,Woillez_sensors_23} for further details),

\begin{equation}\label{complex_U} \tilde{U}=\frac{U_0 e^{-i
(\phi_{cl}+\phi_{el})}}{\cosh (i \pi \frac{2L}{c_2} f-i\phi_{cl} +
\alpha L)}
\end{equation} where $U_0$ is the amplitude of the wave generated by
the emitter, $\phi_{cl}$ the phase shift determined by the boundary
conditions and $\phi_{el}$ the phase shift introduced by the
measurement chain. The damping coefficient $\alpha=\alpha_0+\alpha_\mathcal{L}$ account for both $\alpha_0$, the intrinsic damping of the cavity mostly due to geometrical effects, and $\alpha_\mathcal{L}$ the vortex tangle attenuation introduced previously.

If the excitation frequency $f$ is chosen such that the wavelength
$\lambda = c_2 / f$ is a sub-multiple of the cavity length, the wave
will resonate, accumulating energy and reaching a maximum amplitude
given by
\begin{equation}\label{U_res} U_{res} = \frac{U_0}{\sinh(\alpha L)}.
\end{equation} Given the high sensitivity of the resonant amplitude to
the damping coefficient $\alpha$, the simplest approach to assess the
additional attenuation caused by quantum vortex lines when the motor
is shaking is thus to set the excitation frequency at resonance and
track the decrease in amplitude, $U(t)$. This signal can then be
demodulated using Eqs.~(\ref{U_res}) and (\ref{alpha_vld}) as
\begin{equation}\label{L_no_T} \mathcal{L}(t) = \frac{6 c_2}{B \kappa}
\left[ \frac{1}{L} \sinh^{-1} \left( \frac{U_0}{U(t)} \right) -
\alpha_0 \right],
\end{equation} where $\alpha_0$ and $U_0$ are calibrated by
recording the maximum amplitude with the motor at rest. Note that,
depending on the boundary conditions, the resonant wavelength is given
by $\lambda_{res} = \frac{2L}{n - \frac{1}{2} +
\frac{\phi_{cl}}{\pi}},$ where $\phi_{cl} \in \left[-\frac{\pi}{2},
\frac{\pi}{2} \right]$ but the maximum amplitude given by
Eq.~(\ref{U_res}) remains independent of $\phi_{cl}$.

Equation~(\ref{L_no_T}), or a simpler first-order Taylor expansion
version assuming $\alpha L \ll 1$, leading to
\begin{equation}\label{L_no_T_simple} \mathcal{L}(t) = \frac{6 c_2}{B
\kappa} \alpha_0 \left( \frac{U_{res}}{U(t)} - 1 \right),
\end{equation} has been widely used in the literature to extract mean
vortex line density from second sound attenuation (see, e.g.,
\textcite{Varga19} for a historic review of ``the use of second sound
in experimental investigations of quantized vorticity and quantum
turbulence''). However, as detailed in the
Sec.~\ref{sec:secsoundsetup}, the operating range of our cavity around
the 100th harmonic make it highly sensitive to parasitic temperature
drifts and this simple procedure cannot be applied
directly. More precisely, when the motor oscillates, the temperature
fluctuations may be large enough to affect the value of the second sound
velocity $c_2$. This, in turn, changes the wavelength $\lambda$,
shifting the cavity out of resonance and ``artificially'' reducing the
measured amplitude (i.e., not because of the presence of vortex
lines). One way to handle this effect is to record both the magnitude
$U(t)$ and the phase $\theta(t)$ measured by the receptor and
demodulate the complex amplitude $\tilde{U}(t) = U e^{i\theta(t)}$ as
suggested by Eq.~(\ref{complex_U}):
\begin{equation}\label{demod_alpha_c2} \left\{
\begin{array}{ll} \alpha(t) & =
\Re\text{e}\left[\cosh^{-1}\left(\frac{U_0 e^{-i \phi_{cl}} e^{-i
\phi_{el}}}{\tilde{U}(t)}\right)\right]/L \\ c_2(t) & = {2 \pi L
f}/\left\{\Im\text{m}\left[\cosh^{-1}\left(\frac{U_0 e^{-i \phi_{cl}}
e^{-i \phi_{el}}}{\tilde{U}(t)}\right)\right] + \phi_{cl} + n
\pi\right\}.
\end{array} \right.
\end{equation}

The mean vortex line signal $\mathcal{L}(t)$ can then
straightforwardly be reconstructed as follows,
\begin{equation}\label{best_L} \mathcal{L}(t)=\frac{6 c_2(t)}{B\kappa}
(\alpha(t)-\alpha_0).
\end{equation}

\begin{figure}[ht!]  \centering
\includegraphics[width=\linewidth]{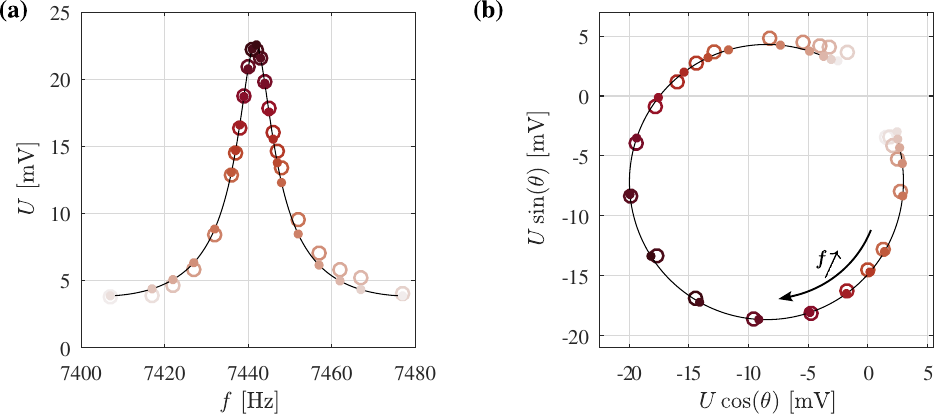}
  \caption{Frequency sweep of the cavity without flow forcing (motor
at rest). Open markers correspond to the measured values. Solid lines
represent Eq.~(\ref{complex_U}) after fitting the parameters
$U_0=\SI{3.9}{mV}$, $\alpha_0=\SI{1.48}{m^{-1}}$,
$\phi_{el}=\SI{134}{\degree}$, $\phi_{cl}=\SI{+96}{\degree}$. Filled
markers indicate the interpolated values at the same frequencies
tested during the frequency sweeps (allowing phase tracking in the
phase-amplitude representations). (a) Amplitude representation. (b)
Polar representation. }
  \label{fig:param_2s_pic}
\end{figure}

To apply this more refined approach, the parameters
$U_0$ and $\alpha_0$ must be calibrated first, as well as $\phi_{cl}$
and $\phi_{el}$. This is done by fitting a frequency sweep of the resonnance peak
at a stable reference temperature with the motor at rest
using Eq.~(\ref{complex_U}). An example of this procedure in our setup
at $T = \SI{2}{K}$ is illustrated in Fig.~\ref{fig:param_2s_pic}.

\subsubsection{\label{sec:secsoundsetup}Set up}

Both to excite the second sound wave within the resonant cavity and
measure its amplitude and phase, we used oscillating superleak
transducers (OST). Those transducers make use of a porous membrane
that allows the superfluid component to pass through while being
nearly impermeable to the normal fluid at times scales of the order of
$\SI{1}{ms}$ and below. This selective permeability enables the membrane to
either induce or detect oscillations in the local density ratio
between the normal and superfluid components, which are an image of
temperature oscillations. To ensure the measurement is as
non-intrusive as possible, each transducer is integrated into a brass
plate that can replace the glass panels of the confinement tunnel
walls (see Fig.~\ref{fig:cryostat}).

\begin{figure}[ht!]  \centering
\includegraphics[width=\linewidth]{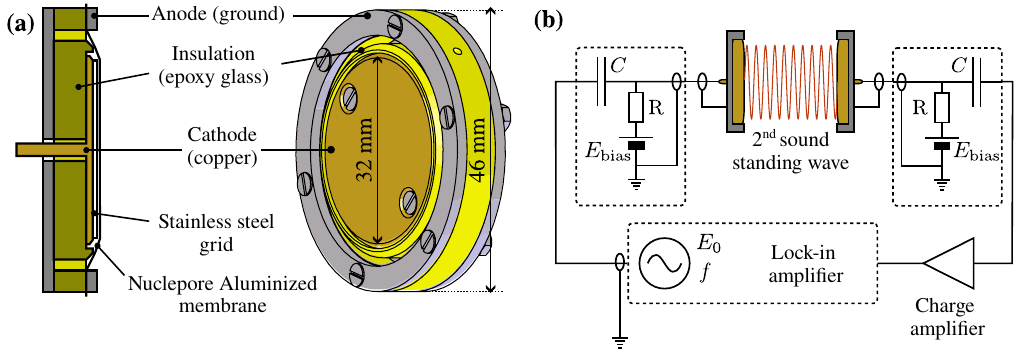}
\caption{Summary of the functioning of the resonant cavity
components. (a) Detailed structure of the OSTs. (b) Control
electronics, including the ``bias box'' composed of ten \SI{9}{V}
batteries and a high-pass filter ($R=\SI{10}{\mega \ohm}$ and
$C=\SI{1}{\micro F}$).}
\label{fig:secson}
\end{figure}

As illustrated in Fig.~\ref{fig:secson}(a), the OSTs are designed
following the Sell principle (see Ref.~\cite{WManthey_1992} for a
detailed review). These capacitive electrostatic transducers are made of 
a fixed cathode (a copper plate located on the rear side) and a mobile
anode (the membrane on the front side, in contact with the fluid).

The porous membranes used are Whatman Nuclepore\textsuperscript{®}
111705 membranes with characteristic pore diameters of
\SI{0.1}{\micro\meter}. The front side of each membrane is coated with
a \SI{30}{\nano\meter} thick layer of aluminum deposited using a
vacuum evaporation system. This method ensures directional deposition,
leaving the pores, with diameters significantly larger than
\SI{30}{\nano\meter}, open. The membranes can then be used as the
anode of a capacitor, connected to the sensor chassis which are
grounded. The rear face of the membranes is separated from the cathode
by a stainless steel mesh with a \SI{100}{\micro\meter} pitch and bar
diameters of \SI{30}{\micro\meter}. The dielectric gap between the
electrodes is structured to segment the sensor surface into multiple
cavities, improving the flatness of the emitted wavefront
\cite{Anke74} and enhancing the directionality of the receptor. This
geometry is chosen to replicate the most efficient configuration
reported in the study by \textcite{Zimmermann86}. The active area of
the membrane has a diameter of \SI{32}{mm}.

The membranes are bent over the grid cavities by applying a bias
voltage of approximately $E_{\mathrm{bias}} \approx \SI{90}{V}$,
generated by ten \SI{9}{V} batteries connected in series.  At the
emitter, as summarized on Fig.~\ref{fig:secson}(b), the membrane
oscillation is driven by a sinusoidal voltage delivered by the output
source of a lock-in amplifier (Stanford Research SR830) with
adjustable frequency $f$ and amplitude $E_0$ (up to \SI{15}{V} by
adding an amplifier) through an RC filter, thereby decoupling the
source from the high voltage $E_{\mathrm{bias}}$. As it emits a
multitude of second sound waves originating from each cavity of the
mesh, these numerous quasi-point waves interfere in such a way that,
to a first approximation, their superposition can be considered
equivalent to a single progressive plane wave. At the receiver, the
second sound wave interacts with the membrane, generating a sinusoidal
voltage response proportional to the amplitude of the
oscillations. More precisely, the interaction causes variations in the
apparent capacitance of the system, as the distance between the anode
and cathode fluctuates. These capacitance variations produce charge
oscillations, which are converted into voltage fluctuations by a
charge amplifier (Stanford Research SR570). The amplifier is
configured with a gain of \SI{1}{V/\micro A} and a \SI{12}{dB/decade}
high-pass filter to reject low-frequency noise, with a cutoff
frequency of \SI{300}{Hz}, as the cavity operates at frequencies above
\SI{1}{kHz} (see Sec. \ref{sec:2S_calib}). The output signal is then
demodulated at frequency $f$ by the lock-in amplifier to retrieve
the amplitude $U(t)$ and phase $\theta(t)$. The time constant is set
to \SI{3}{ms}, which appears sufficient to reject most of the noise.

\subsubsection{\label{sec:2S_calib} Operation procedure of the second
sound resonant cavity}

This section provides an overview of how we practically operate the
cavity, based on the procedure presented in
Sec.~\ref{sec:2S_principle} and illustrated for a bath temperature of
\SI{2}{K}.

\begin{figure}[ht!]  \centering
\includegraphics[width=\linewidth]{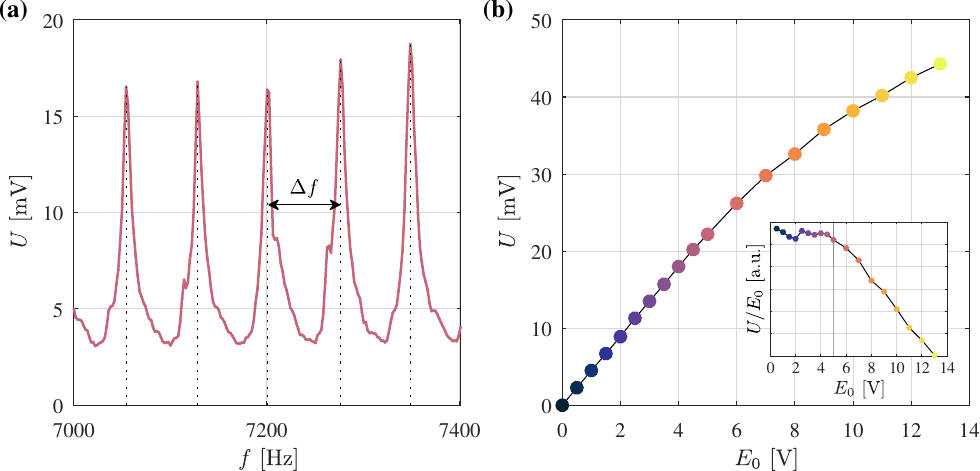}
  \caption{(a) Amplitude $U$ of the wave measured by the receiver as
the excitation frequency of the emitter is swept (resolution of
\SI{2}{Hz}). Successive resonance peaks are observed approximately
every \SI{75}{Hz}, consistent with the expected behavior for a cavity
of length $L=\SI{116}{mm}$ in a saturated bath at \SI{30}{mbar}. The
excitation input amplitude of the emitter is set to $E_0=\SI{5}{V}$
(b) Amplitude $U$ of the wave measured by the receiver  at \SI{2}{K} as a function
of the excitation input amplitude $E_0$ of the emitter.}
  \label{fig:choose_f_E0}
\end{figure}

The first step is to select an operating harmonic by sweeping the
excitation frequency and analyzing the amplitude response measured by
the receptor. Figure~\ref{fig:choose_f_E0}(a) illustrates a frequency
sweep performed with the cavity at rest, revealing successive
second-sound resonant peaks around the 100th harmonic spaced by $
\Delta f = c_2 / 2L \approx \SI{75}{Hz} $ at \SI{2}{K}. By exploring a
wide range of excitation frequencies from \SI{2}{kHz} to \SI{40}{kHz}, we
observe that the sensitivity of the system is optimal in this range.
From this large frequency sweep, we also extract a precise value of the cavity
length $L$ using the following procedure: we identify $n_p = 517$ resonant peaks, with the 
first and last occurring at \SI{2075}{Hz} and \SI{39925}{Hz}, respectively. As the difference 
between these two values is given by $(n_p - 1) c_2 / 2L$, we obtain
$L =\SI{116}{mm}$, using
the saturated value of $c_2=\SI{17}{m/s}$ corresponding to the measured cavity temperature from a 
Cernox\textsuperscript{®} sensor, $T = 1.982 \pm \SI{60e-6}{K}$. The discrepancy between 
this cavity length and the tunnel width (\SI{114}{mm}) is likely due to the slight retraction of 
the OST front faces relative to the brass plates in which they are embedded.

 Note that simple considerations of 2D effects can explain why the
cavity cannot reliably be operated at lower harmonics: as the
emitter diameter, $D = \SI{32}{mm}$, introduces diffraction effects
when the wavelength $\lambda$ is of the same order of magnitude, we can estimate
that to ensure more than 50\% of the emitted beam interacts with the
receptor, the system must operate above the 95th
harmonic. Nevertheless, it is important to note that operating at such
high harmonics makes the system highly sensitive to temperature
fluctuations in the bath. The resonant frequency associated with the
100th harmonic is given by $f_{100} = {c_2}/{\lambda_{100}} =
100{c_2}/{2L}$, implying that fluctuations in $c_2$ due to
temperature variations are amplified by a factor of 100 compared to
the first harmonic. At \SI{2}{K}, temperature changes as small as
\SI{0.1}{mK} -- corresponding to pressure fluctuations of the order of
\SI{8}{\micro bar} -- can shift the resonant frequency of the 100th
harmonic by approximately \SI{1}{Hz}. This shift causes the cavity to
deviate slightly from resonance, as it will no longer be excited
exactly at its resonant frequency. As highlighted in
Sec.~\ref{sec:2S_principle}, it is therefore crucial to record the
full complex amplitude $\tilde{U}(t)=U(t) e^{i\theta(t)}$ of the wave
during turbulence measurements to disentangle amplitude attenuation
caused by vortex lines from fluctuations due to temperature-induced
resonance shifts.

Once the excitation frequency of the cavity is set, to
maximize the signal to noise ratio, one has to choose the largest
possible input amplitude $E_0$ while ensuring that the traveling
counterflow wave does not itself produce parasitic quantum vortex
lines (the so-called T1 transition). As shown in
Fig.~\ref{fig:choose_f_E0}(b), the
characteristic curve $U(E_0)$ remains within the linear response
regime up to $E_0 = \SI{5}{V}$ which is the value used for all the
measurements. Note that due to the unclear boundary conditions imposed
at both ends of the cavity by the OSTs, it is not straightforward to
estimate the associated counter flow velocity $v_{ns}$. A
supplementary analysis using an alternate emitter (a heating plate)
verified that the associated critical heat flux is about
$\varphi_0\approx\SI{2}{W/m^2}$ leading to $v_{ns}={\varphi_0}/{\rho_s
\sigma T}\approx \SI{16}{\micro m/s}$. The hydraulic diameter
associated with the heating plate is $ D_h =
{4\mathcal{S}}/{\mathcal{P}} \approx \SI{50}{mm} $, where $
\mathcal{P} = \SI{214}{mm} $ is the perimeter of the heating and
$\mathcal{S}=42\times \SI{65}{mm^2}$ is its surface. We thus deduce
the critical Reynolds number, $\text{Re}_c = {v_{ns} D_h}/{\kappa}
\approx 8.5$ in excellent agreement with the values compiled in
\textcite{Bertolaccini17}. As the product $\rho_s \sigma T$
monotonically increases when the temperature is lowered, the \SI{5}{V}
voltage value extracted at \SI{2}{K} from
Fig.~\ref{fig:choose_f_E0}(b) is a lower bound of the critical tension
for temperature lower than \SI{2}{K}. As every measurement report in
this paper are made at $T\leq \SI{2}{K}$, we can safely assume that
the second sound standing wave never added parasitic vortex lines in
the flow.

\begin{figure}[ht!]  \centering
\includegraphics[width=\linewidth]{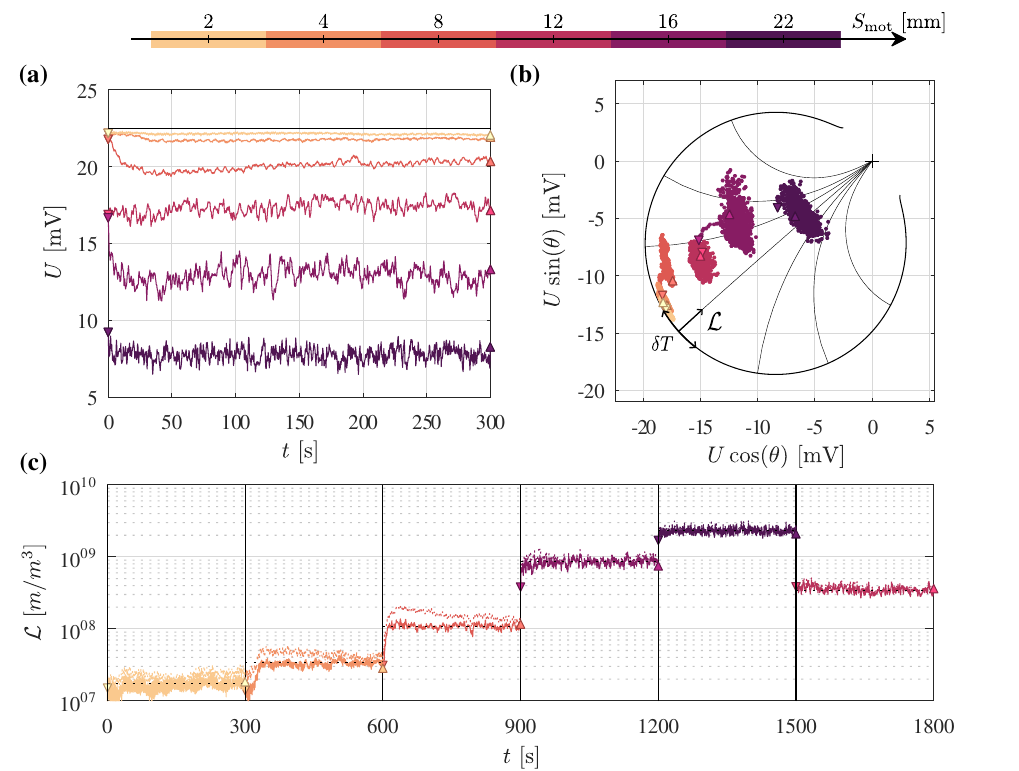}
  \caption{Raw signals (at \SI{2}{K}) measured at the output of the synchronous
detection for a series of flow forcing at a grid oscillation frequency
of $f_{\text{mot}} = \SI{9}{Hz}$ for different stroke amplitudes
$S_{\text{mot}}$. The excitation frequency is set to
$f_{\text{exc}}=\SI{7442}{Hz}$ (a) Amplitude, $U(t)$. The solid black line
represents the maximum amplitude at resonance without vortex lines
[see Eq.~(\ref{U_res})], as deduced from the fit of
Eq.~(\ref{complex_U}) (see Fig.~\ref{fig:param_2s_pic}). (b) Signal
trajectories in the polar plane. The pseudo-circle represents the
complex amplitude of the 102nd harmonic of the cavity without vortex
lines see Eq.~(\ref{complex_U}). as deduced from the fit of
Eq.~(\ref{complex_U}) (see Fig.~\ref{fig:param_2s_pic}). The black
rays show the trajectories of points kept at constant excitation
frequency, while the attenuation coefficient is increased, providing a
reference. (c) Mean vortex line density, $\mathcal{L}$. Solid lines
show values obtained by demodulating both the phase $\theta(t)$ and
amplitude $U(t)$ using Eqs.~(\ref{demod_alpha_c2}) and (\ref{best_L}),
while dotted lines show values obtained by demodulating only the
amplitude $U(t)$ using Eq.~(\ref{L_no_T}).}
  \label{fig:wave_XY}
\end{figure}

The following step is to measure the values of $U_0$, $\alpha_0$,
$\phi_{cl}$ and $\phi_{el}$ to set the reference point at resonance
from which we will follow the evolution in the phase/amplitude plane
to assess both the mean vortex line density of the flow and the
temperature fluctuations. As shown in Fig.~\ref{fig:param_2s_pic},
this is done by sweeping the resonant peak associated with the
selected harmonic while the motor is at rest and fitting the peak
using Eq.~(\ref{complex_U}). The second sound
velocity, $c_2$, is assumed to correspond to its saturation value
under the given regulated pressure bath (\SI{30}{mbar} for
Fig.~\ref{fig:param_2s_pic}) and the cavity length is fixed, $L=\SI{116}{mm}$ as previously discussed. Note
that, as their values are imposed for the fit, the uncertainties 
on $c_2$ and $L$ are encapsulated in the returned $\theta_{cl}$ value that looses its original meaning $\theta_{cl}\in[-\pi/2,\pi/2]$.

We finally choose the excitation frequency of the cavity to resonance and
start the measurement. In the example of
Fig.~\ref{fig:param_2s_pic}(a), the resonant frequency determined from
the fit is \SI{7441.7}{Hz}, and the excitation is set to the nearest
integer, $f = \SI{7442}{Hz}$, as the lock-in allow only integer
frequencies. The proper turbulence measurements can then begin by
recording the output signal [$X = U \cos(\theta)$ and $Y = U
\sin(\theta)$ components from the lock-in detection] for \SI{300}{s}
for different pairs of forcing conditions $(f_{\text{mot}},
S_{\text{mot}})$. As shown in Fig.~\ref{fig:wave_XY}(a), the amplitude
of the wave measured by the receiver decreases with increasing driving
intensity, indicating a progressive rise of the mean vortex line
density of the flow. Short-term fluctuations and long-term trends are
also observed.
 
As already mentioned, these amplitude variations are not solely caused
by fluctuations in vortex line density but are also influenced by
temperature drifts.
Fig.~\ref{fig:wave_XY}(b) shows trajectories
of the complex amplitude $\tilde{U}(t)$ in polar coordinates. 
We observe that the complex amplitude
trajectories move along both the pseudo-diameter, which correlates
with damping fluctuations (i.e. vortex density changes), and the
quasi-circle, which corresponds to phase shifts (i.e. second sound
velocity $c_2$ fluctuations due to temperature changes).

Figure~\ref{fig:wave_XY}(c) shows the deduced mean vortex line density
$\mathcal{L}$, obtained using the two methods presented in
Sec.~\ref{sec:2S_principle}, illustrating the importance of addressing
phase-shift-induced temperature effects. The solid line represents the
result from demodulating both the amplitude $U(t)$ and phase
$\theta(t)$ [as described in Eqs.~(\ref{demod_alpha_c2}) and
(\ref{best_L})], while the dashed line corresponds to demodulating only
the amplitude $U(t)$ [as given by Eq.~(\ref{L_no_T})]. Note that we
take into account both the frequency and temperature dependence of the
$B$ parameter, extracted from Ref.~\cite{Donnelly98}.
Specifically, we use $B = 1.18$ to demodulate the measurements at
$T = \SI{2}{K}$ and $f = \SI{7440}{Hz}$ (as the ones shown in Fig.~\ref{fig:wave_XY}),
and $B = 1.29$ for the dataset acquired at $T = \SI{1.6}{K}$ and $f = \SI{7470}{Hz}$.

For each forcing condition, we observe that the solid-line signals
clearly stabilize at a mean value (represented by a black dotted line)
after a brief transient period of a few tens of seconds (appearing as
``tails'' behind the point cluster in Fig.~\ref{fig:wave_XY}(b)). This
confirms the stationarity of the flow and confirms that the vortex
tangle densifies as the driving intensity increases.

In contrast, the dotted signals of $\mathcal{L}(t)$ are clearly
affected by large temperature-induced drifts that occur when the
forcing conditions change and gradually diminish as the pressure
controller compensates for dissipation changes.

Finally, we observe that the different point clusters in
Fig.~\ref{fig:wave_XY}(b), each associated with a given forcing, are
not centered at the same phase. This is likely due to pressure sensor
uncertainty : the pressure controller adjusts the aperture of the pumping valve according to the dissipated power, with final bath pressure being inside a dead band associated with pressure sensor accuracy~\footnote{As mentioned in
Sec.~\ref{sec:secsoundcavity}, a shift of \SI{8}{\micro bar} is
sufficient to shift the resonance frequency by approximately
\SI{1}{Hz}. The resulting phase shift depends on the damping
coefficient $\alpha$, but for the initial value
$\alpha_0=\SI{1.5}{m^{-1}}$, exciting the cavity \SI{1}{Hz}
off-resonance induces a phase shift of about \SI{15}{\degree} compared
to the phase at exact resonance.}.

Nevertheless, all of these considerations have demonstrated our
ability to measure the mean vortex line density of the flow for
different forcing conditions. Combined with the particle trajectories,
we now have all the ingredients necessary to characterize the
turbulent properties of the flow.

\section{Large-Scale Turbulent Properties of the flow from Particle
Trajectories}\label{sec:large_scale_prop}

In this section, we present a detailed analysis of the particle
velocities that gives us access to the large-scale properties of the
flow, specifically the velocity fluctuations $u_{\text{rms}}$, the
energy dissipation rate $\epsilon$, and the associated integral length
scale $L=u_{\text{rms}}^3/\epsilon$. We first examine the particle
velocity distributions and characterize the evolution of the
standard deviation $u_{\text{rms}}$, as a function of the forcing
parameters $f_{\text{mot}}$ and $S_{\text{mot}}$. Next, we turn to the
characterization of the energy injection rate $\epsilon$ based on the
study of the pseudo-eulerian structure functions of order 2 and 3.

\subsection{\label{sec:distrib2x2D}Velocity distributions and
turbulent velocity fluctuations}

\subsubsection{Velocity distributions}

In the following, we analyze the extensive dataset obtained from the
2$\times$2D system, which includes measurements at three different
temperatures: \SI{1.6}{K} (He~II), \SI{2}{K} (He~II), and \SI{2.3}{K}
(subcooled He~I at \SI{100}{mbar}). The aim is to investigate the
velocity distributions of the particles and to explore the dependence of
their associated standard deviation, $u_{\text{rms}}$, on the forcing
parameters $f_{\text{mot}}$ and $S_{\text{mot}}$.

The definition of particles velocities from their trajectories, which
fundamentally involves selecting a method to differentiate a discrete
and noisy signal, must be approached with caution to avoid amplifying
noise. In our case, we
opt for a simpler and more computationally efficient approach:
convolving each trajectory with a Gaussian derivative kernel, as
described in Ref.~\cite{Mordant04}. To determine the appropriate
Gaussian kernel size, we analyze the power spectral density of the
particle trajectories, which reveals that the noise becomes significant
around $\sim \SI{70}{Hz}$. We thus select a conservative
kernel width of $n_f = 15$ time steps, or $\tau_f = \SI{5}{ms}$
(corresponding to a characteristic frequency of $\sim
\SI{200}{Hz}$). Note that we verify that any kernel width within 
the range 3 to 15 time steps leads to similar values of the standard
deviation of the velocity (less than 5\% difference).

After computing the particle velocities, additional steps are
required to isolate the velocity fluctuations induced by the
turbulence. In particular, parasitic oscillations are observed in the
$\sim 30$-$\SI{60}{Hz}$ range, likely caused by mechanical vibrations,
in some cases. These artifacts, where all particles oscillate 
in phase, can be mitigated by calculating the mean particle velocity
frame by frame and subtracting its quasi-periodic fluctuating
component from the individual trajectories.

Furthermore, preliminary analysis of the 3D dataset revealed the
presence of stationary but inhomogeneous mean recirculating flow. To
isolate the turbulence-induced fluctuations generated by the
oscillating grids, this mean flow must also be subtracted. The
detailed procedure for this critical step is provided in
appendix~\ref{annexe_A}, covering both the 2$\times$2D and 3D
trajectories.

After completing these preprocessing steps, we can confidently analyze
the velocity distributions of the particles. Given that the vertical
component of particle velocities is affected by their settling
velocity, which varies due to differences in particle size and density, our analysis focuses
exclusively on the horizontal velocity components.

\begin{figure}[ht!]  \centering
\includegraphics[width=\linewidth]{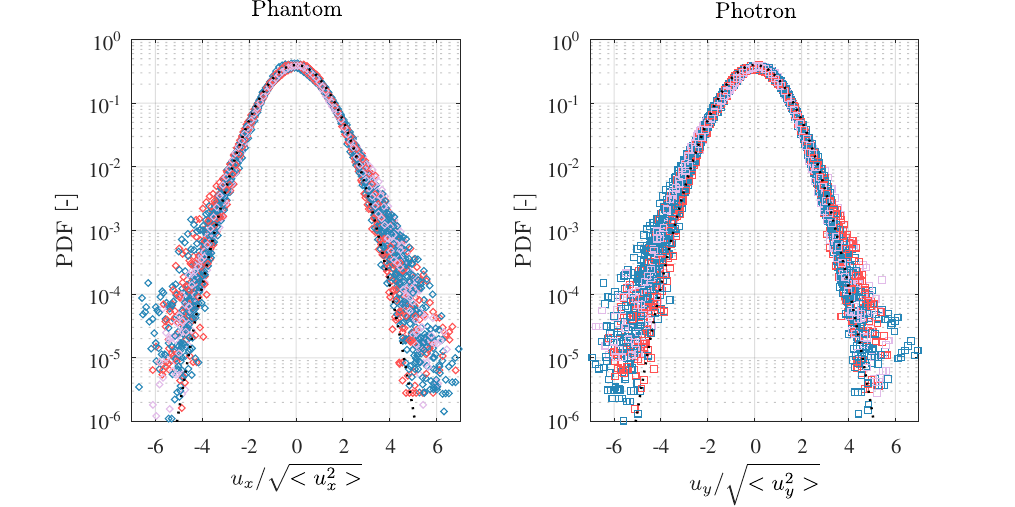}
  \caption{Distributions of horizontal velocities normalized by their
standard deviation. The means are not subtracted as they are nearly
zero (never more than 3\% of the corresponding standard
deviation). The color code corresponds to the bath conditions: red for
subcooled He I, blue for saturated He II at \SI{2}{K}, and purple
for saturated He II at \SI{1.6}{K}. Black dotted lines represent a
Gaussian distribution. (a) Velocity distribution from the horizontal
component of particle trajectories obtained from Phantom camera
images. (b) Same but from Photron camera images.  }
  \label{fig:PDF_vitesse}
\end{figure}

Figure~\ref{fig:PDF_vitesse} shows the horizontal velocity
distributions, normalized by their respective standard deviation. The
results indicate that, under this coflow configuration and across all
three temperatures, the particle velocity distributions are
quasi-Gaussian, in accordance with previous studies
\cite{Svancara2019_grid,Tang20}. Reducing the size of the differentiation 
kernel tends to produce larger tails and increase the flatness 
$F=\langle u^4\rangle /\langle u^2\rangle^2$, especially for low velocities,
from $F\approx 4$ for the base 15 time steps kernel to  4.5 for a 3 time steps kernel.
This is interpreted as a consequence of artifacts occurring during the 
trajectories reconstruction. Whatever the kernel size, no distinction in the shape
of the velocity distributions is observed between He~I and He~II.

\subsubsection{\label{sec:urms_brut}Turbulent velocity fluctuations}

\begin{figure}[ht!]  \centering
\includegraphics[width=\textwidth]{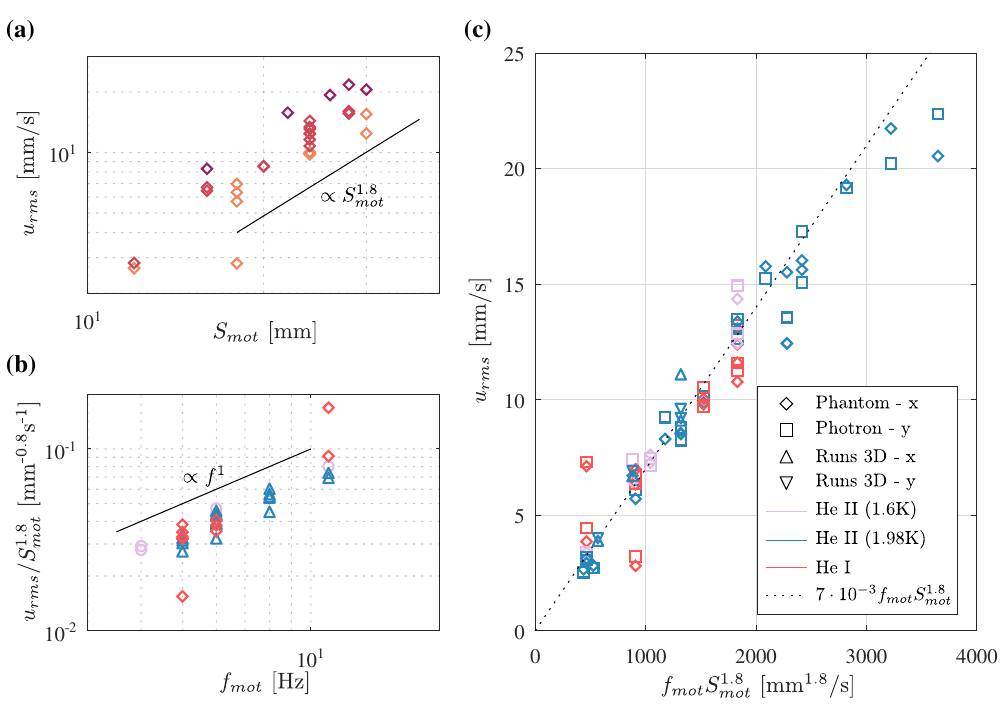}
  \caption{Raw results showing the dependence of the velocity
fluctuation standard deviation $u_{\text{rms}}$ on the flow forcing
parameters ($f_{\text{mot}}$, $S_{\text{mot}}$). Square markers
represent the horizontal velocity standard deviation measured from
Phantom camera images, while diamond markers correspond to
measurements from Photron camera images. (a) $u_{\text{rms}}$ as a
function of $S_{\text{mot}}$ at fixed $f_{\text{mot}}$. Each of the
three frequencies is represented by a different color: \SI{5}{Hz}
(orange), \SI{6}{Hz} (pink), and \SI{8}{Hz} (purple). (b)
$u_{\text{rms}}/S_{\text{mot}}^{1.8}$ as a function of
$f_{\text{mot}}$. The color code indicates the bath conditions:
subcooled He~I (red), saturated He~II at \SI{2}{K} (blue), and
saturated He~II at \SI{1.6}{K} (purple). (c) $u_{\text{rms}}$ as a
function of $f_{\text{mot}} S_{\text{mot}}^{1.8}$. The color code is
the same as in (b).}
  \label{fig:urms_brut}
\end{figure}

We now characterize the dependence of the velocity standard deviation,
$u_{\text{rms}}$ on the forcing parameters: the stroke
$S_{\text{mot}}$ and the oscillation frequency $f_{\text{mot}}$ of the
grids. We first examine the dependence of $u_{\text{rms}}$ on
$S_{\text{mot}}$ for a fixed
$f_{\text{mot}}$. Figure~\ref{fig:urms_brut}(a) shows the data
obtained for a variation of $S_{\text{mot}}$ between 12 and
\SI{30}{mm} for three different frequencies $f_{\text{mot}}$: 5, 6,
and \SI{8}{Hz}. Despite some scatter, a general trend emerges from
which a power law, $u_{\text{rms}} \propto S_{\text{mot}}^{1.8 \pm
0.2}$, can be reasonably extracted.

Having insufficient data to perform a similar analysis of the
dependence of $u_{\text{rms}}$ on $f_{\text{mot}}$ for fixed
$S_{\text{mot}}$, Fig.~\ref{fig:urms_brut}(b) shows the evolution of
$u_{\text{rms}}/S_{\text{mot}}^{1.8}$ as a function of frequency
$f_{\text{mot}}$. Despite the scatter, this representation suggests
that $u_{\text{rms}} \propto f_{\text{mot}}^{1 \pm 0.1}$.

Finally, to provide an overall view of the combined dependence on
these two parameters, Fig.~\ref{fig:urms_brut}(c) shows
$u_{\text{rms}}$ as a function of the product $f_{\text{mot}}
S_{\text{mot}}^{1.8}$. The data seem to be well aligned, and no significant
difference is observed between He~I/He~II. These observations suggest
that, within the explored parameter range, a reasonable approximation
for $u_{\text{rms}}$ can be written as:
\begin{equation}\label{urms_brut_18} u_{\text{rms}} = \gamma
f_{\text{mot}} S_{\text{mot}}^{1.8}
\end{equation} where the prefactor $\gamma = \SI{7e-3}{mm^{-0.8}}$.

Given the scatter in the data, slight adjustments to the exponents
could have been plausible. However, we will show that this scaling is
well supported by theoretical arguments. \textcite{Matsunaga99}
applied the $k$-$\epsilon$ turbulence model to describe the flow
generated by a single oscillating grid. Their derivation of an
analytical expression for the decay of kinetic energy, $k(z) =
\frac{1}{2}u_{\text{rms}}^2(z)$, as a function of the distance
$z$ from the grid, showed good agreement with experimental data
available at the time \cite{Hopfinger76, DeSilva94,
Ura87}, and this result has since been further validated by more
recent studies (e.g., Ref.~\cite{Poulain22}).

The reported solution for the kinetic energy is given by:
\begin{equation}\label{kkmatsu} k = k_0 \left(1 + \frac{z}{1.82 L_0}
\right)^{-5},
\end{equation} where $L_0 = \frac{k_0^{3/2}}{\epsilon_0}$, $k_0$ and
$\epsilon_0$ represent respectively the integral scale, the kinetic
energy, and the dissipation rate at the grid level. The values of $k_0$
and $\epsilon_0$ are empirically determined based on experimental
data, leading to:
\begin{equation}\label{k_0_eps_0} \left\{
\begin{array}{ll} k_0 &= 0.6
\left(\frac{S_{\text{mot}}}{M}\right)^{1/4} f_{\text{mot}}^2
S_{\text{mot}}^2, \\ \epsilon_0 &= 0.45 \frac{S_{\text{mot}}}{M} 
f_{\text{mot}}^3 S_{\text{mot}}^2,
\end{array} \right.
\end{equation} where $M$ is the grid mesh size. These
relations are valid for $\mathrm{Re_g} \geq \num{5.5e3}$, where
$Re_g$ is the Reynolds number calculated from the grid forcing
properties:
$$Re_g = \frac{f_{\text{mot}} S_{\text{mot}}^2}{\nu}.$$
Note that, although the viscosity $\nu$ of He~II cannot be precisely
defined for this calculation, it can be unambiguously computed for the
He~I runs, yielding a minimum value of $\mathrm{Re_g} \approx
\num{3.5e4}$. It thus seems reasonable to assume that $\mathrm{Re_g}
\geq \num{5.5e3}$ holds for all tested ($S_{\text{mot}}$,
$f_{\text{mot}}$)
parameter pairs in both He~I and He~II. It follows that $L_0
=\frac{0.6^{3/2}}{0.45} M_{\text{mot}}^{5/8} S_{\text{mot}}^{3/8}$,
which implies, based on Eq.~(\ref{kkmatsu}), that the dependence of
$u_{\text{rms}}$ on the stroke $S_{\text{mot}}$ does not follow a
simple power-law behavior:
$$u_{\text{rms}} \approx 0.8 f_{\text{mot}} M^{1/8}S_{\text{mot}}^{9/8} \left( 1 + \frac{z}{1.9 M^{5/8} S_{\text{mot}}^{3/8}} \right)^{-5/2}.$$ 
In our parameter range, at the mid-plane between the two grids, $z = H/2$,
this analytical relation can be approximated by the following power law,
$$u_{\text{rms}} \propto f_{\text{mot}} S_{\text{mot}}^{1.82},$$
which corroborates Eq.~(\ref{urms_brut_18}) established from the raw
data and gives an explanation for the somewhat unintuitive
$u_{\text{rms}}\propto S_{\text{mot}}^{1.8}$ scaling.

\subsection{\label{sec:epsilon_L}Injected energy at large scales}

In classical homogeneous and isotropic turbulence, the exact
$\frac{4}{5}$-th law states (see, e.g., \textcite{Pope00}, page 204),
for scales $r$ within the inertial range, that the third-order
longitudinal structure function is expressed as
\begin{equation}\label{k41_45} S_{3,\|}=\langle (\delta_r u_\|)^3
\rangle = -\frac{4}{5} \epsilon r.
\end{equation} This law could thus be used as a tool to estimate
$\epsilon$, the energy transfer rate across scales in the inertial
range, using distributions of longitudinal velocity increments
computed for various scales $r$ within the inertial range. This energy
transfer rate is interpreted, given the stationarity of the flow, as
the flux of energy injected at large scales and dissipated at small
scales. It is known that the $\frac{4}{5}$-th law holds in He~II, at
least in the upper part of the inertial range \cite{Salort12}, and
that, under identical forcing, the energy injection rate and, thus,
the dissipation rate in He~I and He~II are the same \cite{Rousset14}.

In classical turbulence, the large-scale energy injection rate can be
written (see, e.g., \textcite{Pope00}, p. 244) as
\begin{equation}\label{epsilon} \epsilon=\frac{u_{\text{rms}}^3}{L}.
\end{equation} Thus, in what follows, we aim to use the
$\frac{4}{5}$-th law to estimate $\epsilon$ and deduce the value of
the integral scale $L$ from Eq.~(\ref{epsilon}).

Due to the indeterminacy of particle positions in the depth of the
camera fields of view, 2$\times$2D measurements cannot be used to compute l
longitudinal velocity increments. This can be achieved using the 3D 
measurements though. By denoting $\mathbf{v}\left(\mathbf{P}^i(t_k),t_k\right)$
the velocity of the $i$-th
particle at position $\mathbf{P}^i(t_k)$ at time $t_k$, we can define
the set of longitudinal velocity increments at scale $r$
and time $t_k$ as follows:
\begin{equation} \label{dru_pseudo_eul_para} \{ \delta_r u_\| (t_k) \}
\ = \left\{ \left[\mathbf{v}\left(\mathbf{P}^i(t_k),t_k\right) -
\mathbf{v}\left(\mathbf{P}^j(t_k),t_k\right)\right]\cdot \mathbf{e}^{ij} \right\}_{\{i, j\} \,
\mathrm{such \, that} \, ||\mathbf{P}^i\mathbf{P}^j(t_k)|| \in
\Omega_r},
\end{equation} where $\mathbf{e}^{ij}
=\frac{\mathbf{P}^i\mathbf{P}^j(t_k)}{||\mathbf{P}^i\mathbf{P}^j(t_k)||}$
and $ \{ \cdot \}_{\{i, j\} \, \mathrm{such \, that} \,
||\mathbf{P}^i\mathbf{P}^j(t_k)|| \in \Omega_r}$ denote the set of
particle pairs $\{i,j\}$ whose separation distance
$||\mathbf{P}^i\mathbf{P}^j||$ at time $t_k$ is within the interval
$\Omega_r$. The total distance between the two particles, to which we
do not have access in 2$\times$2D measurements, is therefore essential
for calculating longitudinal increments. This approach is referred to
as ``pseudo-Eulerian'' because each particle is considered a velocity
sensor in the flow. Note that the particle velocities along their 3D
trajectories were computed using the same method described earlier for
the 2$\times$2D data in Sec.~\ref{sec:distrib2x2D}, employing an
adjusted kernel width of four images, corresponding to approximately
$\tau_f = \SI{20}{ms}$.

By combining values across different instants, the longitudinal
structure functions of order $p \in \mathbb{N}$ can be defined as:
\begin{equation*} S_{p,\|}(r)=\langle (\delta_r u_\|)^p \rangle.
\end{equation*}

\begin{figure}[ht!]  \centering
\includegraphics[width=\linewidth]{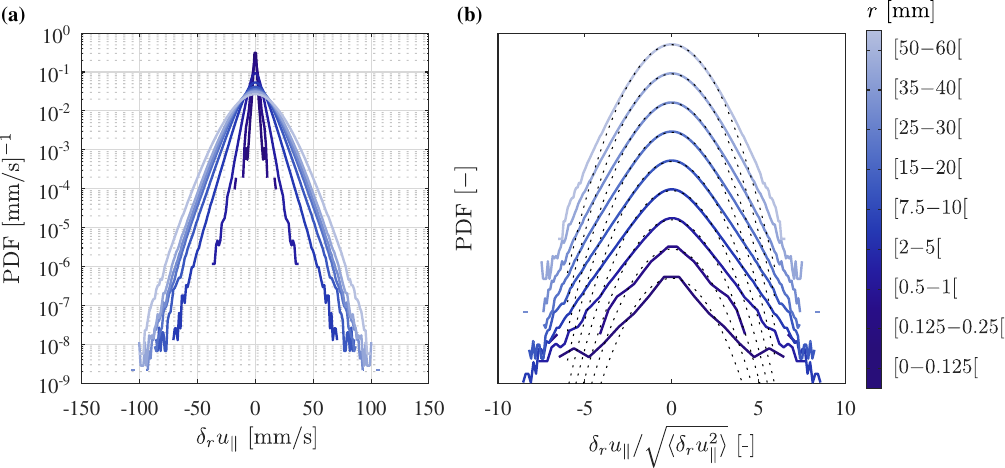}
\caption{a) Distribution of velocity increments $\delta_r u_{\|}$ for
several intervals of distance between particles for the run 3D-2 (He~II at \SI{2}{K}). The
four smallest intervals (one of which is not shown to avoid overloading
the graph) of inter-particle distances below \SI{1}{mm} contain
respectively $730$, $4200$, $38000$, and $350000$ pairs. The others
contain more than a million pairs. b) The same distributions
normalized by their standard deviation and arbitrarily shifted on the
vertical axis.}
\label{fig:dru_para}
\end{figure}

For each instant $t_k$ of the 3D runs, we computed the velocity
differences between pairs of particles and sort the results based on
the distance between the pairs into intervals. Following our
observations in the previous paragraphs, we aim to mitigate the
effects induced by the distribution of sedimentation velocities, which
manifest in the vertical component.  Therefore, we restrict the
calculation of longitudinal velocity increments to pairs of particles
satisfying $\mathbf{e}^{ij} \cdot \mathbf{e_z} \approx 0$. This ensures
that the contribution of velocity differences along the vertical
component is negligible compared to those along the horizontal
components. In practice, we only consider pairs whose vertical
separation distance is less than 10\% of their total separation
distance.  Figure~\ref{fig:dru_para} presents the distributions
obtained from one of the 3D acquisitions for several intervals of
distance between particle pairs.

\begin{figure}[ht!]  \centering
\includegraphics[width=\linewidth]{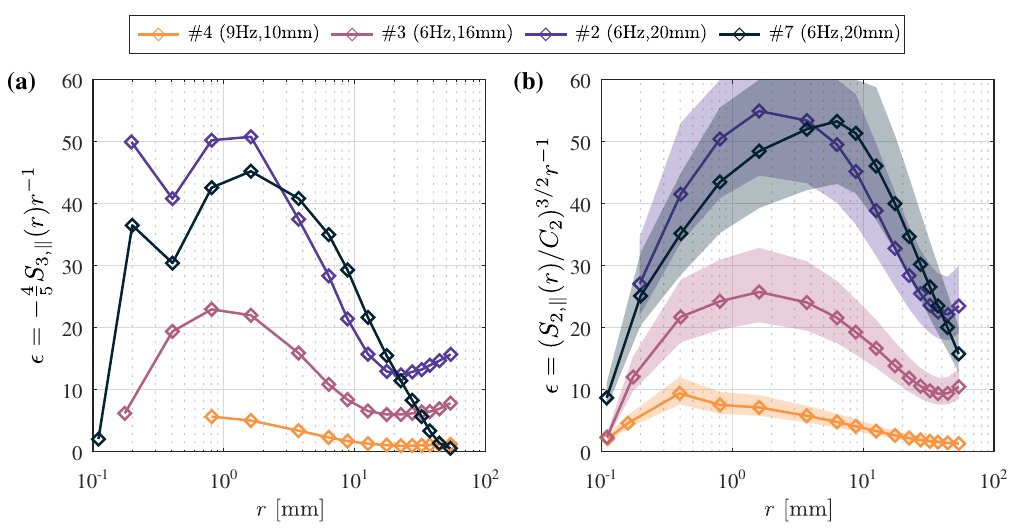}
\caption{a) Pseudo-Eulerian third-order longitudinal structure function of 
particle velocity $S_{3,\|}$, derived from the 3D dataset (He~II at \SI{2}{K}) 
with run IDs referenced in the legend as \#n, compensated as suggested by 
Eq.~(\ref{k41_45}) to directly read $\epsilon$ from the maximum of the
bell-curves maxima. To avoid biased interpretations, the marker positions on
the abscissa 
correspond to the average distance between particle pairs within each interval, 
ensuring consistency even when the inter-particle distance distribution is non-
uniform. Note that certain points at the smallest scales are excluded (for $r < 
\SI{0.125}{mm}$ for runs 2 and 3, and all points for $r < \SI{0.5}{mm}$ for run 
4), where we measure $S_{r,\|} > 0$. b) Pseudo-Eulerian second-order 
longitudinal structure function of particle velocity $S_{2,\|}$, derived from 
the 3D dataset (He~II at \SI{2}{K}). To allow direct comparison with the $
\epsilon$ value deduced from part a), the curves are compensated as suggested by 
Eq.~(\ref{S2_K41}). The shaded area represents the uncertainty on the value 
of $C_2 \in [1.7\text{-} 2.3]$.}
\label{fig:S3}
\end{figure}

 Qualitatively, the distributions are well-converged and
approximately Gaussian at large scales, while exhibiting broader tails
at smaller scales, a marker of turbulence intermittency.  It naturally
follows to compute the third-order moment of the distributions to
construct $S_{3,\|}(r)$, the longitudinal third-order structure
function, and then the energy
transfer rate $\epsilon$ following Eq.~(\ref{k41_45}). In Fig.~\ref{fig:S3}(a)
we present the latter as a function of the considered scale $r$ for the four 3D runs.

First, note that the curves corresponding to runs 2, 3, and 4 reveal
the persistence of a large-scale bias (for $r > \SI{20}{mm}$): at
sufficiently large scales, a Gaussian distribution of velocity
increments is expected, implying 
$\langle (\delta_r u_\|)^3 \rangle= 0$. This behavior can most likely be attributed to a
residual influence of the mean flow.
 
To further interpret the curves in the light of Kolmogorov's theory at
intermediate scales ($r<\SI{20}{mm}$), it is essential to caution the
reader. In practice, the $\frac{4}{5}$-law is strictly valid in the
inertial range only if the Kolmogorov cascade depth is sufficient,
meaning the ratio between the forcing scale $L$ and the dissipative
scale $\eta$ is large enough. According to the investigations of
\textcite{Moisy99}, to observe an unambiguous plateau in the
representation of $|S_{3,\|}(r)|/\left(\frac{4}{5} r\right)$, whose height can be
interpreted as $\epsilon$, the scale-separation condition must satisfy
$L/\eta \gtrsim 4000$ (or equivalently $\mathrm{Re}_\lambda\gtrsim1000$). If this
condition is not met, the expected shape is rather a bell-curve, whose
maximum should be interpreted as $\gamma\epsilon$, where
$\gamma\leq1 $ is a coefficient that increases with the cascade depth for
which the authors propose the following expression:
\begin{equation}\label{gamma} \gamma \approx 1 - 12 \left(
\frac{L}{\eta} \right)^{-4/5}.
\end{equation} 

Finally, note that the position $r_{\text{max}}$ of this
maximum also depends on the cascade depth, and $r_{\text{max}}$ is expected
to decrease as the ratio ${L}/{\eta}$ becomes smaller.

The experimental results in Fig.~\ref{fig:S3}(a) appear to align with
these considerations, suggesting that the scale separation is
insufficient to clearly observe a plateau. The maximum of the curves
in Fig.~\ref{fig:S3}(b) should therefore be interpreted as
$\gamma\epsilon$ and not directly as $\epsilon$. However,
estimating the value of $\gamma$ at this stage is delicate because we
lack estimates of both the integral scale $L$ and the dissipative
scale $\eta$. Temporarily assuming that the scale separation is
sufficient to consider $\gamma \approx 1$, the validity of this
assumption will be verified \textit{a posteriori}.

From the measurement of the bell-curve maxima, we thus obtain an
estimate of the energy dissipation rate $\epsilon$. Combined with
$u_{\text{rms}}$, this allows us to estimate $L$, the integral scale
of the flow, from Eq.~(\ref{epsilon}). Note that, instead of using
Eq.~(\ref{urms_brut_18}) to compute $u_{\text{rms}}$, we directly
measured the standard deviation of the particle velocities in the two
horizontal components, $u_{\text{rms},x}$ and $u_{\text{rms},y}$, and took their
mean.  The resulting values are reported in Table~\ref{recap_res3}.

\begin{table}[htbp]
\begin{center}
\begin{tabular}{|c|c|c|c|c|c|c|c|} \hline Run & Forcing &
$u_{\text{rms}}$ [\si{mm/s}] & $\epsilon_{S3}$ [\si{mm^2/s^3}] & $L$
[\si{mm}] & $\epsilon_{S3}^*$ [\si{mm^2/s^3}] & $L^*$ [\si{mm}] &
$\epsilon_{S2}$ [\si{mm^2/s^3}] \\ \hline 2 & (6Hz,20mm) & 9.2 & 50.8
& 15.3 & 52.9 & 14.7 & 44.6-70.1\\ \hline 3 & (6Hz,16mm) & 6.8 & 23 &
13.7 & 24.2 & 13 & 20.9-32.9\\ \hline 4 & (9Hz,10mm) & 3.9 & 5.7 &
10.4 & 6.2 & 9.5 & 5.9 - 9.3 \\ \hline 7 & (6Hz,20mm) & 10.4 & 45.2 &
24.9 & 46.5 & 24.2 & 43.2 - 68\\ \hline
\end{tabular}
\end{center}
\caption{Estimation of the energy injection rate $\epsilon$ and the integral scale $L$ based on the 3D dataset (He~II at \SI{2}{K}). The values $\epsilon^{*}$ and $L^{*}$ are the corrected estimates of $\epsilon$ and $L$ to account for the finite depth of the Kolmogorov cascade.}
\label{recap_res3}
\end{table}

Given the sensitivity of this method for estimating the integral scale
$L$ to small inaccuracies in the velocity root mean square [it
is raised to the power of 3 in Eq.~(\ref{epsilon})], the obtained
values appear consistent overall. The mean value, $L = \SI{16}{mm}$,
thus serves as a first good estimate.

We can now try to check the validity of the assumption regarding the
value of $\gamma \approx 1$. In the absence of a better approach, we
can use the results of classical turbulence theory to roughly estimate
the dissipative scale, $\eta = (\nu^3 / \epsilon)^{1/4}$, where $\nu =
\mu / \rho$ is the kinematic viscosity calculated from the dynamic
viscosity of the normal component and the total density $\rho = \rho_s
+ \rho_n$.

It follows that, for the four 3D runs, $L / \eta > 700$, and thus,
according to Eq.~(\ref{gamma}), $\gamma > 0.93$. Considering the
measurement uncertainties, the approximation $\gamma \approx 1$
appears relatively reasonable and leads to only a marginal
underestimation of $\epsilon$. More quantitatively, assuming that this
method of estimating $\gamma$ is valid in He~II, the procedure can be
iterated until convergence of the $\gamma$ value is achieved for each
of the four runs. The updated values $\epsilon_{S3}^*$ and the associated
$L^*$ are reported in table \ref{recap_res3}.

The new mean value of the integral scale is then marginally different,
$L=15.3\pm \SI{6}{mm}$.

To further validate the integral scale estimation via the
$\frac{4}{5}$-law, we also analyze the second-order moments of
longitudinal velocity increments. According to classical turbulence
theory, the second-order
longitudinal structure function for scales $r$ in the inertial range
is expressed as:
\begin{equation}\label{S2_K41}
S_{2,\|}(r) =
C_2 \epsilon^{2/3} r^{2/3} 
\end{equation} where $C_2 \in [1.7\text{-}2.3]$ (see,
e.g., \textcite{Pope00}, p.~193–194). This provides another method to
estimate $\epsilon$ using the $r^{2/3}$ scaling regime in the inertial
range.

As illustrated in Fig.~\ref{fig:S3}(b), the expected inertial range
regime is well-defined, extending down to the smallest well-resolved
scales of approximately $\sim 200\text{-}\SI{400}{\micro m}$. By
defining the plateau height on the log-log graph of
$(S_{2,\|}/C_2)^{3/2}/r$ (Fig.~\ref{fig:S3}(b)) as $\epsilon$, we thus
define $\epsilon_{S2}$ for $C_2 = 1.7$ and $C_2 = 2.3$ to capture the
range of uncertainty.

Both approaches, using $S_3$ and $S_2$, show good agreement given
the difficulty of such measurement. These results thus confirm the
validity of the $\frac{4}{5}$-law approach for estimating $\epsilon$
and give credit to the associated integral scale value, $L=15.3\pm
\SI{6}{mm}$.

Combined with Eq.~(\ref{urms_brut_18}), which estimates
$u_{\text{rms}}$, this value of $L$ enables the estimation of the
energy dissipation rate $\epsilon=u_{\text{rms}}^3/L$ for any pair of
forcing parameters ($f_{\text{mot}},S_{\text{mot}}$). This provides a
complete description of the large-scale properties of the flow.

\section{\label{sec:effective_viscosity}Effective viscosity of Quantum
Turbulence}

\subsection{\label{sec:VLD_brut}Experimental results of the mean vortex line density }

After applying the calibration protocol proposed in Sec.~\ref{sec:2S_calib} to
properly interpret the second sound signals in the phase/amplitude
space both at \SI{30.0}{mbar} (\SI{2}{K}), and
at \SI{7.5}{mbar} (\SI{1.6}{K}), we
gathered a database of signals spanning a wide range of flow forcings,
from $u_{\text{rms}} \approx \SI{0.5}{mm/s}$ to $u_{\text{rms}}
\approx \SI{20}{mm/s}$. Each measurement series, in which the motor
frequency $f_{\text{mot}}$ was varied at fixed amplitude or vice
versa, was processed in a manner similar to the previous example to
deduce the average vortex line density of the flow.

\begin{figure}[ht!]  \centering
\includegraphics[width=\textwidth]{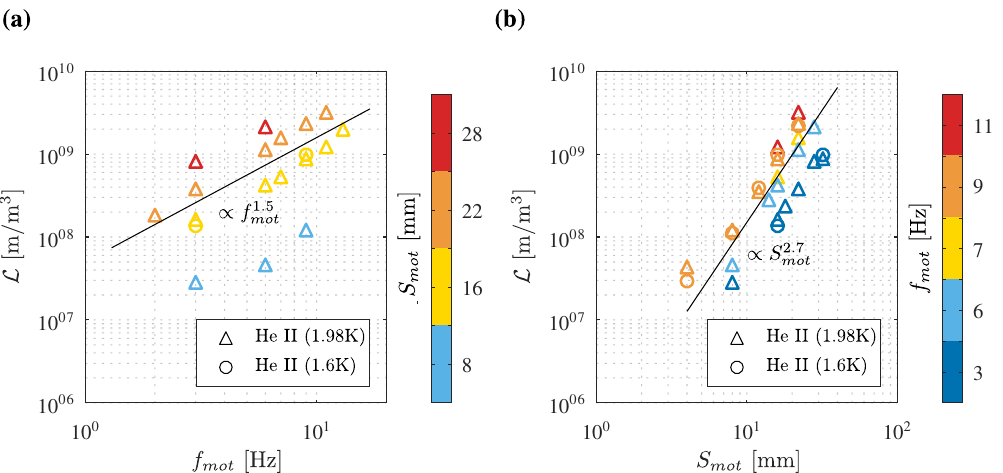}
  \caption{Raw results showing the dependence of the mean vortex line
density $\mathcal{L}$ on the flow forcing parameters
($f_{\text{mot}}$, $S_{\text{mot}}$). The two symbol shapes indicate
the two temperatures ($\bigcirc$~\SI{1.6}{K}, $\bigtriangleup$~\SI{2}{K}). (a) $\mathcal{L}$ as a function of
$f_{\text{mot}}$ at fixed strokes $S_{\text{mot}} = \SI{8}{mm},
\SI{16}{mm}, \SI{22}{mm}, \SI{28}{mm}$.  (b) $\mathcal{L}$ as a
function of $S_{\text{mot}}$ at fixed frequencies $f_{\text{mot}} =
\SI{3}{Hz}, \SI{6}{Hz}, \SI{7}{Hz}, \SI{9}{Hz}$.}
  \label{fig:res_L_vld_mot}
\end{figure}

As with the turbulent velocity fluctuations $u_{\text{rms}}$, we then
investigated the possibility of extracting power
laws. Figure~\ref{fig:res_L_vld_mot}(a) presents the dependence of the
mean vortex line density on $f_{\text{mot}}$, varied
from 2 to \SI{11}{Hz}, for four different strokes $S_{\text{mot}}$: 8,
16, 22, and \SI{28}{mm}. The data suggest a scaling of $\mathcal{L}
\propto f_{\text{mot}}^{1.5}$. Similarly,
Fig.~\ref{fig:res_L_vld_mot}(b) shows $\mathcal{L}$ as a function of
$S_{\text{mot}}$, varied from 4 to \SI{32}{mm}, for four different
motor frequencies $f_{\text{mot}}$: 3, 6, 7, and \SI{9}{Hz}, yielding
$\mathcal{L} \propto S_{\text{mot}}^{2.7}$.

Recalling the scaling Eq.~(\ref{urms_brut_18}) established in
Sec.~\ref{sec:urms_brut}, $u_{\text{rms}} \propto f_{\text{mot}}
S_{\text{mot}}^{1.8}$, the observed scaling $\mathcal{L} \propto
f_{\text{mot}}^{1.5} S_{\text{mot}}^{2.7}$ clearly suggests
$\mathcal{L} \propto u_{\text{rms}}^{3/2}$. Obviously, exponents
could be subject to slight adjustments, especially since the
$u_{\text{rms}} \propto f_{\text{mot}} S_{\text{mot}}^{1.8}$ scaling
may represent only a local trend, given that the broader
Eq.~(\ref{urms_brut_18}) derived from the $k$-$\epsilon$ model cannot
be expressed as a pure power law (see
Sec.~\ref{sec:urms_brut}). 
  
\subsection{\label{sec:estim_eff_visc}Estimation of
  \texorpdfstring{$\nu_{\text{eff}}$}{ν-eff} and comparison with literature}

\begin{figure}[ht!]  \centering
\includegraphics[width=\linewidth]{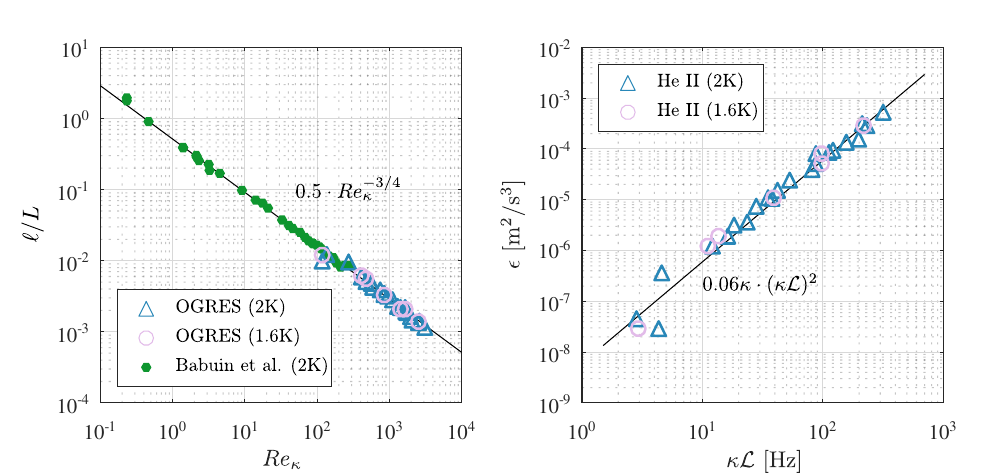}
\caption{Compilation of second sound measurements. Blue markers
correspond to measurements at \SI{2}{K}, while purple markers
correspond to \SI{1.6}{K} at saturation. Green markers show data retrieved
from Ref.~\cite{Babuin14}, for comparison. (a) Mean inter-vortex distance, $\ell =
1/\sqrt{\mathcal{L}}$, normalized by the integral scale $L$, as a
function of the Reynolds number, $\text{Re}_\kappa = u_{\text{rms}} L
/ \kappa$. (b) Dissipation rate, $\epsilon = u_{\text{rms}}^3/L$,
plotted against $(\kappa \mathcal{L})$.}
\label{fig:result_2S}
\end{figure}

Figure~\ref{fig:result_2S} presents all results obtained at \SI{2}{K}
and \SI{1.6}{K}. To aid interpretation, the data are displayed from
two perspectives. In Fig.~\ref{fig:result_2S}(a), we plot the mean
inter-vortex distance, $\ell = 1/\sqrt{\mathcal{L}}$, in the range
$\SI{15}{\micro\meter} \lesssim \ell\lesssim \SI{150}{\micro\meter}$,
normalized by 
the integral length scale $L$, as a function of the Reynolds number,
$\text{Re}_\kappa = u_{\text{rms}} L / \kappa$.
Figure~\ref{fig:result_2S}(b) shows the dissipation rate, $\epsilon =
u_{\text{rms}}^3/L$, as a function of $(\kappa \mathcal{L})$.

Our measurements at both \SI{1.6}{K} and \SI{2}{K} clearly suggest
that
\begin{equation}\label{nu_eff_exp} \epsilon \approx 0.06 \kappa
(\kappa \mathcal{L})^2.
\end{equation} 
This scaling is consistent with Eq.~(\ref{nu_eff}) from which we extract $\nu_{\text{eff}}
\approx 0.06\kappa$, or more precisely by fitting the data:
\begin{equation}\label{nu_eff_estimate}
\nu_{\text{eff}}(T=\SI{1.6}{K}) = 0.061\kappa, \quad
\nu_{\text{eff}}(T=\SI{2}{K}) = 0.054\kappa.
\end{equation}
As we saw that Eq.~(\ref{nu_eff}) is equivalent to
Eq.~(\ref{delta_L}) and it is useful to translate these
$\nu_{\text{eff}}$ values in terms of
$\beta=(\nu_{\text{eff}}/\kappa)^{1/4}$:
\begin{equation}\label{beta_estimate} \beta_{\text{exp}}(T=\SI{1.6}{K}) = 0.5,
\quad \beta_{\text{exp}}(T=\SI{2}{K}) = 0.48.
\end{equation}

We can now compare our results with those of \textcite{Babuin14},
which also come from a steady-state turbulent flow. Their data for 
$\ell/L$ at \SI{1.95}{K} (retrieved manually and plotted in
Fig.~\ref{fig:result_2S}(a)) show excellent agreement with our
measurements. This reinforces the robustness of the remarkable
reported value $\beta \approx 0.5$ from
\textcite{Salort_2011_equipartition}, which combines numerical
simulations and a compilation of values extracted from previously published
studies in pipe flows \cite{Ijsselstein79,Walstrom88b,Holmes92}.

\begin{figure}[ht!]  \centering
\includegraphics[width=\linewidth]{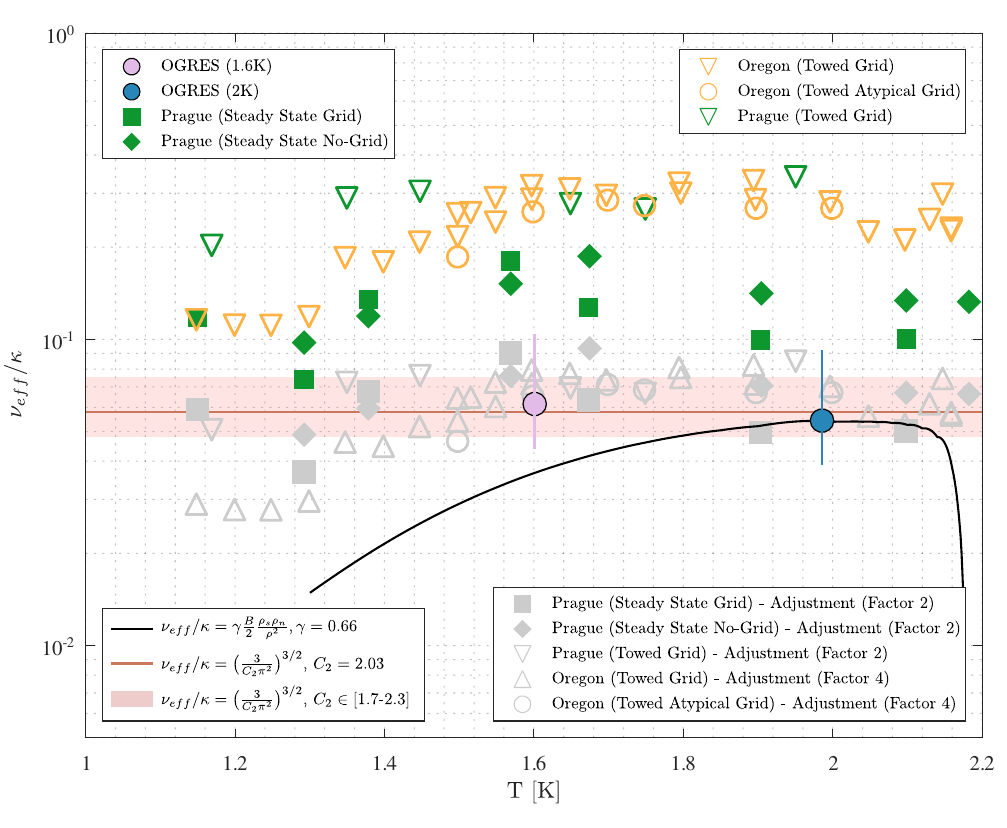}
\caption{Compilation of effective viscosity $\nu_{\text{eff}}$ values as a function of temperature. Our measurements are plotted alongside previous experimental datasets from Prague (Towed Grid \cite{Babuin2013}, Grid and No-Grid Steady State \cite{Babuin14}) and Oregon (Towed Atypical Grid \cite{Oregon_ill_grid} and Towed Grid \cite{Oregon_good_grid}). Note that the Oregon data were corrected by a factor $(3\pi/8)^2$ to account for the sine-squared dependence, as discussed in \cite{Chagovet07}. Also shown are the predicted values of $\nu_{\text{eff}}$ from Eq.~(\ref{nu_eff_babuin_f_mf}) for $\gamma=0.66$ (in black), and from Eq.~(\ref{potential_beta}) (in red).} 
\label{fig:nu_eff_T}
\end{figure}

To further explore the comparison of our data with the one obtained by
\textcite{Babuin14}, Fig.~\ref{fig:nu_eff_T} presents their values of
$\nu_{\text{eff}}$ for various temperatures alongside our
estimates. Since $\nu_{\text{eff}} \propto \beta^4$, even minor
discrepancies in $\beta$ are amplified in this representation.

\textcite{Babuin14} carefully analyzed their results and estimated the uncertainty
on the value of $\nu_{\text{eff}}$ to be of the order of ten
(temperature independent), mainly  due to their indirect estimation of large-scale
flow properties. As shown gray in
Fig.~\ref{fig:nu_eff_T}, with an arbitrary adjustment of a factor of
2, their two steady-state datasets (solid markers) are in excellent agreement with our
measurements. Following their approach, we also plotted the relation
of Eq.~(\ref{nu_eff_babuin_f_mf}), which predicts the temperature
dependence of $\nu_{\text{eff}}$ assuming that dissipation is entirely
due to mutual friction (we adjusted the parameter $\gamma=0.66$ to
match our measurements at \SI{2}{K}). As in \textcite{Babuin14}
measurements, where the temperature likely reflects experimental
noise, our measurements suggest that if a dependence exists, it must
be very weak resulting in a poor agreement of
Eq.~(\ref{nu_eff_babuin_f_mf}) with the measurements. Therefore and as
it is discussed in section \ref{sec:Interpretation}, it seems very unlikely
that mutual friction is responsible for 100\% of the dissipation in the
explored temperature range.

Finally, we also include the available decaying datasets from
\textcite{Oregon_ill_grid}, \textcite{Oregon_good_grid}, and
\textcite{Babuin2013}. As explained in Sec.~\ref{sec:intro}, a key
step in determining the $\nu_{\text{eff}}$ value from these
experiments is assuming that the integral scale $L$ saturates at the
channel diameter $D$. If we adjust this assumption to consider that it
saturates at the channel radius $D/2$, $\nu_{\text{eff}}$ is corrected
by a factor of 4. The corrected data (in grey in
Fig.~\ref{fig:nu_eff_T}) align quite well with our measurements.

To conclude, since we have unambiguously characterized $\epsilon$
using particle velocity measurements  -- something that, to our knowledge,
had never been done before -- we can reasonably consider our values to be
the best current estimates.

\section{\label{sec:Interpretation}Interpretation of the Kolmogorov-like scaling}

As pointed out by \textcite{Babuin14}, although the scaling expressed in Eq.~(\ref{nu_eff}) -- or equivalently Eq.~(\ref{delta_L}) -- has been experimentally observed numerous times across different temperatures and appears remarkably robust, it still lacks a rigorous theoretical justification. Phenomenologically, the introduction of the ``effective viscosity'' $\nu_{\text{eff}}$ is motivated by analogy with the classical expression $\epsilon = \nu \langle \omega^2 \rangle$. This suggests that $\ell$ represents the characteristic scale of a non-viscous dissipation process, namely \textit{mutual friction}. However, as shown on Fig.~\ref{fig:nu_eff_T}, most available data (including our own) in the 1–\SI{2.17}{K} temperature range are incompatible with a scenario in which all dissipation arises solely from mutual friction.

In this section, we aim to revisit a prescient argument proposed by \textcite{Vinen05}, which, despite its potential, does not seem to have led to the conclusive consensus it arguably merits. 

The idea is conceptually straightforward, arising from classical considerations of velocity 
fields in homogeneous and isotropic turbulence within the inertial range of scales, 
bounded by the dissipation scale $\eta$ (whatever the underlying dissipation mechanism)
and the integral scale $L$, i.e. $\eta \ll r \ll L$.
To begin with, we recall that the characteristic circulation $\Gamma(r)$ 
around a loop $\mathcal{C}$ of diameter $r$ within this range of scales is prescribed by
Kolmogorov's 1941 (K41) theory (see Ref.~\cite{Iyer19} for a detailed discussion on the
applicability of K41 in  estimating $\Gamma(r)$), and can be written as:
\begin{equation}\label{circulation}
\Gamma(r) = \oint_{\mathcal{C}} \mathbf{v} \cdot d\mathbf{l} = \pi r \frac{\delta_r v_{\perp}}{2}.
\end{equation}
Since $r$ lies within the inertial range of scales, the transverse velocity increment 
$\delta_r v_{\perp}$ can be estimated as (see, for example, \textcite{Pope00}, pp. 193–194):
\begin{equation*}
(\delta_r v_{\perp})^2 = \frac{4}{3}(\delta_r v_{\parallel})^2 = \frac{4}{3} C_2 (\epsilon r)^{2/3},
\end{equation*}
Therefore, the circulation around any loop $\mathcal{C}$ of diameter
$r$ becomes: 
\begin{equation}\label{circulation_full}
\Gamma(r) = \sqrt{\frac{C_2 \pi^2}{3}}  \epsilon^{1/3} r^{4/3},
\end{equation}
and the intervortex distance $\ell_{K41}$, has to satisfy 
$\Gamma(\ell_{K41}) = \kappa$, yielding
\begin{equation}\label{r_kappa}
\ell_{K41} = \left( \frac{3}{C_2 \pi^2} \right)^{3/8} \left(\frac{\kappa^3}{\epsilon}\right)^{1/4}.
\end{equation}
Using the relation $\epsilon = u_{\text{rms}}^3 / L$, we can rewrite Eq.~(\ref{r_kappa}) in the same form as Eq.~(\ref{delta_L}):
\begin{equation}\label{potential_beta}
\frac{\ell_{K41}}{L} = \underbrace{\left( \frac{3}{C_2 \pi^2} \right)^{3/8}}_{\beta_{K41}} \text{Re}_\kappa^{-3/4}.
\end{equation}
We thus obtain the predicted value of the prefactor $\beta_{K41}$ in the classical K41 framework, which can be used to assess the validity of this interpretation. Given the range $C_2 \in [1.7, 2.3]$ reported in the literature, we find:
\begin{equation}\label{beta_AN}
0.47 < \beta_{K41} < 0.52.
\end{equation}

This value of $\beta_{K41}$, obtained without any adjustable parameters, is in very good 
agreement with the experimentally measured $\beta_{exp}\approx0.5$ and values reported in the 
literature \cite{Salort_2011_equipartition}, yielding:
\begin{equation}\label{ell_equal}
    \ell\approx\ell_{K41}
\end{equation}
This suggests that the Kolmogorov cascade not only 
describes the large-scale structure of quantum turbulence, as was already well established 
\cite{Maurer98,Salort12}, but extends all the way down to the mean intervortex scale $\ell$. 
The temperature independence of the datasets in Fig.~\ref{fig:nu_eff_T} further supports this paradigm, as $\beta_{K41} = \left( \frac{3}{C_2 \pi^2} \right)^{3/8}$ does not depend on temperature.

Pushing this interpretation further, it suggests that the effective viscosity $\nu_{\text{eff}} = \beta^4 \kappa$ does not represent dissipation per se (as proposed by \cite{Babuin14}), but instead characterizes an inertial-scale structural property of the flow.

Nevertheless, it should be noted that this argument appears to break down at very low temperatures. Even though the Kolmogorov-like scaling of Eq.~(\ref{delta_L}) persists, measurements from the Manchester group indicate a reduction in $\beta$ in this regime (see e.g., \cite{Walmsley14}). Identifying the temperature at which Eq.~(\ref{ell_equal}) ceases to be valid remains an open question for future research.

\section{Conclusion and further comments}

In this article, we report an experimental characterization of coflow turbulence in superfluid helium at temperatures of 1.6 and \SI{2}{K}, generated by a double oscillating grid system with adjustable stroke $S_{\text{mot}}$ and frequency $f_{\text{mot}}$ (Sec.~\ref{sec:expsetup}). By seeding the flow with microspheres, we achieved -- for the first time in such a cryogenic experiment -- three-dimensional tracking of particle trajectories (Sec.~\ref{sec:lagrangian_traj}). This allowed a direct measurement of the main flow large scale properties (Sec.~\ref{sec:large_scale_prop}), in particular the dependence of the intensity of velocity fluctuations $u_{\text{rms}}$ on the forcing parameters ($ S_{\text{mot}}$,$f_{\text{mot}}$) as well as the integral length scale $L=u_{\text{rms}}^3/\epsilon$ based on an estimate of $\epsilon$ through pseudo-eulerian velocity structure functions at scales larger than 
$L/20 \approx \SI{1}{mm}$.

In parallel, we determined the mean vortex line density $\mathcal{L}$ by measuring the attenuation of a second-sound standing wave across the flow, and, assuming smooth vortices, we inferred the corresponding intervortex spacing $\ell = 1/\sqrt{\mathcal{L}}$ (Sec.~\ref{sec:secsoundcavity}).
Combining the two approach, our results reveal (Sec.~\ref{sec:effective_viscosity}),
\begin{equation}\label{scalling_ell_beta}
\frac{\ell}{L} = \beta_{\text{exp}} \mathrm{Re_\kappa}^{-3/4},    
\end{equation}
where $\beta_{\text{exp}}\approx0.5$ appears to be a temperature-independent coefficient in our working temperature range (at 1.6 and $\SI{2}{K}$). For the first time, this result is established by fully measuring the properties involved, (in particular, the estimate of the integral scale $L$ derives from direct measurements, and not from a generic assumption). 

We then attempt to interpret this result, particularly what can be learned from the absolute value of $\beta_{\text{exp}}$ and its apparent lack of temperature dependence (Sec.~\ref{sec:Interpretation}).

As presciently predicted by \textcite{Vinen05}, it appears that Eq.~(\ref{scalling_ell_beta}) is in perfect quantitative agreement with the classical scaling laws of a homogeneous and isotropic turbulent cascade: the scale $\ell_{K41}$ in the inertial range at which the circulation around a contour of diameter $\ell_{K41}$ equals $\kappa$ quantitatively corresponds to the intervortex distance, $\ell\approx\ell_{K41}$, and an analytical expression for $\beta$ can be derived:
      $$ \beta_{K41} = \left( \frac{3}{C_2 \pi^2}\right)^{3/8} \approx 0.5 .$$

This interpretation, asserting that classical turbulent scaling laws apply not only to the
upper part of the inertial also range but down to the mean intervortex scale in quantum turbulence,
is further supported by the temperature independence of $\beta$. Indeed,
the fraction of total dissipation attributed to mutual friction is given by:
\begin{equation}\label{prop_eps_mf}
\frac{\epsilon_{\text{mf}}}{\epsilon} = \frac{B}{2} \frac{\rho_s \rho_n}{\rho^2} \frac{\gamma}{\beta^4}.
\end{equation}
In the scenario proposed by \textcite{Babuin14}, where all dissipation is attributed to mutual
friction ($\epsilon_{\text{mf}} \sim \epsilon$), Eq.~(\ref{prop_eps_mf}) implies that $\beta^4 \propto \frac{B}{2} \frac{\rho_s \rho_n}{\rho^2}$, thus predicting a temperature dependence of $\beta$ that we do not observe (see Fig.~\ref{fig:nu_eff_T}).

The coincidence in both the value and temperature independence of $\beta_{exp}$ and $\beta_{K41}$ strongly suggests that the traditional interpretation of the effective viscosity $\nu_{\text{eff}}$ as a manifestation of mutual friction is likely incorrect. Rather, it reflects an inertial-scale structural feature of the flow.

\begin{figure}[ht!]
    \centering
    \includegraphics[width=12cm]{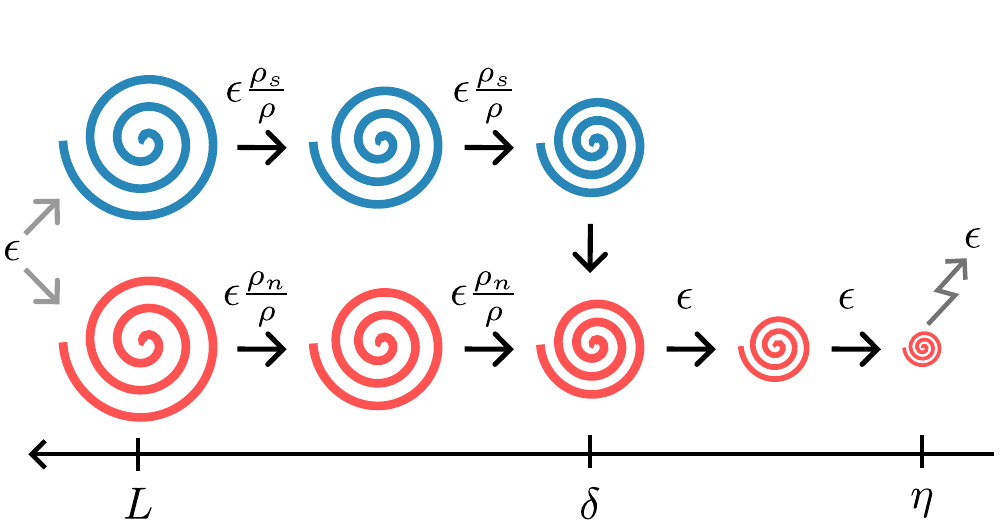}
    \caption{Adaptation of the Richardson Cascade for Quantum Turbulence.}
    \label{fig:Cascade_Quantum}
\end{figure}

All of these considerations support a schematic representation of turbulent cascades in coflow experiments at $T \gtrsim \SI{1}{K}$, as illustrated in Fig.~\ref{fig:Cascade_Quantum}. At large scales, energy is injected into both the normal and superfluid components, weighted by their respective densities. Classical turbulence mechanisms transfer energy down to the scale $\ell$ in the superfluid, and to the viscous dissipation scale $\eta$ in the normal component. Mutual friction couples the two components, enabling energy exchange with negligible dissipation. 
 The energy carried by the superfluid component is ultimately transferred and dissipated by the normal component at smaller scales, $\eta = (\nu^3/\epsilon)^{1/4}$, where $\nu = \mu / \rho$. This further implies that the counterflow velocity remains small compared to the characteristic velocity at scale $\ell$, indicating that the normal and superfluid components remain locked down to 
 that scale. This scenario is confirmed by recent results obtained using a two fluid 
 Hall-Vinen-Bekharevich-Khalatnikov (HVBK) numerical
 study~\cite{Polanco25}. 
A more detailed discussion about the intensity of mutual friction dissipation
can be found in \textcite{Bret24_PHD}. 

Finally, a direct trivial implication of these considerations is that for flows at $T > \SI{1}{K}$ -- where the normal component is still present -- the intervortex distance $\ell$ is greater than the viscous dissipation scale $\eta$, with $$\frac{\ell}{\eta} = \beta \left( \frac{\kappa}{\nu(T)} \right)^{3/4}$$ leading to $1.4<\ell/\eta<3.2$.

\section{Acknowledgment}
The authors warmly thank Jérome Chartier for his technical support in setting up and running the experiments. We also thank Julien Salort, Elian Bernard and Mickaël Bourgoin from the Laboratoire de Physique de l'ENS de Lyon for lending us the MiniShaker and assisting us in its use, which enabled us to capture for the first time 3D particle trajectories in superfluid helium. We are
also thankful to the members of the Nanofab team at Institut Néel, who did the $\mathrm{Al}$ deposits on 
the porous membranes of the second sound sensors. Finally we would like to thank our colleagues
from the informal \textit{TUrbulent SUperfluid GREnoble} group, and especially Juan Ignacio
Polanco, Mathieu Gibert and Philippe-Emmanuel Roche, for their valuable discussions and
step-by-step insights into the interpretation of our experimental results.

\section{Appendix}
\appendix
\section{Mean flow correction \label{annexe_A}}
\subsection{On the 3D Trajectories}

A well-known challenge in oscillating grid experiments is the tendency of the flow to develop a parasitic recirculation, typically sensitive to initial conditions and exhibiting complex spatial structures \cite{McKenna04}. In our setup, the initial conditions are inherently non-reproducible due to slight but continuous variations in the liquid level during the experiment (on the order of $\SI{1}{mm/min}$). Moreover, we suspect that the secondary flow is further amplified by the in-phase oscillation of the grids -- a constraint imposed by the cryostat geometry -- which is likely less favorable than an out-of-phase configuration. The resulting mean flow often manifests as a net vertical drift of the particles, typically downward, though upward drifts are also occasionally observed.

To isolate the turbulent velocity fluctuations from the background mean flow, we apply a straightforward correction procedure. The 3D Eulerian mean velocity field $\mathbf{U}(\mathbf{x})$ is computed by discretizing the measurement volume into cubic cells of size $1 \times 1 \times \SI{1}{mm^3}$ and averaging the particle velocities of all trajectories passing through each cell over the entire duration of the run. Given the long acquisition time of our 3D experiments (over ten seconds), which greatly exceeds the turbulence integral time scale (a few seconds), the temporal averaging effectively suppresses turbulent fluctuations. As a result, to extract the turbulent velocity component of each particle along its trajectory, we simply subtract the interpolated mean flow value at the particle's position.

\begin{figure}[ht!]
\centering
\includegraphics[width=\textwidth]{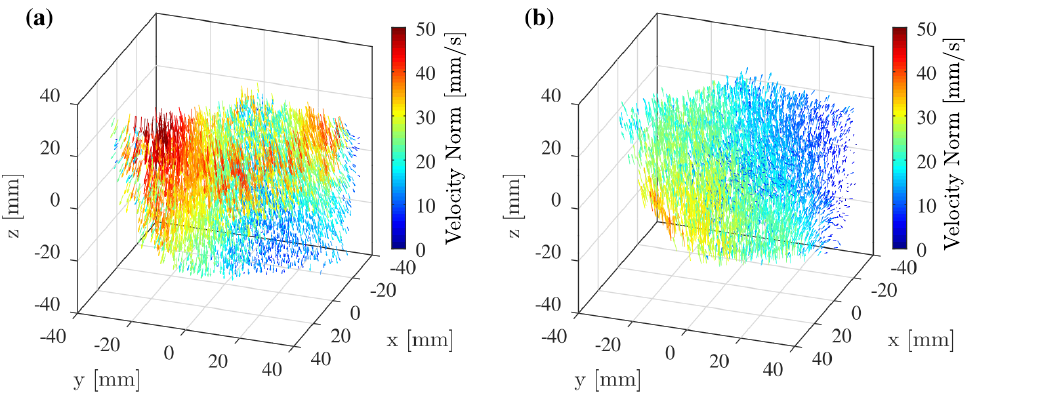}
\caption{Visualization of the recirculating mean flow for two 3D runs. (a) Run3D-2 with $f_{\text{mot}} = \SI{6}{Hz}$ and $S_{\text{mot}} = \SI{20}{mm}$: downward flow of $\sim\SI{25}{mm/s}$. (b) Run3D-7 under identical conditions: upward flow of $\sim\SI{17}{mm/s}$.}
\label{fig:mean_flow}
\end{figure}

Figure~\ref{fig:mean_flow} shows the resulting mean flow field for two of the 3D runs. In addition to the irreproducibility of the mean flow direction under nominally identical forcing conditions, Fig.~\ref{fig:mean_flow} reveals substantial spatial inhomogeneities. If not subtracted, this mean flow would thus introduce a significant bias in the analysis of particle velocities.

Finally, the stationarity of the mean flow has also been verified by comparing the full-run mean field with that computed from only the first half of the acquisition. 

\subsection{On the 2\texorpdfstring{$\times$}{x}2D Trajectories}

In order to suppress the recirculating flow on the 2$\times$2D
trajectories, we apply the same methodology as for the 3D
measurements. The image is divided into cells of $40 \times
\SI{40}{px^2}$ (roughly $0.8 \times 0.8~\si{mm^2}$), in which we
compute the average particle velocity. Since the depth of field of
each camera is approximately \SI{30}{mm}, the resulting mean flow
corresponds to a 2D projection averaged along the optical axis. This
averaging smooths out any depth-dependent inhomogeneities and may
introduce a bias. 

However, Fig.~\ref{fig:mean_flow} suggests that the strongest inhomogeneities in the mean flow occur on spatial scales larger than the depth of field. The typical measurement volume for a single camera in the 2$\times$2D setup is approximately $17 \times 17 \times \SI{30}{mm^3}$, a small fraction of the total 3D volume. The associated bias is thus expected to be limited.

\begin{figure}[ht!]
\centering
\includegraphics[width=\textwidth]{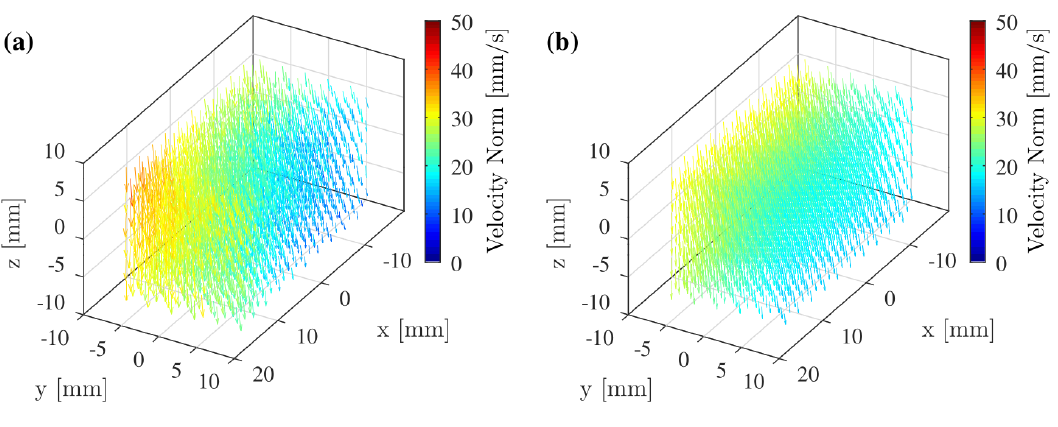}
\caption{Impact of the projection bias in 2$\times$2D measurements, illustrated for Run3D-2. (a) Truncated 3D mean flow field, matching the volume of one 2$\times$2D camera. (b) Field averaged along $x$ and extended from $-L$ to $+L$ for comparison.}
\label{fig:mean_flow_cut2D}
\end{figure}

To assess this quantitatively, we consider $\tilde{\mathbf{U}}$, the 3D mean flow truncated to the volume of a single 2$\times$2D camera, and compute its depth-averaged version:
$$
\tilde{\mathbf{U}}_{\text{2D}}(y,z) = \frac{1}{2L} \int_{-L}^{L} \tilde{\mathbf{U}}(x,y,z)\,\mathrm{d}x,
$$
with $L = \SI{15}{mm}$. This projection is then duplicated along $x \in [-L, L]$ to span the full depth of field and allow pointwise comparison with $\tilde{\mathbf{U}}$ (see Fig.~\ref{fig:mean_flow_cut2D}). Subtracting the two fields yields the residual, whose variance we compare to that of the turbulent velocity fluctuations.

We find that the impact of the projection is negligible (less than 3\%) for the horizontal components, but can reach up to 20\% for the vertical component. For this reason, the interpretation of the 2$\times$2D measurements is restricted to the horizontal velocity components.

\bibliography{OGRES_eff_visc}

\begin{thebibliography}{71}%
\makeatletter
\providecommand \@ifxundefined [1]{%
 \@ifx{#1\undefined}
}%
\providecommand \@ifnum [1]{%
 \ifnum #1\expandafter \@firstoftwo
 \else \expandafter \@secondoftwo
 \fi
}%
\providecommand \@ifx [1]{%
 \ifx #1\expandafter \@firstoftwo
 \else \expandafter \@secondoftwo
 \fi
}%
\providecommand \natexlab [1]{#1}%
\providecommand \enquote  [1]{``#1''}%
\providecommand \bibnamefont  [1]{#1}%
\providecommand \bibfnamefont [1]{#1}%
\providecommand \citenamefont [1]{#1}%
\providecommand \href@noop [0]{\@secondoftwo}%
\providecommand \href [0]{\begingroup \@sanitize@url \@href}%
\providecommand \@href[1]{\@@startlink{#1}\@@href}%
\providecommand \@@href[1]{\endgroup#1\@@endlink}%
\providecommand \@sanitize@url [0]{\catcode `\\12\catcode `\$12\catcode
  `\&12\catcode `\#12\catcode `\^12\catcode `\_12\catcode `\%12\relax}%
\providecommand \@@startlink[1]{}%
\providecommand \@@endlink[0]{}%
\providecommand \url  [0]{\begingroup\@sanitize@url \@url }%
\providecommand \@url [1]{\endgroup\@href {#1}{\urlprefix }}%
\providecommand \urlprefix  [0]{URL }%
\providecommand \Eprint [0]{\href }%
\providecommand \doibase [0]{https://doi.org/}%
\providecommand \selectlanguage [0]{\@gobble}%
\providecommand \bibinfo  [0]{\@secondoftwo}%
\providecommand \bibfield  [0]{\@secondoftwo}%
\providecommand \translation [1]{[#1]}%
\providecommand \BibitemOpen [0]{}%
\providecommand \bibitemStop [0]{}%
\providecommand \bibitemNoStop [0]{.\EOS\space}%
\providecommand \EOS [0]{\spacefactor3000\relax}%
\providecommand \BibitemShut  [1]{\csname bibitem#1\endcsname}%
\let\auto@bib@innerbib\@empty
\bibitem [{\citenamefont {Dana}\ and\ \citenamefont {Onnes}(1926)}]{Dana26}%
  \BibitemOpen
  \bibfield  {author} {\bibinfo {author} {\bibfnamefont {L.~I.}\ \bibnamefont
  {Dana}}\ and\ \bibinfo {author} {\bibfnamefont {K.~H.}\ \bibnamefont
  {Onnes}},\ }\bibfield  {title} {\bibinfo {title} {Further experiments with
  liquid helium. b. a. preliminary determinations of the latent heat of
  vaporization of liquid helium},\ }in\ \href
  {https://dwc.knaw.nl/DL/publications/PU00014608.pdf} {\emph {\bibinfo
  {booktitle} {Proceedings of the Royal Academy of Amsterdam}}},\ Vol.~\bibinfo
  {volume} {29}\ (\bibinfo  {publisher} {Proceedings of the Royal Academy of
  Amsterdam},\ \bibinfo {year} {1926})\ pp.\ \bibinfo {pages}
  {1051--1061}\BibitemShut {NoStop}%
\bibitem [{\citenamefont {Wolfke}\ and\ \citenamefont
  {Keesom}(1927)}]{Wolfke27}%
  \BibitemOpen
  \bibfield  {author} {\bibinfo {author} {\bibfnamefont {M.}~\bibnamefont
  {Wolfke}}\ and\ \bibinfo {author} {\bibfnamefont {W.~H.}\ \bibnamefont
  {Keesom}},\ }\bibfield  {title} {\bibinfo {title} {On the change of the
  dielectric constant of liquid helium with the temperature. provisional
  measurements.},\ }in\ \href
  {https://dwc.knaw.nl/DL/publications/PU00014611.pdf} {\emph {\bibinfo
  {booktitle} {Proceedings of the Royal Academy of Amsterdam}}},\ Vol.~\bibinfo
  {volume} {31}\ (\bibinfo {year} {1927})\ pp.\ \bibinfo {pages}
  {81--89}\BibitemShut {NoStop}%
\bibitem [{\citenamefont {Keesom}\ and\ \citenamefont
  {Wolfke}(1927)}]{Keesom27}%
  \BibitemOpen
  \bibfield  {author} {\bibinfo {author} {\bibfnamefont {W.~H.}\ \bibnamefont
  {Keesom}}\ and\ \bibinfo {author} {\bibfnamefont {M.}~\bibnamefont
  {Wolfke}},\ }\bibfield  {title} {\bibinfo {title} {Two different liquid
  states af helium},\ }in\ \href
  {https://dwc.knaw.nl/DL/publications/PU00014611.pdf} {\emph {\bibinfo
  {booktitle} {Proceedings of the Royal Academy of Amsterdam}}},\ Vol.~\bibinfo
  {volume} {31}\ (\bibinfo {year} {1927})\ pp.\ \bibinfo {pages}
  {90--94}\BibitemShut {NoStop}%
\bibitem [{\citenamefont {Kapitza}(1938)}]{Kapitza38}%
  \BibitemOpen
  \bibfield  {author} {\bibinfo {author} {\bibfnamefont {P.}~\bibnamefont
  {Kapitza}},\ }\bibfield  {title} {\bibinfo {title} {{Viscosity of Liquid
  Helium below the {\ensuremath{\lambda}}-Point}},\ }\href
  {https://doi.org/10.1038/141074a0} {\bibfield  {journal} {\bibinfo  {journal}
  {nat}\ }\textbf {\bibinfo {volume} {141}},\ \bibinfo {pages} {74} (\bibinfo
  {year} {1938})}\BibitemShut {NoStop}%
\bibitem [{\citenamefont {Allen}\ and\ \citenamefont
  {Misener}(1938)}]{Allen38}%
  \BibitemOpen
  \bibfield  {author} {\bibinfo {author} {\bibfnamefont {J.~F.}\ \bibnamefont
  {Allen}}\ and\ \bibinfo {author} {\bibfnamefont {A.~D.}\ \bibnamefont
  {Misener}},\ }\bibfield  {title} {\bibinfo {title} {Flow of liquid helium
  ii},\ }\href {https://doi.org/10.1038/141075a0} {\bibfield  {journal}
  {\bibinfo  {journal} {Nature}\ }\textbf {\bibinfo {volume} {141}},\ \bibinfo
  {pages} {75} (\bibinfo {year} {1938})}\BibitemShut {NoStop}%
\bibitem [{\citenamefont {London}(1938)}]{London38}%
  \BibitemOpen
  \bibfield  {author} {\bibinfo {author} {\bibfnamefont {F.}~\bibnamefont
  {London}},\ }\bibfield  {title} {\bibinfo {title} {On the bose-einstein
  condensation},\ }\href {https://doi.org/10.1103/PhysRev.54.947} {\bibfield
  {journal} {\bibinfo  {journal} {Phys. Rev.}\ }\textbf {\bibinfo {volume}
  {54}},\ \bibinfo {pages} {947} (\bibinfo {year} {1938})}\BibitemShut
  {NoStop}%
\bibitem [{\citenamefont {Tisza}(1938)}]{Tisza_two_fluids_38}%
  \BibitemOpen
  \bibfield  {author} {\bibinfo {author} {\bibfnamefont {L.}~\bibnamefont
  {Tisza}},\ }\bibfield  {title} {\bibinfo {title} {Transport phenomena in
  helium ii},\ }\href {https://doi.org/10.1038/141913a0} {\bibfield  {journal}
  {\bibinfo  {journal} {Nature}\ }\textbf {\bibinfo {volume} {141}},\ \bibinfo
  {pages} {913} (\bibinfo {year} {1938})}\BibitemShut {NoStop}%
\bibitem [{\citenamefont {Landau}(1941)}]{Landau_two_fluids_41}%
  \BibitemOpen
  \bibfield  {author} {\bibinfo {author} {\bibfnamefont {L.}~\bibnamefont
  {Landau}},\ }\bibfield  {title} {\bibinfo {title} {Theory of the
  superfluidity of helium ii},\ }\href {https://doi.org/10.1103/PhysRev.60.356}
  {\bibfield  {journal} {\bibinfo  {journal} {Phys. Rev.}\ }\textbf {\bibinfo
  {volume} {60}},\ \bibinfo {pages} {356} (\bibinfo {year} {1941})}\BibitemShut
  {NoStop}%
\bibitem [{\citenamefont {Osborne}(1950)}]{Osborne50}%
  \BibitemOpen
  \bibfield  {author} {\bibinfo {author} {\bibfnamefont {D.~V.}\ \bibnamefont
  {Osborne}},\ }\bibfield  {title} {\bibinfo {title} {The rotation of liquid
  helium ii},\ }\href {https://doi.org/10.1088/0370-1298/63/8/315} {\bibfield
  {journal} {\bibinfo  {journal} {Proceedings of the Physical Society. Section
  A}\ }\textbf {\bibinfo {volume} {63}},\ \bibinfo {pages} {909} (\bibinfo
  {year} {1950})}\BibitemShut {NoStop}%
\bibitem [{\citenamefont {Onsager}(1949)}]{Onsager1949}%
  \BibitemOpen
  \bibfield  {author} {\bibinfo {author} {\bibfnamefont {L.}~\bibnamefont
  {Onsager}},\ }\bibfield  {title} {\bibinfo {title} {Statistical
  hydrodynamics},\ }\href {https://doi.org/10.1007/BF02780991} {\bibfield
  {journal} {\bibinfo  {journal} {Il Nuovo Cimento (1943-1954)}\ }\textbf
  {\bibinfo {volume} {6}},\ \bibinfo {pages} {279} (\bibinfo {year}
  {1949})}\BibitemShut {NoStop}%
\bibitem [{\citenamefont {Feynman}(1955)}]{Feynman55}%
  \BibitemOpen
  \bibfield  {author} {\bibinfo {author} {\bibfnamefont {R.}~\bibnamefont
  {Feynman}},\ }\bibfield  {title} {\bibinfo {title} {Chapter ii application of
  quantum mechanics to liquid helium}\ }(\bibinfo  {publisher} {Elsevier},\
  \bibinfo {year} {1955})\ pp.\ \bibinfo {pages} {17--53}\BibitemShut {NoStop}%
\bibitem [{\citenamefont {Hall}\ \emph
  {et~al.}(1956{\natexlab{a}})\citenamefont {Hall}, \citenamefont {Vinen},\
  and\ \citenamefont {Shoenberg}}]{Hall56a}%
  \BibitemOpen
  \bibfield  {author} {\bibinfo {author} {\bibfnamefont {H.~E.}\ \bibnamefont
  {Hall}}, \bibinfo {author} {\bibfnamefont {W.~F.}\ \bibnamefont {Vinen}},\
  and\ \bibinfo {author} {\bibfnamefont {D.}~\bibnamefont {Shoenberg}},\
  }\bibfield  {title} {\bibinfo {title} {The rotation of liquid helium ii i.
  experiments on the propagation of second sound in uniformly rotating helium
  ii},\ }\href {https://doi.org/10.1098/rspa.1956.0214} {\bibfield  {journal}
  {\bibinfo  {journal} {Proceedings of the Royal Society of London. Series A.
  Mathematical and Physical Sciences}\ }\textbf {\bibinfo {volume} {238}},\
  \bibinfo {pages} {204} (\bibinfo {year} {1956}{\natexlab{a}})}\BibitemShut
  {NoStop}%
\bibitem [{\citenamefont {Hall}\ \emph
  {et~al.}(1956{\natexlab{b}})\citenamefont {Hall}, \citenamefont {Vinen},\
  and\ \citenamefont {Shoenberg}}]{Hall56b}%
  \BibitemOpen
  \bibfield  {author} {\bibinfo {author} {\bibfnamefont {H.~E.}\ \bibnamefont
  {Hall}}, \bibinfo {author} {\bibfnamefont {W.~F.}\ \bibnamefont {Vinen}},\
  and\ \bibinfo {author} {\bibfnamefont {D.}~\bibnamefont {Shoenberg}},\
  }\bibfield  {title} {\bibinfo {title} {The rotation of liquid helium ii ii.
  the theory of mutual friction in uniformly rotating helium ii},\ }\href
  {https://doi.org/10.1098/rspa.1956.0215} {\bibfield  {journal} {\bibinfo
  {journal} {Proceedings of the Royal Society of London. Series A. Mathematical
  and Physical Sciences}\ }\textbf {\bibinfo {volume} {238}},\ \bibinfo {pages}
  {215} (\bibinfo {year} {1956}{\natexlab{b}})}\BibitemShut {NoStop}%
\bibitem [{\citenamefont {Bekarevich}\ and\ \citenamefont
  {Khalatnikov}(1961)}]{Bekarevich61}%
  \BibitemOpen
  \bibfield  {author} {\bibinfo {author} {\bibfnamefont {I.}~\bibnamefont
  {Bekarevich}}\ and\ \bibinfo {author} {\bibfnamefont {I.~M.}\ \bibnamefont
  {Khalatnikov}},\ }\bibfield  {title} {\bibinfo {title} {Phenomenological
  derivation of the equations of vortex motion in he ii},\ }\href@noop {}
  {\bibfield  {journal} {\bibinfo  {journal} {Sov. Phys. JETP}\ }\textbf
  {\bibinfo {volume} {13}},\ \bibinfo {pages} {643} (\bibinfo {year}
  {1961})}\BibitemShut {NoStop}%
\bibitem [{\citenamefont {{Walstrom}}\ \emph {et~al.}(1988)\citenamefont
  {{Walstrom}}, \citenamefont {{Weisend}}, \citenamefont {{Maddocks}},\ and\
  \citenamefont {{van Sciver}}}]{Walstrom88a}%
  \BibitemOpen
  \bibfield  {author} {\bibinfo {author} {\bibfnamefont {P.~L.}\ \bibnamefont
  {{Walstrom}}}, \bibinfo {author} {\bibfnamefont {I.}~\bibnamefont
  {{Weisend}}, \bibfnamefont {J.~G.}}, \bibinfo {author} {\bibfnamefont
  {J.~R.}\ \bibnamefont {{Maddocks}}},\ and\ \bibinfo {author} {\bibfnamefont
  {S.~W.}\ \bibnamefont {{van Sciver}}},\ }\bibfield  {title} {\bibinfo {title}
  {{Turbulent flow pressure drop in various He II transfer system
  components}},\ }\href {https://doi.org/10.1016/0011-2275(88)90054-9}
  {\bibfield  {journal} {\bibinfo  {journal} {Cryogenics}\ }\textbf {\bibinfo
  {volume} {28}},\ \bibinfo {pages} {101} (\bibinfo {year} {1988})}\BibitemShut
  {NoStop}%
\bibitem [{\citenamefont {Rousset}\ \emph {et~al.}(1994)\citenamefont
  {Rousset}, \citenamefont {Claudet}, \citenamefont {Gauthier}, \citenamefont
  {Seyfert}, \citenamefont {Martinez}, \citenamefont {Lebrun}, \citenamefont
  {Marquet},\ and\ \citenamefont {{Van Weelderen}}}]{Rousset94}%
  \BibitemOpen
  \bibfield  {author} {\bibinfo {author} {\bibfnamefont {B.}~\bibnamefont
  {Rousset}}, \bibinfo {author} {\bibfnamefont {G.}~\bibnamefont {Claudet}},
  \bibinfo {author} {\bibfnamefont {A.}~\bibnamefont {Gauthier}}, \bibinfo
  {author} {\bibfnamefont {P.}~\bibnamefont {Seyfert}}, \bibinfo {author}
  {\bibfnamefont {A.}~\bibnamefont {Martinez}}, \bibinfo {author}
  {\bibfnamefont {P.}~\bibnamefont {Lebrun}}, \bibinfo {author} {\bibfnamefont
  {M.}~\bibnamefont {Marquet}},\ and\ \bibinfo {author} {\bibfnamefont
  {R.}~\bibnamefont {{Van Weelderen}}},\ }\bibfield  {title} {\bibinfo {title}
  {Pressure drop and transient heat transport in forced flow single phase
  helium ii at high reynoldsnumbers},\ }\href
  {https://doi.org/https://doi.org/10.1016/S0011-2275(05)80070-0} {\bibfield
  {journal} {\bibinfo  {journal} {Cryogenics}\ }\textbf {\bibinfo {volume}
  {34}},\ \bibinfo {pages} {317} (\bibinfo {year} {1994})},\ \bibinfo {note}
  {fifteenth International Cryogenic Engineering Conference}\BibitemShut
  {NoStop}%
\bibitem [{\citenamefont {Fuzier}\ \emph {et~al.}(2001)\citenamefont {Fuzier},
  \citenamefont {Baudouy},\ and\ \citenamefont {{Van Sciver}}}]{Fuzier01}%
  \BibitemOpen
  \bibfield  {author} {\bibinfo {author} {\bibfnamefont {S.}~\bibnamefont
  {Fuzier}}, \bibinfo {author} {\bibfnamefont {B.}~\bibnamefont {Baudouy}},\
  and\ \bibinfo {author} {\bibfnamefont {S.}~\bibnamefont {{Van Sciver}}},\
  }\bibfield  {title} {\bibinfo {title} {Steady-state pressure drop and heat
  transfer in he ii forced flow at high reynolds number},\ }\href
  {https://doi.org/https://doi.org/10.1016/S0011-2275(01)00090-X} {\bibfield
  {journal} {\bibinfo  {journal} {Cryogenics}\ }\textbf {\bibinfo {volume}
  {41}},\ \bibinfo {pages} {453} (\bibinfo {year} {2001})}\BibitemShut
  {NoStop}%
\bibitem [{\citenamefont {Rousset}\ \emph {et~al.}(2014)\citenamefont
  {Rousset}, \citenamefont {Bonnay}, \citenamefont {Diribarne}, \citenamefont
  {Girard}, \citenamefont {Poncet}, \citenamefont {Herbert}, \citenamefont
  {Salort}, \citenamefont {Baudet}, \citenamefont {Castaing}, \citenamefont
  {Chevillard}, \citenamefont {Daviaud}, \citenamefont {Dubrulle},
  \citenamefont {Gagne}, \citenamefont {Gibert}, \citenamefont {Hébral},
  \citenamefont {Lehner}, \citenamefont {Roche}, \citenamefont {Saint-Michel},\
  and\ \citenamefont {Bon~Mardion}}]{Rousset14}%
  \BibitemOpen
  \bibfield  {author} {\bibinfo {author} {\bibfnamefont {B.}~\bibnamefont
  {Rousset}}, \bibinfo {author} {\bibfnamefont {P.}~\bibnamefont {Bonnay}},
  \bibinfo {author} {\bibfnamefont {P.}~\bibnamefont {Diribarne}}, \bibinfo
  {author} {\bibfnamefont {A.}~\bibnamefont {Girard}}, \bibinfo {author}
  {\bibfnamefont {J.~M.}\ \bibnamefont {Poncet}}, \bibinfo {author}
  {\bibfnamefont {E.}~\bibnamefont {Herbert}}, \bibinfo {author} {\bibfnamefont
  {J.}~\bibnamefont {Salort}}, \bibinfo {author} {\bibfnamefont
  {C.}~\bibnamefont {Baudet}}, \bibinfo {author} {\bibfnamefont
  {B.}~\bibnamefont {Castaing}}, \bibinfo {author} {\bibfnamefont
  {L.}~\bibnamefont {Chevillard}}, \bibinfo {author} {\bibfnamefont
  {F.}~\bibnamefont {Daviaud}}, \bibinfo {author} {\bibfnamefont
  {B.}~\bibnamefont {Dubrulle}}, \bibinfo {author} {\bibfnamefont
  {Y.}~\bibnamefont {Gagne}}, \bibinfo {author} {\bibfnamefont
  {M.}~\bibnamefont {Gibert}}, \bibinfo {author} {\bibfnamefont
  {B.}~\bibnamefont {Hébral}}, \bibinfo {author} {\bibfnamefont
  {T.}~\bibnamefont {Lehner}}, \bibinfo {author} {\bibfnamefont {P.-E.}\
  \bibnamefont {Roche}}, \bibinfo {author} {\bibfnamefont {B.}~\bibnamefont
  {Saint-Michel}},\ and\ \bibinfo {author} {\bibfnamefont {M.}~\bibnamefont
  {Bon~Mardion}},\ }\bibfield  {title} {\bibinfo {title} {{Superfluid high
  REynolds von Kármán experiment}},\ }\href
  {https://doi.org/10.1063/1.4897542} {\bibfield  {journal} {\bibinfo
  {journal} {Review of Scientific Instruments}\ }\textbf {\bibinfo {volume}
  {85}},\ \bibinfo {pages} {103908} (\bibinfo {year} {2014})}\BibitemShut
  {NoStop}%
\bibitem [{\citenamefont {Saint-Michel}\ \emph {et~al.}(2014)\citenamefont
  {Saint-Michel}, \citenamefont {Herbert}, \citenamefont {Salort},
  \citenamefont {Baudet}, \citenamefont {Bon~Mardion}, \citenamefont {Bonnay},
  \citenamefont {Bourgoin}, \citenamefont {Castaing}, \citenamefont
  {Chevillard}, \citenamefont {Daviaud}, \citenamefont {Diribarne},
  \citenamefont {Dubrulle}, \citenamefont {Gagne}, \citenamefont {Gibert},
  \citenamefont {Girard}, \citenamefont {Hébral}, \citenamefont {Lehner},
  \citenamefont {Rousset},\ and\ \citenamefont
  {Collaboration}}]{SaintMichel14_Kp}%
  \BibitemOpen
  \bibfield  {author} {\bibinfo {author} {\bibfnamefont {B.}~\bibnamefont
  {Saint-Michel}}, \bibinfo {author} {\bibfnamefont {E.}~\bibnamefont
  {Herbert}}, \bibinfo {author} {\bibfnamefont {J.}~\bibnamefont {Salort}},
  \bibinfo {author} {\bibfnamefont {C.}~\bibnamefont {Baudet}}, \bibinfo
  {author} {\bibfnamefont {M.}~\bibnamefont {Bon~Mardion}}, \bibinfo {author}
  {\bibfnamefont {P.}~\bibnamefont {Bonnay}}, \bibinfo {author} {\bibfnamefont
  {M.}~\bibnamefont {Bourgoin}}, \bibinfo {author} {\bibfnamefont
  {B.}~\bibnamefont {Castaing}}, \bibinfo {author} {\bibfnamefont
  {L.}~\bibnamefont {Chevillard}}, \bibinfo {author} {\bibfnamefont
  {F.}~\bibnamefont {Daviaud}}, \bibinfo {author} {\bibfnamefont
  {P.}~\bibnamefont {Diribarne}}, \bibinfo {author} {\bibfnamefont
  {B.}~\bibnamefont {Dubrulle}}, \bibinfo {author} {\bibfnamefont
  {Y.}~\bibnamefont {Gagne}}, \bibinfo {author} {\bibfnamefont
  {M.}~\bibnamefont {Gibert}}, \bibinfo {author} {\bibfnamefont
  {A.}~\bibnamefont {Girard}}, \bibinfo {author} {\bibfnamefont
  {B.}~\bibnamefont {Hébral}}, \bibinfo {author} {\bibfnamefont
  {T.}~\bibnamefont {Lehner}}, \bibinfo {author} {\bibfnamefont
  {B.}~\bibnamefont {Rousset}},\ and\ \bibinfo {author} {\bibfnamefont
  {S.}~\bibnamefont {Collaboration}},\ }\bibfield  {title} {\bibinfo {title}
  {{Probing quantum and classical turbulence analogy in von Kármán liquid
  helium, nitrogen, and water experiments}},\ }\href
  {https://doi.org/10.1063/1.4904378} {\bibfield  {journal} {\bibinfo
  {journal} {Physics of Fluids}\ }\textbf {\bibinfo {volume} {26}},\ \bibinfo
  {pages} {125109} (\bibinfo {year} {2014})}\BibitemShut {NoStop}%
\bibitem [{\citenamefont {Maurer}\ and\ \citenamefont
  {Tabeling}(1998)}]{Maurer98}%
  \BibitemOpen
  \bibfield  {author} {\bibinfo {author} {\bibfnamefont {J.}~\bibnamefont
  {Maurer}}\ and\ \bibinfo {author} {\bibfnamefont {P.}~\bibnamefont
  {Tabeling}},\ }\bibfield  {title} {\bibinfo {title} {Local investigation of
  superfluid turbulence},\ }\href {https://doi.org/10.1209/epl/i1998-00314-9}
  {\bibfield  {journal} {\bibinfo  {journal} {Europhysics Letters ({EPL})}\
  }\textbf {\bibinfo {volume} {43}},\ \bibinfo {pages} {29} (\bibinfo {year}
  {1998})}\BibitemShut {NoStop}%
\bibitem [{\citenamefont {Salort}\ \emph {et~al.}(2010)\citenamefont {Salort},
  \citenamefont {Baudet}, \citenamefont {Castaing}, \citenamefont {Chabaud},
  \citenamefont {Daviaud}, \citenamefont {Didelot}, \citenamefont {Diribarne},
  \citenamefont {Dubrulle}, \citenamefont {Gagne}, \citenamefont {Gauthier},
  \citenamefont {Girard}, \citenamefont {Hébral}, \citenamefont {Rousset},
  \citenamefont {Thibault},\ and\ \citenamefont {Roche}}]{Salort10}%
  \BibitemOpen
  \bibfield  {author} {\bibinfo {author} {\bibfnamefont {J.}~\bibnamefont
  {Salort}}, \bibinfo {author} {\bibfnamefont {C.}~\bibnamefont {Baudet}},
  \bibinfo {author} {\bibfnamefont {B.}~\bibnamefont {Castaing}}, \bibinfo
  {author} {\bibfnamefont {B.}~\bibnamefont {Chabaud}}, \bibinfo {author}
  {\bibfnamefont {F.}~\bibnamefont {Daviaud}}, \bibinfo {author} {\bibfnamefont
  {T.}~\bibnamefont {Didelot}}, \bibinfo {author} {\bibfnamefont
  {P.}~\bibnamefont {Diribarne}}, \bibinfo {author} {\bibfnamefont
  {B.}~\bibnamefont {Dubrulle}}, \bibinfo {author} {\bibfnamefont
  {Y.}~\bibnamefont {Gagne}}, \bibinfo {author} {\bibfnamefont
  {F.}~\bibnamefont {Gauthier}}, \bibinfo {author} {\bibfnamefont
  {A.}~\bibnamefont {Girard}}, \bibinfo {author} {\bibfnamefont
  {B.}~\bibnamefont {Hébral}}, \bibinfo {author} {\bibfnamefont
  {B.}~\bibnamefont {Rousset}}, \bibinfo {author} {\bibfnamefont
  {P.}~\bibnamefont {Thibault}},\ and\ \bibinfo {author} {\bibfnamefont
  {P.-E.}\ \bibnamefont {Roche}},\ }\bibfield  {title} {\bibinfo {title}
  {Turbulent velocity spectra in superfluid flows},\ }\href
  {https://doi.org/10.1063/1.3504375} {\bibfield  {journal} {\bibinfo
  {journal} {Physics of Fluids}\ }\textbf {\bibinfo {volume} {22}},\ \bibinfo
  {pages} {125102} (\bibinfo {year} {2010})}\BibitemShut {NoStop}%
\bibitem [{\citenamefont {Salort}\ \emph {et~al.}(2012)\citenamefont {Salort},
  \citenamefont {Chabaud}, \citenamefont {Lévêque},\ and\ \citenamefont
  {Roche}}]{Salort12}%
  \BibitemOpen
  \bibfield  {author} {\bibinfo {author} {\bibfnamefont {J.}~\bibnamefont
  {Salort}}, \bibinfo {author} {\bibfnamefont {B.}~\bibnamefont {Chabaud}},
  \bibinfo {author} {\bibfnamefont {E.}~\bibnamefont {Lévêque}},\ and\
  \bibinfo {author} {\bibfnamefont {P.-E.}\ \bibnamefont {Roche}},\ }\bibfield
  {title} {\bibinfo {title} {Energy cascade and the four-fifths law in
  superfluid turbulence},\ }\href {https://doi.org/10.1209/0295-5075/97/34006}
  {\bibfield  {journal} {\bibinfo  {journal} {Europhysics Letters}\ }\textbf
  {\bibinfo {volume} {97}},\ \bibinfo {pages} {34006} (\bibinfo {year}
  {2012})}\BibitemShut {NoStop}%
\bibitem [{\citenamefont {Roche}\ \emph {et~al.}(2009)\citenamefont {Roche},
  \citenamefont {Barenghi},\ and\ \citenamefont {Leveque}}]{Roche2009}%
  \BibitemOpen
  \bibfield  {author} {\bibinfo {author} {\bibfnamefont {P.-E.}\ \bibnamefont
  {Roche}}, \bibinfo {author} {\bibfnamefont {C.~F.}\ \bibnamefont
  {Barenghi}},\ and\ \bibinfo {author} {\bibfnamefont {E.}~\bibnamefont
  {Leveque}},\ }\bibfield  {title} {\bibinfo {title} {Quantum turbulence at
  finite temperature: The two-fluids cascade},\ }\href
  {https://doi.org/10.1209/0295-5075/87/54006} {\bibfield  {journal} {\bibinfo
  {journal} {Europhysics Letters}\ }\textbf {\bibinfo {volume} {87}},\ \bibinfo
  {pages} {54006} (\bibinfo {year} {2009})}\BibitemShut {NoStop}%
\bibitem [{\citenamefont {Salort}\ \emph {et~al.}(2011)\citenamefont {Salort},
  \citenamefont {Roche},\ and\ \citenamefont
  {Leveque}}]{Salort_2011_equipartition}%
  \BibitemOpen
  \bibfield  {author} {\bibinfo {author} {\bibfnamefont {J.}~\bibnamefont
  {Salort}}, \bibinfo {author} {\bibfnamefont {P.-E.}\ \bibnamefont {Roche}},\
  and\ \bibinfo {author} {\bibfnamefont {E.}~\bibnamefont {Leveque}},\
  }\bibfield  {title} {\bibinfo {title} {Mesoscale equipartition of kinetic
  energy in quantum turbulence},\ }\href
  {https://doi.org/10.1209/0295-5075/94/24001} {\bibfield  {journal} {\bibinfo
  {journal} {Europhysics Letters}\ }\textbf {\bibinfo {volume} {94}},\ \bibinfo
  {pages} {24001} (\bibinfo {year} {2011})}\BibitemShut {NoStop}%
\bibitem [{\citenamefont {Vinen}\ and\ \citenamefont
  {Niemela}(2002)}]{Vinen2002}%
  \BibitemOpen
  \bibfield  {author} {\bibinfo {author} {\bibfnamefont {W.~F.}\ \bibnamefont
  {Vinen}}\ and\ \bibinfo {author} {\bibfnamefont {J.~J.}\ \bibnamefont
  {Niemela}},\ }\bibfield  {title} {\bibinfo {title} {Quantum turbulence},\
  }\href {https://doi.org/10.1023/A:1019695418590} {\bibfield  {journal}
  {\bibinfo  {journal} {Journal of Low Temperature Physics}\ }\textbf {\bibinfo
  {volume} {128}},\ \bibinfo {pages} {167} (\bibinfo {year}
  {2002})}\BibitemShut {NoStop}%
\bibitem [{\citenamefont {Barenghi}\ \emph {et~al.}(2014)\citenamefont
  {Barenghi}, \citenamefont {L’vov},\ and\ \citenamefont
  {Roche}}]{Barenghi14}%
  \BibitemOpen
  \bibfield  {author} {\bibinfo {author} {\bibfnamefont {C.~F.}\ \bibnamefont
  {Barenghi}}, \bibinfo {author} {\bibfnamefont {V.~S.}\ \bibnamefont
  {L’vov}},\ and\ \bibinfo {author} {\bibfnamefont {P.-E.}\ \bibnamefont
  {Roche}},\ }\bibfield  {title} {\bibinfo {title} {Experimental, numerical,
  and analytical velocity spectra in turbulent quantum fluid},\ }\href
  {https://doi.org/10.1073/pnas.1312548111} {\bibfield  {journal} {\bibinfo
  {journal} {Proceedings of the National Academy of Sciences}\ }\textbf
  {\bibinfo {volume} {111}},\ \bibinfo {pages} {4683} (\bibinfo {year}
  {2014})}\BibitemShut {NoStop}%
\bibitem [{\citenamefont {Stalp}\ \emph {et~al.}(1999)\citenamefont {Stalp},
  \citenamefont {Skrbek},\ and\ \citenamefont {Donnelly}}]{Stalp99}%
  \BibitemOpen
  \bibfield  {author} {\bibinfo {author} {\bibfnamefont {S.~R.}\ \bibnamefont
  {Stalp}}, \bibinfo {author} {\bibfnamefont {L.}~\bibnamefont {Skrbek}},\ and\
  \bibinfo {author} {\bibfnamefont {R.~J.}\ \bibnamefont {Donnelly}},\
  }\bibfield  {title} {\bibinfo {title} {Decay of grid turbulence in a finite
  channel},\ }\href {https://doi.org/10.1103/PhysRevLett.82.4831} {\bibfield
  {journal} {\bibinfo  {journal} {Phys. Rev. Lett.}\ }\textbf {\bibinfo
  {volume} {82}},\ \bibinfo {pages} {4831} (\bibinfo {year}
  {1999})}\BibitemShut {NoStop}%
\bibitem [{\citenamefont {Stalp}\ \emph {et~al.}(2002)\citenamefont {Stalp},
  \citenamefont {Niemela}, \citenamefont {Vinen},\ and\ \citenamefont
  {Donnelly}}]{Oregon_ill_grid}%
  \BibitemOpen
  \bibfield  {author} {\bibinfo {author} {\bibfnamefont {S.~R.}\ \bibnamefont
  {Stalp}}, \bibinfo {author} {\bibfnamefont {J.~J.}\ \bibnamefont {Niemela}},
  \bibinfo {author} {\bibfnamefont {W.~F.}\ \bibnamefont {Vinen}},\ and\
  \bibinfo {author} {\bibfnamefont {R.~J.}\ \bibnamefont {Donnelly}},\
  }\bibfield  {title} {\bibinfo {title} {Dissipation of grid turbulence in
  helium ii},\ }\href {https://doi.org/10.1063/1.1449902} {\bibfield  {journal}
  {\bibinfo  {journal} {Physics of Fluids}\ }\textbf {\bibinfo {volume} {14}},\
  \bibinfo {pages} {1377} (\bibinfo {year} {2002})}\BibitemShut {NoStop}%
\bibitem [{\citenamefont {Niemela}\ \emph {et~al.}(2005)\citenamefont
  {Niemela}, \citenamefont {Sreenivasan},\ and\ \citenamefont
  {Donnelly}}]{Oregon_good_grid}%
  \BibitemOpen
  \bibfield  {author} {\bibinfo {author} {\bibfnamefont {J.}~\bibnamefont
  {Niemela}}, \bibinfo {author} {\bibfnamefont {K.}~\bibnamefont
  {Sreenivasan}},\ and\ \bibinfo {author} {\bibfnamefont {R.}~\bibnamefont
  {Donnelly}},\ }\bibfield  {title} {\bibinfo {title} {Grid generated
  turbulence in helium ii},\ }\href {https://doi.org/10.1007/s10909-005-2253-8}
  {\bibfield  {journal} {\bibinfo  {journal} {Journal of Low Temperature
  Physics}\ }\textbf {\bibinfo {volume} {138}},\ \bibinfo {pages} {537}
  (\bibinfo {year} {2005})}\BibitemShut {NoStop}%
\bibitem [{\citenamefont {Babuin}\ \emph
  {et~al.}(2014{\natexlab{a}})\citenamefont {Babuin}, \citenamefont {Varga},\
  and\ \citenamefont {Skrbek}}]{Babuin2013}%
  \BibitemOpen
  \bibfield  {author} {\bibinfo {author} {\bibfnamefont {S.}~\bibnamefont
  {Babuin}}, \bibinfo {author} {\bibfnamefont {E.}~\bibnamefont {Varga}},\ and\
  \bibinfo {author} {\bibfnamefont {L.}~\bibnamefont {Skrbek}},\ }\bibfield
  {title} {\bibinfo {title} {The decay of forced turbulent coflow of he ii past
  a grid},\ }\href {https://doi.org/10.1007/s10909-013-0938-y} {\bibfield
  {journal} {\bibinfo  {journal} {Journal of Low Temperature Physics}\ }\textbf
  {\bibinfo {volume} {175}},\ \bibinfo {pages} {324} (\bibinfo {year}
  {2014}{\natexlab{a}})}\BibitemShut {NoStop}%
\bibitem [{\citenamefont {Walmsley}\ \emph {et~al.}(2014)\citenamefont
  {Walmsley}, \citenamefont {Zmeev}, \citenamefont {Pakpour},\ and\
  \citenamefont {Golov}}]{Walmsley14}%
  \BibitemOpen
  \bibfield  {author} {\bibinfo {author} {\bibfnamefont {P.}~\bibnamefont
  {Walmsley}}, \bibinfo {author} {\bibfnamefont {D.}~\bibnamefont {Zmeev}},
  \bibinfo {author} {\bibfnamefont {F.}~\bibnamefont {Pakpour}},\ and\ \bibinfo
  {author} {\bibfnamefont {A.}~\bibnamefont {Golov}},\ }\bibfield  {title}
  {\bibinfo {title} {Dynamics of quantum turbulence of different spectra},\
  }\href {https://doi.org/10.1073/pnas.1312544110} {\bibfield  {journal}
  {\bibinfo  {journal} {Proceedings of the National Academy of Sciences}\
  }\textbf {\bibinfo {volume} {111}},\ \bibinfo {pages} {4691} (\bibinfo {year}
  {2014})}\BibitemShut {NoStop}%
\bibitem [{\citenamefont {Babuin}\ \emph
  {et~al.}(2014{\natexlab{b}})\citenamefont {Babuin}, \citenamefont {Varga},
  \citenamefont {Skrbek}, \citenamefont {L{\'e}v{\^e}que},\ and\ \citenamefont
  {Roche}}]{Babuin14}%
  \BibitemOpen
  \bibfield  {author} {\bibinfo {author} {\bibfnamefont {S.}~\bibnamefont
  {Babuin}}, \bibinfo {author} {\bibfnamefont {E.}~\bibnamefont {Varga}},
  \bibinfo {author} {\bibfnamefont {L.}~\bibnamefont {Skrbek}}, \bibinfo
  {author} {\bibfnamefont {E.}~\bibnamefont {L{\'e}v{\^e}que}},\ and\ \bibinfo
  {author} {\bibfnamefont {P.-E.}\ \bibnamefont {Roche}},\ }\bibfield  {title}
  {\bibinfo {title} {Effective viscosity in quantum turbulence: A steady-state
  approach},\ }\href {https://doi.org/10.1209/0295-5075/106/24006} {\bibfield
  {journal} {\bibinfo  {journal} {Europhysics Letters}\ }\textbf {\bibinfo
  {volume} {106}},\ \bibinfo {pages} {24006} (\bibinfo {year}
  {2014}{\natexlab{b}})}\BibitemShut {NoStop}%
\bibitem [{\citenamefont {Guo}\ \emph {et~al.}(2014)\citenamefont {Guo},
  \citenamefont {Mantia}, \citenamefont {Lathrop},\ and\ \citenamefont
  {Sciver}}]{Guo14}%
  \BibitemOpen
  \bibfield  {author} {\bibinfo {author} {\bibfnamefont {W.}~\bibnamefont
  {Guo}}, \bibinfo {author} {\bibfnamefont {M.~L.}\ \bibnamefont {Mantia}},
  \bibinfo {author} {\bibfnamefont {D.~P.}\ \bibnamefont {Lathrop}},\ and\
  \bibinfo {author} {\bibfnamefont {S.~W.~V.}\ \bibnamefont {Sciver}},\
  }\bibfield  {title} {\bibinfo {title} {Visualization of two-fluid flows of
  superfluid helium-4},\ }\href {https://doi.org/10.1073/pnas.1312546111}
  {\bibfield  {journal} {\bibinfo  {journal} {Proceedings of the National
  Academy of Sciences}\ }\textbf {\bibinfo {volume} {111}},\ \bibinfo {pages}
  {4653} (\bibinfo {year} {2014})}\BibitemShut {NoStop}%
\bibitem [{\citenamefont {Švančara}\ and\ \citenamefont
  {La~Mantia}(2017)}]{Svancara17}%
  \BibitemOpen
  \bibfield  {author} {\bibinfo {author} {\bibfnamefont {P.}~\bibnamefont
  {Švančara}}\ and\ \bibinfo {author} {\bibfnamefont {M.}~\bibnamefont
  {La~Mantia}},\ }\bibfield  {title} {\bibinfo {title} {Flows of liquid 4he due
  to oscillating grids},\ }\href {https://doi.org/10.1017/jfm.2017.703}
  {\bibfield  {journal} {\bibinfo  {journal} {Journal of Fluid Mechanics}\
  }\textbf {\bibinfo {volume} {832}},\ \bibinfo {pages} {578–599} (\bibinfo
  {year} {2017})}\BibitemShut {NoStop}%
\bibitem [{\citenamefont {Peretti}\ \emph {et~al.}(2023)\citenamefont
  {Peretti}, \citenamefont {Vessaire}, \citenamefont {Durozoy},\ and\
  \citenamefont {Gibert}}]{Peretti23}%
  \BibitemOpen
  \bibfield  {author} {\bibinfo {author} {\bibfnamefont {C.}~\bibnamefont
  {Peretti}}, \bibinfo {author} {\bibfnamefont {J.}~\bibnamefont {Vessaire}},
  \bibinfo {author} {\bibfnamefont {E.}~\bibnamefont {Durozoy}},\ and\ \bibinfo
  {author} {\bibfnamefont {M.}~\bibnamefont {Gibert}},\ }\bibfield  {title}
  {\bibinfo {title} {Direct visualization of the quantum vortex lattice
  structure, oscillations, and destabilization in rotating 4he},\ }\href
  {https://doi.org/10.1126/sciadv.adh2899} {\bibfield  {journal} {\bibinfo
  {journal} {Science Advances}\ }\textbf {\bibinfo {volume} {9}},\ \bibinfo
  {pages} {eadh2899} (\bibinfo {year} {2023})}\BibitemShut {NoStop}%
\bibitem [{\citenamefont {Rousset}\ \emph {et~al.}(2015)\citenamefont
  {Rousset}, \citenamefont {Baudet}, \citenamefont {Bon-Mardion}, \citenamefont
  {Bourgoin}, \citenamefont {Braslau}, \citenamefont {Daviaud}, \citenamefont
  {Diribarne}, \citenamefont {Dubrulle}, \citenamefont {Gagne}, \citenamefont
  {Gallet}, \citenamefont {Gibert}, \citenamefont {Girard}, \citenamefont
  {Lehner}, \citenamefont {Moukharski},\ and\ \citenamefont {Sy}}]{Rousset15}%
  \BibitemOpen
  \bibfield  {author} {\bibinfo {author} {\bibfnamefont {B.}~\bibnamefont
  {Rousset}}, \bibinfo {author} {\bibfnamefont {C.}~\bibnamefont {Baudet}},
  \bibinfo {author} {\bibfnamefont {M.}~\bibnamefont {Bon-Mardion}}, \bibinfo
  {author} {\bibfnamefont {M.}~\bibnamefont {Bourgoin}}, \bibinfo {author}
  {\bibfnamefont {A.}~\bibnamefont {Braslau}}, \bibinfo {author} {\bibfnamefont
  {F.}~\bibnamefont {Daviaud}}, \bibinfo {author} {\bibfnamefont
  {P.}~\bibnamefont {Diribarne}}, \bibinfo {author} {\bibfnamefont
  {B.}~\bibnamefont {Dubrulle}}, \bibinfo {author} {\bibfnamefont
  {Y.}~\bibnamefont {Gagne}}, \bibinfo {author} {\bibfnamefont
  {B.}~\bibnamefont {Gallet}}, \bibinfo {author} {\bibfnamefont
  {M.}~\bibnamefont {Gibert}}, \bibinfo {author} {\bibfnamefont
  {A.}~\bibnamefont {Girard}}, \bibinfo {author} {\bibfnamefont
  {T.}~\bibnamefont {Lehner}}, \bibinfo {author} {\bibfnamefont
  {I.}~\bibnamefont {Moukharski}},\ and\ \bibinfo {author} {\bibfnamefont
  {F.}~\bibnamefont {Sy}},\ }\bibfield  {title} {\bibinfo {title} {{Cryogenic
  turbulence test facilities at CEA/SBT}},\ }\href
  {https://doi.org/10.1088/1757-899X/101/1/012187} {\bibfield  {journal}
  {\bibinfo  {journal} {{IOP Conference Series: Materials Science and
  Engineering}}\ }\textbf {\bibinfo {volume} {101}},\ \bibinfo {pages} {012187}
  (\bibinfo {year} {2015})}\BibitemShut {NoStop}%
\bibitem [{\citenamefont {Sy}(2016)}]{Sy16_phd}%
  \BibitemOpen
  \bibfield  {author} {\bibinfo {author} {\bibfnamefont {N.~F.~I.}\
  \bibnamefont {Sy}},\ }\emph {\bibinfo {title} {{Turbulence de grille
  oscillante {\`a} basses temp{\'e}ratures}}},\ \href
  {https://theses.hal.science/tel-01599365} {Ph.D. thesis},\ \bibinfo  {school}
  {{Universit{\'e} Grenoble Alpes}} (\bibinfo {year} {2016})\BibitemShut
  {NoStop}%
\bibitem [{\citenamefont {Sy}\ \emph {et~al.}(2021)\citenamefont {Sy},
  \citenamefont {Diribarne}, \citenamefont {Rousset}, \citenamefont {Gibert},\
  and\ \citenamefont {Bourgoin}}]{Sy21_article}%
  \BibitemOpen
  \bibfield  {author} {\bibinfo {author} {\bibfnamefont {F.}~\bibnamefont
  {Sy}}, \bibinfo {author} {\bibfnamefont {P.}~\bibnamefont {Diribarne}},
  \bibinfo {author} {\bibfnamefont {B.}~\bibnamefont {Rousset}}, \bibinfo
  {author} {\bibfnamefont {M.}~\bibnamefont {Gibert}},\ and\ \bibinfo {author}
  {\bibfnamefont {M.}~\bibnamefont {Bourgoin}},\ }\bibfield  {title} {\bibinfo
  {title} {Multiscale energy budget of inertially driven turbulence in normal
  and superfluid helium},\ }\href
  {https://doi.org/10.1103/PhysRevFluids.6.064604} {\bibfield  {journal}
  {\bibinfo  {journal} {Phys. Rev. Fluids}\ }\textbf {\bibinfo {volume} {6}},\
  \bibinfo {pages} {064604} (\bibinfo {year} {2021})}\BibitemShut {NoStop}%
\bibitem [{\citenamefont {Srdic}\ \emph {et~al.}(1996)\citenamefont {Srdic},
  \citenamefont {Fernando},\ and\ \citenamefont {Montenegro}}]{Srdic96}%
  \BibitemOpen
  \bibfield  {author} {\bibinfo {author} {\bibfnamefont {A.}~\bibnamefont
  {Srdic}}, \bibinfo {author} {\bibfnamefont {H.~J.~S.}\ \bibnamefont
  {Fernando}},\ and\ \bibinfo {author} {\bibfnamefont {L.}~\bibnamefont
  {Montenegro}},\ }\bibfield  {title} {\bibinfo {title} {Generation of nearly
  isotropic turbulence using two oscillating grids},\ }\href
  {https://doi.org/10.1007/BF00191022} {\bibfield  {journal} {\bibinfo
  {journal} {Experiments in Fluids}\ }\textbf {\bibinfo {volume} {20}},\
  \bibinfo {pages} {395} (\bibinfo {year} {1996})}\BibitemShut {NoStop}%
\bibitem [{\citenamefont {Shy}\ \emph {et~al.}(1997)\citenamefont {Shy},
  \citenamefont {Tang},\ and\ \citenamefont {Fann}}]{Shy97}%
  \BibitemOpen
  \bibfield  {author} {\bibinfo {author} {\bibfnamefont {S.}~\bibnamefont
  {Shy}}, \bibinfo {author} {\bibfnamefont {C.}~\bibnamefont {Tang}},\ and\
  \bibinfo {author} {\bibfnamefont {S.}~\bibnamefont {Fann}},\ }\bibfield
  {title} {\bibinfo {title} {A nearly isotropic turbulence generated by a pair
  of vibrating grids},\ }\href
  {https://doi.org/https://doi.org/10.1016/S0894-1777(96)00111-2} {\bibfield
  {journal} {\bibinfo  {journal} {Experimental Thermal and Fluid Science}\
  }\textbf {\bibinfo {volume} {14}},\ \bibinfo {pages} {251} (\bibinfo {year}
  {1997})}\BibitemShut {NoStop}%
\bibitem [{\citenamefont {McLennan}\ \emph {et~al.}(1932)\citenamefont
  {McLennan}, \citenamefont {Smith},\ and\ \citenamefont
  {Wilhelm}}]{McLennan32}%
  \BibitemOpen
  \bibfield  {author} {\bibinfo {author} {\bibfnamefont {J.~C.}\ \bibnamefont
  {McLennan}}, \bibinfo {author} {\bibfnamefont {H.~D.}\ \bibnamefont
  {Smith}},\ and\ \bibinfo {author} {\bibfnamefont {J.~O.}\ \bibnamefont
  {Wilhelm}},\ }\bibfield  {title} {\bibinfo {title} {Xiv. the scattering of
  light by liquid helium},\ }\href {https://doi.org/10.1080/14786443209462044}
  {\bibfield  {journal} {\bibinfo  {journal} {The London, Edinburgh, and Dublin
  Philosophical Magazine and Journal of Science}\ }\textbf {\bibinfo {volume}
  {14}},\ \bibinfo {pages} {161} (\bibinfo {year} {1932})}\BibitemShut
  {NoStop}%
\bibitem [{\citenamefont {Bret}\ \emph {et~al.}(2023)\citenamefont {Bret},
  \citenamefont {Chartier}, \citenamefont {Diribarne}, \citenamefont {Duplat},\
  and\ \citenamefont {Rousset}}]{Bret2023}%
  \BibitemOpen
  \bibfield  {author} {\bibinfo {author} {\bibfnamefont {C.}~\bibnamefont
  {Bret}}, \bibinfo {author} {\bibfnamefont {J.}~\bibnamefont {Chartier}},
  \bibinfo {author} {\bibfnamefont {P.}~\bibnamefont {Diribarne}}, \bibinfo
  {author} {\bibfnamefont {J.}~\bibnamefont {Duplat}},\ and\ \bibinfo {author}
  {\bibfnamefont {B.}~\bibnamefont {Rousset}},\ }\bibfield  {title} {\bibinfo
  {title} {{Controlled spherical deuterium droplets as Lagrangian tracers for
  cryogenic turbulence experiments}},\ }\href
  {https://doi.org/10.1063/5.0167132} {\bibfield  {journal} {\bibinfo
  {journal} {Review of Scientific Instruments}\ }\textbf {\bibinfo {volume}
  {94}},\ \bibinfo {pages} {105114} (\bibinfo {year} {2023})}\BibitemShut
  {NoStop}%
\bibitem [{\citenamefont {Bret}(2024)}]{Bret24_PHD}%
  \BibitemOpen
  \bibfield  {author} {\bibinfo {author} {\bibfnamefont {C.}~\bibnamefont
  {Bret}},\ }\emph {\bibinfo {title} {{Turbulence classique et quantique :
  Exp{\'e}riences et analyses comparatives}}},\ \href
  {https://hal.science/tel-04883645} {Ph.D. thesis},\ \bibinfo  {school}
  {{Universit{\'e} Grenoble Alpe}} (\bibinfo {year} {2024})\BibitemShut
  {NoStop}%
\bibitem [{\citenamefont {Helmli}\ and\ \citenamefont
  {Scherer}(2001)}]{Helmli01}%
  \BibitemOpen
  \bibfield  {author} {\bibinfo {author} {\bibfnamefont {F.}~\bibnamefont
  {Helmli}}\ and\ \bibinfo {author} {\bibfnamefont {S.}~\bibnamefont
  {Scherer}},\ }\bibfield  {title} {\bibinfo {title} {Adaptive shape from focus
  with an error estimation in light microscopy},\ }in\ \href
  {https://doi.org/10.1109/ISPA.2001.938626} {\emph {\bibinfo {booktitle} {ISPA
  2001. Proceedings of the 2nd International Symposium on Image and Signal
  Processing and Analysis. In conjunction with 23rd International Conference on
  Information Technology Interfaces (IEEE Cat.)}}}\ (\bibinfo  {publisher}
  {IEEE},\ \bibinfo {address} {New York},\ \bibinfo {year} {2001})\ pp.\
  \bibinfo {pages} {188--193}\BibitemShut {NoStop}%
\bibitem [{\citenamefont {Babuin}\ \emph {et~al.}(2012)\citenamefont {Babuin},
  \citenamefont {Stammeier}, \citenamefont {Varga}, \citenamefont {Rotter},\
  and\ \citenamefont {Skrbek}}]{Babuin12}%
  \BibitemOpen
  \bibfield  {author} {\bibinfo {author} {\bibfnamefont {S.}~\bibnamefont
  {Babuin}}, \bibinfo {author} {\bibfnamefont {M.}~\bibnamefont {Stammeier}},
  \bibinfo {author} {\bibfnamefont {E.}~\bibnamefont {Varga}}, \bibinfo
  {author} {\bibfnamefont {M.}~\bibnamefont {Rotter}},\ and\ \bibinfo {author}
  {\bibfnamefont {L.}~\bibnamefont {Skrbek}},\ }\bibfield  {title} {\bibinfo
  {title} {Quantum turbulence of bellows-driven ${}^{4}$he superflow: Steady
  state},\ }\href {https://doi.org/10.1103/PhysRevB.86.134515} {\bibfield
  {journal} {\bibinfo  {journal} {Phys. Rev. B}\ }\textbf {\bibinfo {volume}
  {86}},\ \bibinfo {pages} {134515} (\bibinfo {year} {2012})}\BibitemShut
  {NoStop}%
\bibitem [{\citenamefont {Varga}\ \emph {et~al.}(2019)\citenamefont {Varga},
  \citenamefont {Jackson}, \citenamefont {Schmoranzer},\ and\ \citenamefont
  {Skrbek}}]{Varga19}%
  \BibitemOpen
  \bibfield  {author} {\bibinfo {author} {\bibfnamefont {E.}~\bibnamefont
  {Varga}}, \bibinfo {author} {\bibfnamefont {M.}~\bibnamefont {Jackson}},
  \bibinfo {author} {\bibfnamefont {D.}~\bibnamefont {Schmoranzer}},\ and\
  \bibinfo {author} {\bibfnamefont {L.}~\bibnamefont {Skrbek}},\ }\bibfield
  {title} {\bibinfo {title} {The use of second sound in investigations of
  quantum turbulence in he ii},\ }\href
  {https://doi.org/10.1007/s10909-019-02208-4} {\bibfield  {journal} {\bibinfo
  {journal} {Journal of Low Temperature Physics}\ }\textbf {\bibinfo {volume}
  {197}} (\bibinfo {year} {2019})}\BibitemShut {NoStop}%
\bibitem [{\citenamefont {Woillez}\ \emph {et~al.}(2023)\citenamefont
  {Woillez}, \citenamefont {Valentin},\ and\ \citenamefont
  {Roche}}]{Woillez_sensors_23}%
  \BibitemOpen
  \bibfield  {author} {\bibinfo {author} {\bibfnamefont {E.}~\bibnamefont
  {Woillez}}, \bibinfo {author} {\bibfnamefont {J.}~\bibnamefont {Valentin}},\
  and\ \bibinfo {author} {\bibfnamefont {P.-E.}\ \bibnamefont {Roche}},\
  }\bibfield  {title} {\bibinfo {title} {{Second sound resonators and tweezers
  as vorticity or velocity probes: Fabrication, model, and method}},\ }\href
  {https://doi.org/10.1063/5.0152091} {\bibfield  {journal} {\bibinfo
  {journal} {Review of Scientific Instruments}\ }\textbf {\bibinfo {volume}
  {94}},\ \bibinfo {pages} {105109} (\bibinfo {year} {2023})}\BibitemShut
  {NoStop}%
\bibitem [{\citenamefont {Manthey}\ \emph {et~al.}(1992)\citenamefont
  {Manthey}, \citenamefont {Kroemer},\ and\ \citenamefont
  {Magori}}]{WManthey_1992}%
  \BibitemOpen
  \bibfield  {author} {\bibinfo {author} {\bibfnamefont {W.}~\bibnamefont
  {Manthey}}, \bibinfo {author} {\bibfnamefont {N.}~\bibnamefont {Kroemer}},\
  and\ \bibinfo {author} {\bibfnamefont {V.}~\bibnamefont {Magori}},\
  }\bibfield  {title} {\bibinfo {title} {Ultrasonic transducers and transducer
  arrays for applications in air},\ }\href
  {https://doi.org/10.1088/0957-0233/3/3/001} {\bibfield  {journal} {\bibinfo
  {journal} {Measurement Science and Technology}\ }\textbf {\bibinfo {volume}
  {3}},\ \bibinfo {pages} {249} (\bibinfo {year} {1992})}\BibitemShut {NoStop}%
\bibitem [{\citenamefont {Anke}(1974)}]{Anke74}%
  \BibitemOpen
  \bibfield  {author} {\bibinfo {author} {\bibfnamefont {D.}~\bibnamefont
  {Anke}},\ }\bibfield  {title} {\bibinfo {title} {An air sound transducer for
  frequencies in the range 50 khz to 100 khz using the sell principle},\ }\href
  {https://www.ingentaconnect.com/content/dav/aaua/1974/00000030/00000001/art00006}
  {\bibfield  {journal} {\bibinfo  {journal} {Acta Acustica united with
  Acustica}\ }\textbf {\bibinfo {volume} {30}},\ \bibinfo {pages} {30}
  (\bibinfo {year} {1974})}\BibitemShut {NoStop}%
\bibitem [{\citenamefont {Zimmermann}(1986)}]{Zimmermann86}%
  \BibitemOpen
  \bibfield  {author} {\bibinfo {author} {\bibfnamefont {W.}~\bibnamefont
  {Zimmermann}},\ }\bibfield  {title} {\bibinfo {title} {Porous-membrane
  second-sound transducers for superfluid $^{4}\mathrm{He}$},\ }\href
  {https://doi.org/10.1103/PhysRevB.33.139} {\bibfield  {journal} {\bibinfo
  {journal} {Phys. Rev. B}\ }\textbf {\bibinfo {volume} {33}},\ \bibinfo
  {pages} {139} (\bibinfo {year} {1986})}\BibitemShut {NoStop}%
\bibitem [{\citenamefont {Bertolaccini}\ \emph {et~al.}(2017)\citenamefont
  {Bertolaccini}, \citenamefont {L\'ev\^eque},\ and\ \citenamefont
  {Roche}}]{Bertolaccini17}%
  \BibitemOpen
  \bibfield  {author} {\bibinfo {author} {\bibfnamefont {J.}~\bibnamefont
  {Bertolaccini}}, \bibinfo {author} {\bibfnamefont {E.}~\bibnamefont
  {L\'ev\^eque}},\ and\ \bibinfo {author} {\bibfnamefont {P.-E.}\ \bibnamefont
  {Roche}},\ }\bibfield  {title} {\bibinfo {title} {Disproportionate entrance
  length in superfluid flows and the puzzle of counterflow instabilities},\
  }\href {https://doi.org/10.1103/PhysRevFluids.2.123902} {\bibfield  {journal}
  {\bibinfo  {journal} {Phys. Rev. Fluids}\ }\textbf {\bibinfo {volume} {2}},\
  \bibinfo {pages} {123902} (\bibinfo {year} {2017})}\BibitemShut {NoStop}%
\bibitem [{\citenamefont {Donnelly}\ and\ \citenamefont
  {Barenghi}(1998)}]{Donnelly98}%
  \BibitemOpen
  \bibfield  {author} {\bibinfo {author} {\bibfnamefont {R.~J.}\ \bibnamefont
  {Donnelly}}\ and\ \bibinfo {author} {\bibfnamefont {C.~F.}\ \bibnamefont
  {Barenghi}},\ }\bibfield  {title} {\bibinfo {title} {{The Observed Properties
  of Liquid Helium at the Saturated Vapor Pressure}},\ }\href
  {https://doi.org/10.1063/1.556028} {\bibfield  {journal} {\bibinfo  {journal}
  {Journal of Physical and Chemical Reference Data}\ }\textbf {\bibinfo
  {volume} {27}},\ \bibinfo {pages} {1217} (\bibinfo {year}
  {1998})}\BibitemShut {NoStop}%
\bibitem [{Note1()}]{Note1}%
  \BibitemOpen
  \bibinfo {note} {As mentioned in Sec.~\ref {sec:secsoundcavity}, a shift of
  \SI {8}{\micro bar} is sufficient to shift the resonance frequency by
  approximately \SI {1}{Hz}. The resulting phase shift depends on the damping
  coefficient $\alpha $, but for the initial value $\alpha _0=\SI
  {1.5}{m^{-1}}$, exciting the cavity \SI {1}{Hz} off-resonance induces a phase
  shift of about \SI {15}{\degree } compared to the phase at exact
  resonance.}\BibitemShut {Stop}%
\bibitem [{\citenamefont {Mordant}\ \emph {et~al.}(2004)\citenamefont
  {Mordant}, \citenamefont {Crawford},\ and\ \citenamefont
  {Bodenschatz}}]{Mordant04}%
  \BibitemOpen
  \bibfield  {author} {\bibinfo {author} {\bibfnamefont {N.}~\bibnamefont
  {Mordant}}, \bibinfo {author} {\bibfnamefont {A.}~\bibnamefont {Crawford}},\
  and\ \bibinfo {author} {\bibfnamefont {E.}~\bibnamefont {Bodenschatz}},\
  }\bibfield  {title} {\bibinfo {title} {Experimental lagrangian acceleration
  probability density function measurement},\ }\href
  {https://doi.org/https://doi.org/10.1016/j.physd.2004.01.041} {\bibfield
  {journal} {\bibinfo  {journal} {Physica D: Nonlinear Phenomena}\ }\textbf
  {\bibinfo {volume} {193}},\ \bibinfo {pages} {245} (\bibinfo {year}
  {2004})},\ \bibinfo {note} {anomalous distributions, nonlinear dynamics, and
  nonextensivity}\BibitemShut {NoStop}%
\bibitem [{\citenamefont {Švančara}\ and\ \citenamefont
  {La~Mantia}(2019)}]{Svancara2019_grid}%
  \BibitemOpen
  \bibfield  {author} {\bibinfo {author} {\bibfnamefont {P.}~\bibnamefont
  {Švančara}}\ and\ \bibinfo {author} {\bibfnamefont {M.}~\bibnamefont
  {La~Mantia}},\ }\bibfield  {title} {\bibinfo {title} {Flight-crash events in
  superfluid turbulence},\ }\href {https://doi.org/10.1017/jfm.2019.586}
  {\bibfield  {journal} {\bibinfo  {journal} {Journal of Fluid Mechanics}\
  }\textbf {\bibinfo {volume} {876}},\ \bibinfo {pages} {R2} (\bibinfo {year}
  {2019})}\BibitemShut {NoStop}%
\bibitem [{\citenamefont {Tang}\ \emph {et~al.}(2020)\citenamefont {Tang},
  \citenamefont {Bao}, \citenamefont {Kanai},\ and\ \citenamefont
  {Guo}}]{Tang20}%
  \BibitemOpen
  \bibfield  {author} {\bibinfo {author} {\bibfnamefont {Y.}~\bibnamefont
  {Tang}}, \bibinfo {author} {\bibfnamefont {S.}~\bibnamefont {Bao}}, \bibinfo
  {author} {\bibfnamefont {T.}~\bibnamefont {Kanai}},\ and\ \bibinfo {author}
  {\bibfnamefont {W.}~\bibnamefont {Guo}},\ }\bibfield  {title} {\bibinfo
  {title} {Statistical properties of homogeneous and isotropic turbulence in he
  ii measured via particle tracking velocimetry},\ }\href
  {https://doi.org/10.1103/PhysRevFluids.5.084602} {\bibfield  {journal}
  {\bibinfo  {journal} {Phys. Rev. Fluids}\ }\textbf {\bibinfo {volume} {5}},\
  \bibinfo {pages} {084602} (\bibinfo {year} {2020})}\BibitemShut {NoStop}%
\bibitem [{\citenamefont {Matsunaga}\ \emph {et~al.}(1999)\citenamefont
  {Matsunaga}, \citenamefont {Sugihara}, \citenamefont {Komatsu},\ and\
  \citenamefont {Masuda}}]{Matsunaga99}%
  \BibitemOpen
  \bibfield  {author} {\bibinfo {author} {\bibfnamefont {N.}~\bibnamefont
  {Matsunaga}}, \bibinfo {author} {\bibfnamefont {Y.}~\bibnamefont {Sugihara}},
  \bibinfo {author} {\bibfnamefont {T.}~\bibnamefont {Komatsu}},\ and\ \bibinfo
  {author} {\bibfnamefont {A.}~\bibnamefont {Masuda}},\ }\bibfield  {title}
  {\bibinfo {title} {Quantitative properties of oscillating-grid turbulence in
  a homogeneous fluid},\ }\href {https://doi.org/10.1016/S0169-5983(98)00034-3}
  {\bibfield  {journal} {\bibinfo  {journal} {Fluid Dynamics Research}\
  }\textbf {\bibinfo {volume} {25}},\ \bibinfo {pages} {147} (\bibinfo {year}
  {1999})}\BibitemShut {NoStop}%
\bibitem [{\citenamefont {Hopfinger}\ and\ \citenamefont
  {Toly}(1976)}]{Hopfinger76}%
  \BibitemOpen
  \bibfield  {author} {\bibinfo {author} {\bibfnamefont {E.~J.}\ \bibnamefont
  {Hopfinger}}\ and\ \bibinfo {author} {\bibfnamefont {J.-A.}\ \bibnamefont
  {Toly}},\ }\bibfield  {title} {\bibinfo {title} {Spatially decaying
  turbulence and its relation to mixing across density interfaces},\ }\href
  {https://doi.org/10.1017/S0022112076002371} {\bibfield  {journal} {\bibinfo
  {journal} {Journal of Fluid Mechanics}\ }\textbf {\bibinfo {volume} {78}},\
  \bibinfo {pages} {155–175} (\bibinfo {year} {1976})}\BibitemShut {NoStop}%
\bibitem [{\citenamefont {De~Silva}\ and\ \citenamefont
  {Fernando}(1994)}]{DeSilva94}%
  \BibitemOpen
  \bibfield  {author} {\bibinfo {author} {\bibfnamefont {I.~P.~D.}\
  \bibnamefont {De~Silva}}\ and\ \bibinfo {author} {\bibfnamefont {H.~J.~S.}\
  \bibnamefont {Fernando}},\ }\bibfield  {title} {\bibinfo {title}
  {{Oscillating grids as a source of nearly isotropic turbulence}},\ }\href
  {https://doi.org/10.1063/1.868193} {\bibfield  {journal} {\bibinfo  {journal}
  {Physics of Fluids}\ }\textbf {\bibinfo {volume} {6}},\ \bibinfo {pages}
  {2455} (\bibinfo {year} {1994})}\BibitemShut {NoStop}%
\bibitem [{\citenamefont {Ura}\ \emph {et~al.}(1987)\citenamefont {Ura},
  \citenamefont {Komatsu},\ and\ \citenamefont {Matsunaga}}]{Ura87}%
  \BibitemOpen
  \bibfield  {author} {\bibinfo {author} {\bibfnamefont {M.}~\bibnamefont
  {Ura}}, \bibinfo {author} {\bibfnamefont {T.}~\bibnamefont {Komatsu}},\ and\
  \bibinfo {author} {\bibfnamefont {N.}~\bibnamefont {Matsunaga}},\ }\bibfield
  {title} {\bibinfo {title} {Entrainment due to oscillating-grid turbulence in
  two-layered fluid},\ }in\ \href@noop {} {\emph {\bibinfo {booktitle}
  {Turbulence Measurements and Flow Modeling}}}\ (\bibinfo  {publisher}
  {Hemisphere Publishing Corporation},\ \bibinfo {address} {Washington},\
  \bibinfo {year} {1987})\ pp.\ \bibinfo {pages} {109--118}\BibitemShut
  {NoStop}%
\bibitem [{\citenamefont {Poulain-Zarcos}\ \emph {et~al.}(2022)\citenamefont
  {Poulain-Zarcos}, \citenamefont {Mercier},\ and\ \citenamefont {ter
  Halle}}]{Poulain22}%
  \BibitemOpen
  \bibfield  {author} {\bibinfo {author} {\bibfnamefont {M.}~\bibnamefont
  {Poulain-Zarcos}}, \bibinfo {author} {\bibfnamefont {M.~J.}\ \bibnamefont
  {Mercier}},\ and\ \bibinfo {author} {\bibfnamefont {A.}~\bibnamefont {ter
  Halle}},\ }\bibfield  {title} {\bibinfo {title} {Global characterization of
  oscillating grid turbulence in homogeneous and two-layer fluids, and its
  implication for mixing at high peclet number},\ }\href
  {https://doi.org/10.1103/PhysRevFluids.7.054606} {\bibfield  {journal}
  {\bibinfo  {journal} {Phys. Rev. Fluids}\ }\textbf {\bibinfo {volume} {7}},\
  \bibinfo {pages} {054606} (\bibinfo {year} {2022})}\BibitemShut {NoStop}%
\bibitem [{\citenamefont {Pope}(2000)}]{Pope00}%
  \BibitemOpen
  \bibfield  {author} {\bibinfo {author} {\bibfnamefont {S.~B.}\ \bibnamefont
  {Pope}},\ }\href@noop {} {\emph {\bibinfo {title} {Turbulent Flows}}}\
  (\bibinfo  {publisher} {Cambridge University Press},\ \bibinfo {address}
  {Cambridge, England},\ \bibinfo {year} {2000})\BibitemShut {NoStop}%
\bibitem [{\citenamefont {Moisy}\ \emph {et~al.}(1999)\citenamefont {Moisy},
  \citenamefont {Tabeling},\ and\ \citenamefont {Willaime}}]{Moisy99}%
  \BibitemOpen
  \bibfield  {author} {\bibinfo {author} {\bibfnamefont {F.}~\bibnamefont
  {Moisy}}, \bibinfo {author} {\bibfnamefont {P.}~\bibnamefont {Tabeling}},\
  and\ \bibinfo {author} {\bibfnamefont {H.}~\bibnamefont {Willaime}},\
  }\bibfield  {title} {\bibinfo {title} {Kolmogorov equation in a fully
  developed turbulence experiment},\ }\href
  {https://doi.org/10.1103/PhysRevLett.82.3994} {\bibfield  {journal} {\bibinfo
   {journal} {Phys. Rev. Lett.}\ }\textbf {\bibinfo {volume} {82}},\ \bibinfo
  {pages} {3994} (\bibinfo {year} {1999})}\BibitemShut {NoStop}%
\bibitem [{\citenamefont {Ijsselstein}\ \emph {et~al.}(1979)\citenamefont
  {Ijsselstein}, \citenamefont {{de Goeje}},\ and\ \citenamefont
  {Kramers}}]{Ijsselstein79}%
  \BibitemOpen
  \bibfield  {author} {\bibinfo {author} {\bibfnamefont {R.}~\bibnamefont
  {Ijsselstein}}, \bibinfo {author} {\bibfnamefont {M.}~\bibnamefont {{de
  Goeje}}},\ and\ \bibinfo {author} {\bibfnamefont {H.}~\bibnamefont
  {Kramers}},\ }\bibfield  {title} {\bibinfo {title} {Combined measurements on
  various types of stationary flow of superfluid 4he},\ }\href
  {https://doi.org/https://doi.org/10.1016/0378-4363(79)90014-7} {\bibfield
  {journal} {\bibinfo  {journal} {Physica B+C}\ }\textbf {\bibinfo {volume}
  {96}},\ \bibinfo {pages} {312} (\bibinfo {year} {1979})}\BibitemShut
  {NoStop}%
\bibitem [{\citenamefont {Walstrom}(1988)}]{Walstrom88b}%
  \BibitemOpen
  \bibfield  {author} {\bibinfo {author} {\bibfnamefont {P.~L.}\ \bibnamefont
  {Walstrom}},\ }\bibfield  {title} {\bibinfo {title} {Heat transfer by
  internal convection in turbulent he ii forced flow},\ }\href
  {https://doi.org/10.1007/BF00683569} {\bibfield  {journal} {\bibinfo
  {journal} {Journal of Low Temperature Physics}\ }\textbf {\bibinfo {volume}
  {73}},\ \bibinfo {pages} {391} (\bibinfo {year} {1988})}\BibitemShut
  {NoStop}%
\bibitem [{\citenamefont {Holmes}\ and\ \citenamefont
  {Van~Sciver}(1992)}]{Holmes92}%
  \BibitemOpen
  \bibfield  {author} {\bibinfo {author} {\bibfnamefont {D.~S.}\ \bibnamefont
  {Holmes}}\ and\ \bibinfo {author} {\bibfnamefont {S.~W.}\ \bibnamefont
  {Van~Sciver}},\ }\bibfield  {title} {\bibinfo {title} {Attenuation of second
  sound in bulk-flowing he ii},\ }\bibfield  {journal} {\bibinfo  {journal}
  {Journal of Low Temperature Physics; (United States)}\ }\textbf {\bibinfo
  {volume} {87:1-2}},\ \href {https://doi.org/10.1007/BF00141568}
  {10.1007/BF00141568} (\bibinfo {year} {1992})\BibitemShut {NoStop}%
\bibitem [{\citenamefont {Chagovets}\ \emph {et~al.}(2007)\citenamefont
  {Chagovets}, \citenamefont {Gordeev},\ and\ \citenamefont
  {Skrbek}}]{Chagovet07}%
  \BibitemOpen
  \bibfield  {author} {\bibinfo {author} {\bibfnamefont {T.~V.}\ \bibnamefont
  {Chagovets}}, \bibinfo {author} {\bibfnamefont {A.~V.}\ \bibnamefont
  {Gordeev}},\ and\ \bibinfo {author} {\bibfnamefont {L.}~\bibnamefont
  {Skrbek}},\ }\bibfield  {title} {\bibinfo {title} {Effective kinematic
  viscosity of turbulent $\mathrm{He}\phantom{\rule{0.3em}{0ex}}\mathrm{II}$},\
  }\href {https://doi.org/10.1103/PhysRevE.76.027301} {\bibfield  {journal}
  {\bibinfo  {journal} {Phys. Rev. E}\ }\textbf {\bibinfo {volume} {76}},\
  \bibinfo {pages} {027301} (\bibinfo {year} {2007})}\BibitemShut {NoStop}%
\bibitem [{\citenamefont {Vinen}(2005)}]{Vinen05}%
  \BibitemOpen
  \bibfield  {author} {\bibinfo {author} {\bibfnamefont {W.~F.}\ \bibnamefont
  {Vinen}},\ }\bibfield  {title} {\bibinfo {title} {How is turbulent energy
  dissipated in a superfluid?},\ }\href
  {https://doi.org/10.1088/0953-8984/17/45/006} {\bibfield  {journal} {\bibinfo
   {journal} {Journal of Physics: Condensed Matter}\ }\textbf {\bibinfo
  {volume} {17}},\ \bibinfo {pages} {S3231} (\bibinfo {year}
  {2005})}\BibitemShut {NoStop}%
\bibitem [{\citenamefont {Iyer}\ \emph {et~al.}(2019)\citenamefont {Iyer},
  \citenamefont {Sreenivasan},\ and\ \citenamefont {Yeung}}]{Iyer19}%
  \BibitemOpen
  \bibfield  {author} {\bibinfo {author} {\bibfnamefont {K.~P.}\ \bibnamefont
  {Iyer}}, \bibinfo {author} {\bibfnamefont {K.~R.}\ \bibnamefont
  {Sreenivasan}},\ and\ \bibinfo {author} {\bibfnamefont {P.~K.}\ \bibnamefont
  {Yeung}},\ }\bibfield  {title} {\bibinfo {title} {Circulation in high
  reynolds number isotropic turbulence is a bifractal},\ }\href
  {https://doi.org/10.1103/PhysRevX.9.041006} {\bibfield  {journal} {\bibinfo
  {journal} {Phys. Rev. X}\ }\textbf {\bibinfo {volume} {9}},\ \bibinfo {pages}
  {041006} (\bibinfo {year} {2019})}\BibitemShut {NoStop}%
\bibitem [{\citenamefont {Polanco}\ \emph {et~al.}(2025)\citenamefont
  {Polanco}, \citenamefont {Roche}, \citenamefont {Danaila},\ and\
  \citenamefont {Lévêque}}]{Polanco25}%
  \BibitemOpen
  \bibfield  {author} {\bibinfo {author} {\bibfnamefont {J.~I.}\ \bibnamefont
  {Polanco}}, \bibinfo {author} {\bibfnamefont {P.-E.}\ \bibnamefont {Roche}},
  \bibinfo {author} {\bibfnamefont {L.}~\bibnamefont {Danaila}},\ and\ \bibinfo
  {author} {\bibfnamefont {E.}~\bibnamefont {Lévêque}},\ }\bibfield  {title}
  {\bibinfo {title} {Disentangling temperature and reynolds number effects in
  quantum turbulence},\ }\href {https://doi.org/10.1073/pnas.2426598122}
  {\bibfield  {journal} {\bibinfo  {journal} {Proceedings of the National
  Academy of Sciences}\ }\textbf {\bibinfo {volume} {122}},\ \bibinfo {pages}
  {e2426598122} (\bibinfo {year} {2025})}\BibitemShut {NoStop}%
\bibitem [{\citenamefont {McKenna}\ and\ \citenamefont
  {McGillis}(2004)}]{McKenna04}%
  \BibitemOpen
  \bibfield  {author} {\bibinfo {author} {\bibfnamefont {S.~P.}\ \bibnamefont
  {McKenna}}\ and\ \bibinfo {author} {\bibfnamefont {W.~R.}\ \bibnamefont
  {McGillis}},\ }\bibfield  {title} {\bibinfo {title} {{Observations of flow
  repeatability and secondary circulation in an oscillating grid-stirred
  tank}},\ }\href {https://doi.org/10.1063/1.1779671} {\bibfield  {journal}
  {\bibinfo  {journal} {Physics of Fluids}\ }\textbf {\bibinfo {volume} {16}},\
  \bibinfo {pages} {3499} (\bibinfo {year} {2004})}\BibitemShut {NoStop}%
\end{thebibliography}%
\end{document}